\documentclass[a4paper,12pt]{article}

\usepackage{verbatim}
\usepackage{graphicx} 
\usepackage{epsfig}
\usepackage{dcolumn} 
\usepackage{bm} 
\usepackage{amsmath}
\usepackage{amssymb}
\usepackage{fancyhdr}

\usepackage{natbib}
\usepackage{har2nat}

\newcommand{\ep}{\varepsilon}
\newcommand{\up}{\uparrow}
\newcommand{\dn}{\downarrow}
\newcommand{\vectg}[1]{\mbox{\boldmath ${#1}$}}
\newcommand{\vect}[1]{{\bf #1}}

\begin{document}

\begin{center}
{\Large \textbf{SPIN CURRENTS IN SEMICONDUCTOR NANOSTRUCTURES: A NONEQUILIBRIUM GREEN-FUNCTION APPROACH}\\}

\vspace{0.5cm}
Branislav K. Nikoli\'{c},$^1$ Liviu P. Z\^{a}rbo,$^2$ and Satofumi Souma$^3$

\vspace{0.3cm}
$^1${\em Department of Physics and Astronomy, University of Delaware, Newark DE 19716, USA} \\
$^2${\em Department of Physics, Texas A\&M University, College Station TX 77843-4242, USA} \\
$^3${\em Department of Electronics and Electrical Engineering, Kobe University, 1-1 Rokkodai, Nada, Kobe 657-8501, Japan}

\vspace{1.5cm}

Chapter 24, page 814--866 in Volume I of \\ {\em The Oxford Handbook on Nanoscience and Technology: Frontiers and Advances} \\ Eds. A. V. Narlikar and Y. Y. Fu (Oxford University Press, Oxford, 2010)
\end{center}

\vspace{1.5cm}

\thispagestyle{empty}

\tableofcontents

\newpage

\section{INTRODUCTION} \label{sec:intro}
Over the past two decades, {\em spintronics}~\cite{vZuti'2004} has emerged as one of the most
vigorously pursued areas of condensed matter physics, materials science, and nanotechnology. A glimpse
at the Oxford English Dictionary reveals the following attempt to define the field succinctly: {\em ``spintronics
is a branch of physics concerned with the storage and transfer of information by means of electron spins
in addition to electron charge as in conventional electronics.''} The rise of spintronics was ignited
by basic research on ferromagnet/normal-metal multilayers in late 1980s~\cite{Maekawa2002}, as recognized by the
Nobel Prize in Physics for 2007 being awarded to A. Fert and P. Gr\" unberg for the discovery of giant magnetoresistance
(GMR). The GMR phenomenon also exemplifies one of  the fastest transfers of basic research results in condensed matter physics
into applications where in less then ten years since its discovery it has revolutionized information storage
technologies by enabling 100 times increase in hard disk storage capacity.

In recent years frontiers of spintronics have been reshaped through several intertwined lines of research:
({\em i}) ferromagnetic metal devices where the main theme is manipulation of magnetization
via electric currents and vice versa~\cite{Ralph2008}; ({\em ii}) ferromagnetic semiconductors which, unlike metal ferromagnets,
offer additional possibilities to manipulate their magnetic ordering (such as Curie temperature, coercive fields, and magnetic
dopants), but are still below optimal operating temperature~\cite{Jungwirth2006}; ({\em iii}) paramagnetic semiconductor
spintronics~\cite{Awschalom2007} largely focused on {\em all-electrical} manipulation of spins via spin-orbit (SO) coupling effects
in solids~\cite{Fabian2007}; and ({\em iv}) spins in semiconductors as building blocks of futuristic solid-state-based quantum
computers~\cite{Hanson2007a}. Unlike early non-coherent spintronics phenomena (such as GMR),
the major themes of the ``second-generation'' spintronics~\cite{Awschalom2007} are moving toward the spin coherent realm where spin component persists
in the direction transverse to external or effective internal magnetic fields.
%
\begin{figure}[t]
\centerline{\psfig{file=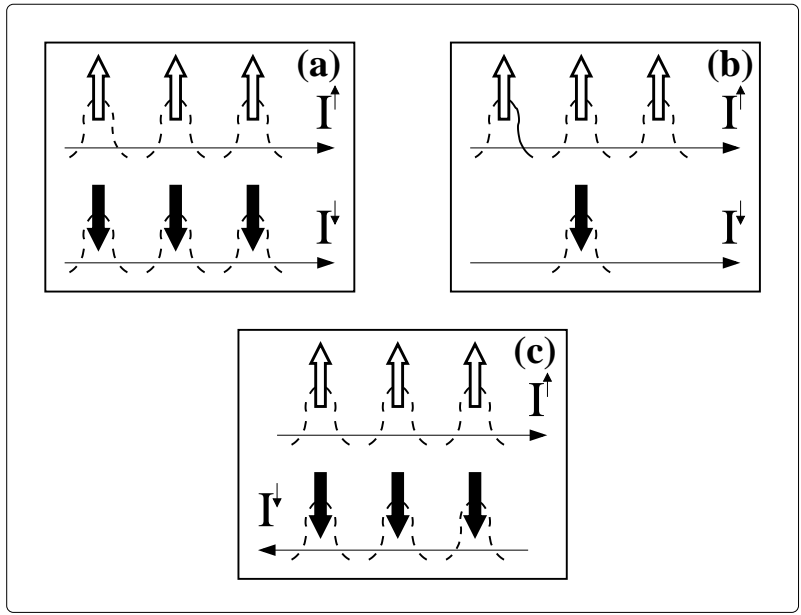,scale=0.4,angle=0} }
\caption{The classification of spin $I^S$ and charge $I$ currents
in metal and semiconductor spintronic systems corresponding to
spatial propagation of spin-$\up$ and spin-$\dn$ electronic wave packets
carrying spin-resolved currents $I^\up$ and $I^\dn$: (a) conventional charge current $I=I^\up + I^\dn \neq 0$ is
spin-unpolarized $I^S=\frac{\hbar}{2e}(I^\up -I^\dn) \equiv 0$; (b)
spin-polarized  charge current $I \neq 0$ is accompanied also by
spin current $I^S \neq 0$; and (c) {\em pure} spin current
$I^S=\frac{\hbar}{2e}(I^\up -I^\dn) \neq 0$ arising
when spin-$\up$ electrons move in one direction, while an
equal number of spin-$\dn$ electrons move in the opposite direction,
so that total charge current is $I \equiv 0$.}
\label{fig:spin_currents}
\end{figure}
%
Recent experiments exploring  such phenomena include: spin-transfer torque where spin current of large enough density injected into a ferromagnetic layer either switches its magnetization from one static configuration to another or generates a dynamical situation with steady-state precessing magnetization~\cite{Ralph2008}; spin pumping as the ``inverse'' effect of spin-transfer torque in which precessing magnetization of a ferromagnetic layer emits pure spin currents into adjacent normal metal layers in the absence of any bias voltage~\cite{Saitoh2006,Costache2006,Moriyama2008}; transport of coherent  spins (able to precess in the external magnetic fields) across $\sim 100$ $\mu$m thick silicon wafers~\cite{Huang2007}; and the direct and inverse spin-Hall effects (SHE) in bulk~\cite{Kato2004b} and low-dimensional~\cite{Wunderlich2005,Sih2005a} semiconductors and metals~\cite{Valenzuela2006,Saitoh2006,Kimura2007} [SHE in both metals and semiconductors has been observed even at room temperature~\cite{Stern2006,Kimura2007}]. A closely related effort that permeates these subfields is the generation and detection of {\em pure spin currents}~\cite{Nagaosa2008} which do not transport any net charge, as illustrated in Fig.~\ref{fig:spin_currents}. Their harnessing is expected to offer both new functionality and greatly reduced power dissipation.\footnote{The Joule heat losses induced by the current flow set the most important limits~\cite{Keyes2005} for conventional electronics, as well as for hybrid electronic-spintronic or purely spintronic devices envisioned to perform both storage and information processing on a single chip.}

In this chapter we discuss how different tools of quantum transport theory, based on the nonequilibrium Green  function (NEGF) techniques,\footnote{The NEGF techniques for finite-size devices have been developed over the past three decades mainly through the studies of charge currents of non-interacting quasiparticles in mesoscopic semiconductor~\cite{Datta1995} and nanoscopic molecular~\cite{Koentopp2008} systems where quantum-coherent effects on electron transport are the dominant mechanism because of the smallness of their size. Current frontiers of the NEGF theory are also concerned with the inclusion of many-particle interaction effects responsible for dephasing~\cite{Okamoto2007,Thygesen2008}. For a lucid introduction to the general scope of NEGF formalism applied to finite-size devices see the chapter 1 by S. Datta in this Volume, as well as the chapter 23 by K. S. Thygesen and A. Rubio in the same Volume focusing on the inclusion of electronic correlations in NEGF applied to molecular junctions.} can be extended to treat spin currents and spin densities in {\em realistic} open paramagnetic semiconductor devices out of equilibrium. In such devices, the most important spin-dependent interaction is the SO coupling stemming from relativistic corrections to the 
Pauli-Schr\"{o}dinger dynamics of spin-$\frac{1}{2}$ electrons. Recent theoretical efforts to understand spin transport in the presence of SO couplings by using conventional approaches, such as the bulk conductivity of infinite homogeneous systems [computed via the Kubo formula~\cite{Murakami2003a,Sinova2004} or the kinetic equation~\cite{Mishchenko2004}] or spin-density\footnote{Spin densities within the sample and spin accumulation along the sample boundaries are typically probed in recent SHE experiments on semiconductors~\cite{Kato2004b,Wunderlich2005,Sih2005a,Sih2006a}.} diffusion equations for bounded systems~\cite{Burkov2004,Bleibaum2006,Galitski2006}, have encountered enormous challenges even when treating non-interacting quasiparticles. Such intricacies  in systems with the intrinsic SO couplings, that act homogeneously throughout the sample, can be traced to spin non-conservation due to spin precession which leads to ambiguity in defining spin currents~\cite{Shi2006,Sugimoto2006} in the bulk  or ambiguity in supplying the boundary conditions for the diffusion equations~\cite{Bleibaum2006,Galitski2006}.

On the other hand, spin-resolved NEGF techniques discussed in this chapter offer consistent description of both phase-coherent (at low temperatures) and semiclassical (at finite temperatures where dephasing takes place) coupled spin and charge transport in both clean and disordered realistic finite-size devices attached to external current and voltage probes, as encountered in experiments. The physical quantities that can be computed within this framework yield  experimentally testable predictions for outflowing spin currents, induced voltages by their flow, and spin densities within the device.  The presentation is tailored to be mostly of a tutorial style, introducing the essential theoretical formalism and practical computational techniques at an accessible level that should make it possible for graduate students and non-specialists in physics and engineering to engaged in theoretical and computational modeling of nanospintronic devices. We illustrate formal developments with examples drawn from the filed of the  {\em mesoscopic SHE}~\cite{Nikoli'c2005b,Sheng2006b,Hankiewicz2004a,Ren2006,Bardarson2007,Silvestrov2009} in low-dimensional SO-coupled semiconductor nanostructures.\footnote{After learning about the NEGF techniques for spin transport discussed in this chapter, an interested reader might enter the field by trying to reproduce many other examples treated within this framework in our journal articles~\cite{Souma2004,Nikoli'c2005,Nikoli'c2005b,Souma2005a,Nikoli'c2005d,Nikoli'c2006,Nikoli'c2007,Dragomirova2007,Zarbo2007,Dragomirova2008,Nikoli'c2009,Chen2009}.}

\section{WHAT IS PURE SPIN CURRENT?}\label{sec:pure}

Pure spin current represents flow of spin angular momentum which is not accompanied by any
net charge transport [Fig.~\ref{fig:spin_currents}(c)]. They can be contrasted with traditional electronic circuits where equal number of spin-$\up$ and spin-$\dn$ electrons propagate in the same direction, so that total charge current in that direction $I=I^\up + I^\dn$ is unpolarized $I^S = 0$ [Fig.~\ref{fig:spin_currents}(a)]. Spin currents are substantially different from familiar charge currents in two key aspects: they are time-reversal invariant and they transport a vector quantity. In metal spintronic devices, ferromagnetic elements polarize electron spin thereby leading to a difference in charge currents of spin-$\up$ and spin-$\dn$ electrons [Fig.~\ref{fig:spin_currents}(b)]. Such spin-polarized charge currents are accompanied by a net spin current $I^S = \frac{\hbar}{2e}(I^\up - I^\dn) \neq 0$, as created and detected
in magnetic multilayers~\cite{Maekawa2002}. Figure~\ref{fig:spin_currents} also provides a transparent illustration of one of the major advantages of pure spin currents over the spin-polarized charge currents employed by the ``first-generation'' spintronic devices. For example, to transport information via $2 \times \hbar/2$ spin angular momenta, Fig.~\ref{fig:spin_currents}(b)
utilizes four electrons. In Fig.~\ref{fig:spin_currents}(c) the same transport of $2 \times \hbar/2$ is achieved using only two electrons moving in opposite direction. The Joule heat loss in the latter situation is only 25\% of the dissipative losses in the former case.

\section{HOW CAN PURE SPIN CURRENTS BE GENERATED AND DETECTED?}

Among a plethora of imaginative theoretical proposals~\cite{Sharma2005,Tserkovnyak2005,Tang2006,Nagaosa2008} to generate pure spin currents using quantum effects in ferromagnet, semiconductor, and superconductor systems and their hybrids, only few have received continuous experimental attention. These include: non-local spin injection in lateral spin valves~\cite{Valenzuela2006,Kimura2007}; adiabatic quantum spin pumps based on semiconductor quantum dots~\cite{Watson2003}; spin current pumping by precessing magnetization of a ferromagnetic layer driven by microwaves under the ferromagnetic resonance conditions~\cite{Saitoh2006,Costache2006,Moriyama2008}; optical pump-probe experiments on semiconductors~\cite{Stevens2003}; and the SHE~\cite{Sih2006a}.

Even if spin currents are induced easily, their detection can be quite challenging since transport of electron spin between two locations in real space is alien to Maxwell electrodynamics and no ``spin current ammeter'' exists~\cite{Adagideli2006}. In metal spintronic devices spin currents can be converted into voltage signal~\cite{Jung2005} by injection into ferromagnetic electrode (as achieved in lateral spin valves). On the other hand, for semiconductor spintronic devices, which do not couple well to metallic ferromagnets~\cite{Fabian2007}, it is important to avoid ferromagnetic elements (and their stray fields) in both the spin injection and the spin detection processes. Multifarious theoretical ideas have been contemplated to solve this fundamental problem, ranging from the detection of tiny electric fields induced by the flow of magnetic dipoles associated with spins~\cite{Meier2003} to nanomechanical detection of oscillations induced by spin currents in suspended rods~\cite{Mal'shukov2005}. Desirable schemes to detect pure spin current in semiconductors should exploit fundamental quantum-mechanical effects that can transform its flux into conventionally measurable voltage drops and charge currents within the same circuit through which the spin current is flowing. The recently discovered SHE  holds a great promise to revolutionize  generation, control, and detection of pure spin fluxes within the setting of all-electrical circuits.

\section{WHAT IS THE SPIN-HALL EFFECT?}

\begin{figure}[t]
\centerline{\psfig{file=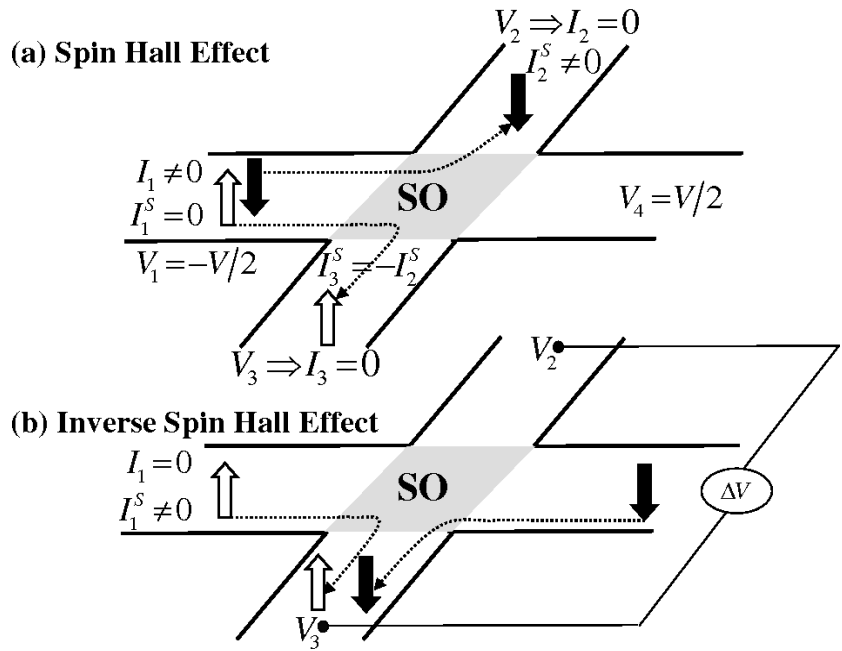,scale=0.45,angle=-90}}
\caption{Basic phenomenology of the direct and inverse SHEs: (a) conventional
(unpolarized) charge current flowing longitudinally through the sample experiences
transverse deflection of opposite spins in opposite direction due to SO coupling induced ``forces''.
This generates pure spin current in the transverse direction or spin accumulation (when transverse electrodes
are removed) of opposite sign at the lateral sample edges; (b) pure spin current flowing through the same sample
governed by SO interactions will induce transverse charge current or voltage drop $\Delta V=V_2 - V_3$ (when transverse leads are
removed). Note that to ensure purity ($I_2=I_3 \equiv 0$) of the transverse spin-Hall current in (a),
employed to define manifestations of the mesoscopic SHE in ballistic or disordered SO-coupled multiterminal nanostructures,
one has to apply proper voltages $V_2$ and $V_3$.}
\label{fig:she_inverse}
\end{figure}

The SHE actually denotes a {\em collection} of phenomena manifesting as transverse separation of spin-$\up$ and spin-$\dn$ states
driven by longitudinally injected standard unpolarized charge current or longitudinal external electric field~\cite{Murakami2006,Schliemann2006,Sinova2006,Engel2007a,Nagaosa2008}. The spins separated in this fashion comprise either a pure spin current or accumulate at the lateral sample boundaries. Its Onsager reciprocal phenomenon---the inverse SHE~\cite{Hirsch1999,Hankiewicz2004a,Hankiewicz2005,Li2006} where longitudinal pure spin current generates transverse charge current or voltage between the lateral boundaries---offers one of the most efficient schemes to detect elusive pure spin currents by converting them into electrical quantities~\cite{Valenzuela2006,Saitoh2006,Kimura2007}. For an illustration of possible experimental manifestations of the direct and inverse SHE in multiterminal nanostructures see Fig.~\ref{fig:she_inverse}.

While SHE is analogous to the classical Hall effect of charges, it occurs in the absence of any externally
applied magnetic fields or magnetic ordering in the equilibrium state. Instead, both the direct and the inverse SHE
essentially require the presence of some type of SO interactions in solids. Although SO couplings are a tiny relativistic effect
for electrons in vacuum, they can be enhanced by several orders of magnitude for itinerant electrons in semiconductors due to
the interplay of crystal symmetry and strong electric fields of atom cores~\cite{Winkler2003}. They have recently emerged as
one of the central paradigms~\cite{Fabian2007} of semiconductor spintronics---unlike cumbersome magnetic fields, they make
possible spin control on very short length and time scales via electric fields, and could, therefore, enable smooth integration
with conventional electronics.

Some of the observed SHE manifestations~\cite{Kato2004b} have been explained~\cite{Engel2007a} by the SO coupling effects localized to the region around impurities which bring interplay of skew-scattering (asymmetric SO-dependent scattering which deflects spin-$\up$ and spin-$\dn$ electrons of an unpolarized flux in opposite directions transverse to the flux) and side jump [due to the noncanonical nature of the physical position and velocity operators in the presence of the SO coupling around an  impurity~\cite{Sinitsyn2008}]. These impurity-driven mechanisms were crucial ingredient
of  the seminal arguments in early 1970s~\cite{D'yakonov1971c} predicting theoretically the existence of [in modern terminology~\cite{Hirsch1999}] the {\em extrinsic} SHE. The extrinsic SO effects are fixed by the materials properties and the corresponding SHE is hardly controllable, except through charge density and mobility~\cite{Awschalom2007}.

A strong impetus for the revival of interest into the realm of SHE has ascended from speculations~\cite{Murakami2003a,Sinova2004} that transverse
pure spin currents, several orders of magnitude larger than in the case of extrinsic SHE,  can be driven by longitudinal electric fields in systems with {\em intrinsic} SO couplings. Such SO couplings manifest in materials with bulk inversion asymmetry or semiconductor heterostructures where
inversion symmetry is broken  structurally. They act homogeneously throughout the sample inducing the spin-splitting of the quasiparticle energy
bands. The {\em intrinsic} (or band structure-driven) SHE could account for large spin current signals observed in 2D hole gases~\cite{Wunderlich2005} or huge SHE response in some metals~\cite{Guo2008}. In addition, since the strength of the Rashba SO coupling~\cite{Winkler2003} in two-dimensional electron gases (2DEGs) within heterostructures with strong structural inversion asymmetry can be  controlled experimentally by a gate electrode~\cite{Nitta1997,Grundler2000}, intrinsically driven SHEs are amenable to easy all-electrical manipulation in realistic nanoscale multiterminal devices~\cite{Nikoli'c2005b}.

The magnitude of both the extrinsic and intrinsic SHE also depends on the impurities, charge density, geometry, and dimensionality. Such a variety of SHE manifestations  poses immense challenge for attempts at a unified theoretical description of spin transport in the presence of relativistic effects. This has not been resolved by early hopes~\cite{Murakami2003a,Sinova2004} that auxiliary spin current  density computed within infinite homogeneous systems could be elevated to universally applicable and experimentally measurable quantity (for more technical discussion of these issues see Sec.~\ref{sec:spin-curr-oper}).  Thus, theoretical analysis has increasingly been shifted toward experimentally relevant quantities in confined geometries and predictions on how to control parameters that can enhance them~\cite{Onoda2005}. Examples of such quantities are edge spin accumulation~\cite{Nikoli'c2005d,Onoda2005,Nomura2005b,Zyuzin2007,Silvestrov2009} and bulk spin density~\cite{Nikoli'c2006,Reynoso2006,Finkler2007,Chen2007,Liu2007}, or outflowing spin currents driven by them~\cite{Nikoli'c2006}.

\section{WHAT IS THE MESOSCOPIC SPIN-HALL EFFECT?}

Realistic devices on which SHE experiments are performed are always in contact with external electrodes and circuits which typically
inject the charge current (rather than applying ``longitudinal electric field'') or perform measurement of the resulting voltage drops and spatial
distribution of spins and charges. In the seminal arguments~\cite{Sinova2004} for the intrinsic SHE in infinite Rashba spin-split 2DEG (in the ``clean'' limit), electric-field-driven acceleration of electron momenta and associated precession of spins plays a crucial role. On the other hand, mesoscopic SHE was introduced~\cite{Nikoli'c2005b} for ballistic finite-size 2DEGs attached to multiple current and voltage probes where electric field is absent in the SO-coupled central sample (on the proviso that surrounding leads are reflectionless). Another stunning difference~\cite{Sheng2006b} between intrinsically driven SHE in the bulk and finite-size 2DEGs is extreme sensitivity to disorder in the former case~\cite{Inoue2004} which, for linear in momentum SO couplings (such as the Rashba one), is able to completely destroy the spin-Hall current density in unbounded systems~\cite{Mishchenko2004,Adagideli2005}. Unlike in three-dimensional semiconductor  and metallic devices, which are always disordered and where extrinsic contribution to the SHE is therefore present or dominant, ballistic conditions for
the mesoscopic SHE can be achieved in low-dimensional semiconductor systems. In fact, the very recent experiment on nanoscale H-shaped structures realized using high mobility HgTe/HgCdTe quantum wells has reported for the first time the detection of mesoscopic SHE via non-local and purely electrical measurements~\cite{Brune2008}.

The magnitude of pure spin currents flowing out of mesoscopic SHE device (illustrated in Fig.~\ref{fig:she_inverse}) through ideal (spin and charge interaction free) electrodes is governed by the spin precession length $L_{\rm SO}$. This mesoscopic length scale (e.g., $L_{\rm SO} \sim 100$ nm in recently fabricated 2DEGs), on which the vector of the expectation values of spin precesses by an angle $\pi$, has been identified through intuitive physical arguments~\cite{Engel2007a} as an important parameter for spin distributions (e.g., in clean systems the spin response to inhomogeneous field diverges at the wave vector $q=2/L_{\rm SO}$). In fact, the mesoscopic SHE analysis predicts~\cite{Nikoli'c2005b} via numerically exact calculations (see Fig.~\ref{fig:meso_she}) that optimal device size for achieving large  spin
currents is indeed $L \simeq L_{\rm SO}$. This is further confirmed by alternative analyses of the SHE response  in finite-size systems~\cite{Moca2007}. In the general cases~\cite{Sih2005a,Hankiewicz2008}, where both the extrinsic and intrinsic SO interaction effects are present, the intrinsically driven contribution to SHE in finite-size devices dominates~\cite{Nikoli'c2007} when the ratio of characteristic energy scales~\cite{Nagaosa2008} for the disorder and SO coupling effects satisfies $\Delta_{\rm SO}\tau/\hbar \gtrsim 1$ ($\Delta_{\rm SO}$ is the spin-splitting of quasiparticle energies and $\hbar/\tau$ is the disorder induced broadening of energy levels due to transport scattering time $\tau$).

For mesoscopic SHE devices in the phase-coherent transport regime (device smaller than the dephasing length), one can also observe the effects of quantum confinement and quantum interferences in spin-related quantities that counterpart familiar examples from mesoscopic charge transport~\cite{Datta1995}. They include: SHE conductance fluctuations~\cite{Ren2006,Bardarson2007}; resonances in SHE conductance due to opening of new conducting channels~\cite{Nikoli'c2005b,Sheng2006b} or mixing of bound~\cite{Bulgakov1999} and propagating states due to SO couplings; and constructive or destructive quantum interference-based control~\cite{Souma2005a} of spin-Hall current in multiterminal Aharonov-Casher rings (as the electromagnetic dual of Aharonov-Bohm rings where SO coupling, rather than magnetic field, permeates the ring). The charge and spin dephasing  can be included~\cite{Golizadeh-Mojarad2007a} within the same NEGF transport formalism to allow for comparison with experiments performed at finite  temperatures where quantum coherence effects are smeared out~\cite{Golizadeh-Mojarad2007}.

In general, the presence of SO couplings requires to treat the whole device geometry when studying the dynamics of transported spin densities. For example, the decay of nonequilibrium spin polarizations in ballistic or disordered quantum wires is highly dependent on the transverse confinement effects~\cite{Nikoli'c2005,Holleitner2006} or chaotic vs. regular boundaries of quantum dots~\cite{Chang2004}. Since SO couplings in SHE devices manifest through both of their aspects---creation of spin currents and concurrently relaxation of spins---it is a nontrivial task to understand how spin currents and edge spin accumulations scale with increasing the strength of the SO couplings~\cite{Onoda2005}.

The analysis of the whole device setup, where central SO-coupled sample is treated together with the surrounding electrodes, also simplifies
the discussion of esoteric SHE concepts, such as the SHE in insulators~\cite{Onoda2005a} [where electrodes introduce dissipation
necessary to obtain nonzero value of spin accumulation as time-reversal odd quantity] or quantum SHE~\cite{Konig2008} whose quantized
spin-Hall conductance is due to chiral spin-filtered (or helical) edge states (i.e., Kramers doublets of states forcing electrons of opposite
spin  to flow in opposite directions along the edges of the sample) in a multiterminal SO-coupled bridge with energy gap in the central sample~\cite{Sheng2005b}.
For example, recent direct experimental evidence~\cite{Roth2009} for nonlocal transport in HgTe quantum wells in the quantum spin-Hall regime, which
shows how non-dissipative quantum transport occurs through chiral spin-filtered edge states while the contacts lead to equilibration between the
counter-propagating spin states at the edge, can be analyzed only by the device-oriented quantum transport techniques discussed in this chapter.

\section{SO COUPLINGS IN LOW-DIMENSIONAL SEMICONDUCTORS}\label{sec:so_2deg}

The coupling between the orbital and the spin degree of freedom of electrons is a relativistic effect described formally
by the nonrelativistic expansion of the Dirac equation in external electric and magnetic fields (for which exact solutions do not exist)
in powers of the inverse speed of light $c$. In the second order $v^2/c^2$, one identifies~\cite{Zawadzki2005} the SO term\footnote{Although a topic of numerous textbooks on relativistic quantum mechanics and quantum field theory, the  $v^2/c^2$ expansion has recently been carefully reexamined~\cite{Zawadzki2005} to find all terms at this order of approximation in manifestly gauge invariant form,
thereby revealing various inconsistencies in the textbook literature.}
\begin{equation} \label{eq:so}
\hat{H}_{\rm SO}=\frac{\hbar}{4 m_0^2c^2} \hat{\bf p} \cdot (\hat{\bm \sigma} \times \nabla V({\bf r})),
\end{equation}
responsible for the entanglement of the spin and orbital degrees of freedom in the two-component nonrelativistic Pauli Hamiltonian for spin-$\frac{1}{2}$ electron. Here $m_0$ is the free electron mass, $\hat{\bm \sigma}=(\hat{\sigma}_x,\hat{\sigma}_y,\hat{\sigma}_z)$ is the vector of the Pauli matrices, and $V({\bf r})$ is the electric potential. The SO coupling term can also be extracted from the semiclassical analysis that usually invokes the interaction of the electron magnetic dipole moment (associated with spin) with magnetic field in the frame moving with electron~\cite{Jackson1998}. In the instantaneous rest frame of an electron, magnetic field is  obtained by Lorentz transforming electric field from the
laboratory frame. It is actually more efficient for intuitive analysis of different experimental situations to
remain~\cite{Fisher1971} in the laboratory frame where a magnetic dipole ${\bm \mu}$  moving with velocity ${\bf v}$ generates
electric dipole moment
\begin{equation}\label{eq:plab}
{\bf P}_{\rm lab}={\bf v} \times {\bm \mu}/c^2.
\end{equation}
Here the right-hand side is evaluated in the electron rest frame and ${\bf P}_{\rm lab}$ is measured in the lab (both sides can be evaluated
in the lab yielding the same result to first order in $v/c$). The potential energy of the interaction of the electric
dipole with the external electric field ${\bf E}_{\rm lab}$ in the lab frame, $U_{\rm dipole}=-{\bf P}_{\rm lab} \cdot {\bf E}_{\rm lab}$, corrected to the Thomas precession\footnote{The Thomas precession takes into account change in rotational kinetic energy due to the precession of the accelerated electron as seen by laboratory observer in the ``extended'' special relativity of accelerated objects.} $U_{\rm Thomas}=-U_{\rm dipole}/2$ leads to the SO coupling term $U_{\rm SO}=U_{\rm dipole}+U_{\rm Thomas}=-{\bf P}_{\rm lab}\cdot {\bf E}_{\rm lab}/2$.

The lab frame analysis allows one to quickly depict the Mott skew-scattering off a target whose Coulomb field deflects
a beam of spin-$\up$ and spin-$\dn$ particles in opposite directions, thereby, e.g., polarizing the beam of neutrons~\cite{Fisher1971}
or generating skew-scattering contribution~\cite{Engel2007a,Hankiewicz2008} to the extrinsic SHE. For example, if we look at spin-$\up$ electron from behind moving along the $y$-axis, whose expectation value of the spin vector is oriented along the positive $z$-axis so that the corresponding magnetic dipole moment lies along the negative  $z$-axis, then in the lab frame we also see its Lorentz transformed electric dipole moment ${\bf P}_{\rm lab}$ oriented along the negative $x$-axis. The electric dipole feels the force ${\bf F} = ({\bf P}_{\rm lab} \cdot \nabla) {\bf E}_{\rm lab}$, oriented in this case along the positive $x$-axis (right transverse direction with respect to the motion of the incoming electron) since gradient of the electric field ${\bf E}_{\rm lab}$ generated by the target is always negative outside of it. Note that this simple-minded classical picture  only explains one aspect of the SO-dependent interaction with impurity. The other one---the so-called side jump (i.e., sideways shift of the scattering wave packet)---requires more quantum mechanical analysis~\cite{Sinitsyn2008} to extract additional contribution to the velocity operator due to impurity potential $V_{\rm disorder}({\bf r})$ in the SO Hamiltonian Eq.~(\ref{eq:so}).

The heuristic discussion based on the Lorentz transformations gives only a minuscule effect and the influence of electronic band structure is
essential~\cite{Engel2007a} to make these effects experimentally observable in solids. In the case of atoms, SO coupling is due to interaction
of electron spin with the average Coulomb field of the nuclei and other electrons. In solids, $V({\bf r})$ is the sum of periodic crystalline potential and an aperiodic part containing potentials due to impurities, confinement, boundaries and external electric fields. The nonrelativistic expansion of the Dirac equation can be viewed as a method of systematically including the effects of the negative-energy solutions on the positive energy states  starting from their nonrelativistic limit~\cite{Zawadzki2005}. This effect in vacuum is small due to huge gap $2m_0c^2$ between positive and negative energy states. In solids, strong nuclear potential competes with this huge denominator in Eq.~(\ref{eq:so}), so that much smaller band gap between conduction and valence band (playing the role of electron positive energy sea and positron negative energy sea, respectively) replaces $2m_0c^2$, thereby illustrating the origin of strong enhancement of the SO couplings in solids~\cite{Winkler2003}.

Although intrinsic SO couplings can always be written in the Zeeman form $\hat{\bm \sigma} \cdot {\bf B}_{\rm SO}({\bf p})$,
their effective magnetic field ${\bf B}_{\rm SO}({\bf p})$ is momentum-dependent and, therefore,
does not break the time-reversal invariance. The Kramers theorem~\cite{Ballentine1998} for time-reversal
invariant quantum systems requires that the energy bands $\varepsilon_n({\bf k})$ of an electron in a periodic
potential satisfy $\varepsilon_n ({\bf k},\up) = \varepsilon_n (-{\bf k},\dn)$ since ${\bf k} \mapsto -{\bf k}$ and $\sigma\!\!=\up   \mapsto \sigma\!\!=\dn$ upon time reversal ($\hbar{\bf k}$ is crystal momentum). Therefore, in semiconductors invariant under spatial inversion ${\bf k} \mapsto -{\bf k}$ (such as silicon) the
Kramers theorem gives double degenerate spin states for any ${\bf k}$ value, $\varepsilon_n({\bf k},\up) = \varepsilon_n({\bf k},\dn)$. To obtain a non-zero ${\bf B}_{\rm SO}({\bf p})$ that breaks the spin degeneracy in three-dimensional crystals $\varepsilon_n({\bf k},\up) \neq \varepsilon_n({\bf k},\dn)$ the host crystal has to be inversion asymmetric.
In bulk semiconductors with zinc-blende symmetry, the conduction band of III-V compounds will split into two
subbands where anisotropic spin-splitting is proportional to $k^3$. Such SO-induced splitting is termed
cubic Dresselhaus SO coupling, and it is associated with bulk inversion asymmetry (BIA). In semiconductor heterostructures~\cite{Winkler2003},
ideal symmetry of the 3D host crystal is broken by the interface where 2DEG or two-dimensional hole gas (2DHG) is confined
within a quantum well. In such reduced effective dimensionality, the symmetry of the underlying crystal lattice is
lowered, so that an additional linear in $k$ Dresselhaus term becomes relevant. Besides microscopic crystalline potential
$V_{\rm crystal}({\bf r})$ as the source of the electric field in the SO coupling term in Eq.~(\ref{eq:so}), an interface electric
field accompanying the quantum well structural inversion asymmetry (SIA) gives rise to the Rashba spin-splitting of conduction
band electrons
\begin{equation}\label{eq:hso_rashba}
\hat{H}_{\rm Rashba} = \frac{\alpha}{\hbar} (\hat{\bm \sigma} \times \hat{\bf p}) \cdot \hat{\bf e}_z,
\end{equation}
for 2DEG in the $xy$-plane and $\hat{\bf e}_z$ as the unit vector along the $z$-axis. In narrow gap semiconductors the Rashba effect linear
in $k$ should dominate over the bulk $k^3$ Dresselhaus term. Moreover, it has been experimentally demonstrated that Rashba coupling can be changed by as much as 50\% by external gate electrode~\cite{Nitta1997,Grundler2000} covering 2DEG, which has become on of the key concepts in semiconductor spintronics. Nevertheless, there is a lengthy theoretical debate on the importance of different electric field contributions  to the value of experimentally observed $\alpha$ and the parameters which  are effectively manipulated via the gate electrode to cause its increase~\cite{Grundler2000}---we refer to a comprehensive overview of these issues by \citeasnoun{Fabian2007} and \citeasnoun{Winkler2003}.
Note that impurity determined $V_{\rm disorder}({\bf r})$ contribution to Eq.~(\ref{eq:so}) does not require broken inversion
asymmetry of the pure crystal or of the structure.

The coupling between electron momentum and spin correlates charge currents and spin densities in SO-coupled
semiconductors~\cite{Silsbee2004} leading to highly non-trivial effects in nonequilibrium situations. Some of these
have been observed in recent magneto-electric experiments~\cite{Ganichev2006a,Silov2004,Kato2004a}, where charge
current induces spin density, as well as in the spin galvanic experiments~\cite{Ganichev2003}, where
nonequilibrium spin density drives a charge current. It is also the key ingredient of intrinsically driven
SHEs [where the induced spin density is oriented out-of-plane, rather than in-plane as in the magneto-electric effects~\cite{Silsbee2004}].

In this Section we focus on the description of the Rashba SO coupling\footnote{The same analysis applies to linear Dresselhaus
coupling since they can be transformed into each other by a unitary matrix.} in the form suitable for spin-dependent NEGF
calculations using local orbital basis and the lattice Hamiltonian defined by it. The Rashba Hamiltonian prepared in this
fashion will be used as the starting point for illustrating SHE-related spin and charge transport calculations in
Sec.~\ref{sec:negf}. We also add the treatment of the extrinsic SO coupling in 2DEG within the same local orbital basis
framework in Sec.~\ref{sec:discr-extr-so} to enable the description of the most general experimental situations~\cite{Sih2005a}
in low-dimensional devices where both extrinsic and intrinsic mechanisms can act concurrently~\cite{Nikoli'c2007,Hankiewicz2008,Dragomirova2008}.

\subsection{Rashba coupling in bulk 2DEG}\label{sec:rasbaSO2deg}

The effective single-particle SO Hamiltonian for a clean infinite homogeneous 2DEG
with the Rashba coupling Eq.~(\ref{eq:hso_rashba}) can be formally rewritten as
\begin{equation}\label{eq:rashbaclean}
\hat{H}^{\rm 2D}_{\rm R} = \frac{\hat{\bf p}^2}{2m^*} \otimes \mathbf{I}_{\rm S} + \frac{\alpha}{\hbar}
\left( \hat{p}_y \otimes \hat{\sigma}_x  - \hat{p}_x  \otimes \hat{\sigma}_y  \right).
\end{equation}
Here $\otimes$ stands for the tensor product of two operators acting in the tensor product $\mathcal{H}_{\rm O} \otimes \mathcal{H}_{\rm S}$
of the orbital and spin Hilbert spaces, and $m^*$ is the effective mass. To enforce pedagogical notation,  we also use $\mathbf{I}_{\rm O}$ as the unit operator in  $\mathcal{H}_{\rm O}$ and $\mathbf{I}_{\rm S}$ for the unit operator in $\mathcal{H}_{\rm S}$. The internal {\em momentum-dependent} magnetic field corresponding to the Rashba coupling is extracted from Eq.~(\ref{eq:hso_rashba}), recast in the form of the Zeeman term $-g\mu_B\hat{\bm \sigma} \cdot \mathbf{B}_{\rm R}({\bf p})/2$, as $\mathbf{B}_{\rm R} (\mathbf{p})=(2\alpha/g\mu_B)\left(\mathbf{p} \times  \mathbf{e}_z \right)$. The Hamiltonian commutes with the momentum operator $\hat{\mathbf{p}}$, time-reversal operator $\hat{T}$, and the chirality operator $(\hat{\bm \sigma} \times \hat{\bf p}/|{\bf p}|) \cdot \mathbf{e}_z$. The zero commutator  $\left[ \hat{H}, \hat{\mathbf{p}} \right]=0$ due to the translation invariance of infinite 2DEG implies  that solutions of the Schr\"odinger equation are of the form
\begin{equation}\label{eq:solrashba}
\Psi(\mathbf{r})=C{\rm e}^{i\mathbf{kr}}\left| \chi \right>_s,
\end{equation}
where $\left| \chi \right>_s$ is a two-component spinor. Due to time-reversal invariance
$\left[ \hat{H},\hat{T} \right] =0$, the two eigenstates of opposite momenta and
spins are degenerate, $E(\mathbf{k},\sigma) = E(-\mathbf{k},-\sigma)$. The time-reversal
operator can be written as $\hat{T}=\hat{K} \otimes \exp (i \pi \hat{\sigma}_y/2)$, where $\hat{K}$ is
the complex conjugation operator~\cite{Ballentine1998}. If the Hamiltonian is invariant to chirality
transformation, the spin state is momentum-dependent. In the Rashba case the
spin and momentum of an eigenstate are always perpendicular to each other.

The Rashba Hamiltonian Eq.~(\ref{eq:rashbaclean}) in the case of a 1D electron gas (1DEG) reduces to
\begin{equation}\label{eq:rashba1D}
\hat{H}^{\rm 1D}_{\rm R}(k_x)=\frac{\hbar^2k_x^2}{2m^*}-\alpha k_x\hat{\sigma}_y.
\end{equation}
Its eigenvalues define the energy-momentum dispersion
\begin{equation}\label{eq:rashba1Deigen}
E_{\pm}(k_x)=\frac{\hbar^2 k_x^2}{2 m^*} \pm \alpha |k_x|,
\end{equation}
with the corresponding eigenvectors
\begin{equation}\label{eq:rashba1deigvec}
\psi_{k_x,\pm }=\frac{e^{i k_x x}}{\sqrt{2\pi\hbar}}\cdot \frac{1}{\sqrt{2}}
\left (
\begin{array}{c}
1 \\
\mp \displaystyle i \frac{k_x}{|k_x|}
\end{array} \right ).
\end{equation}
These eigenstates are labeled with the momentum operator eigenvalue $\hbar k_x$ and the chirality operator eigenvalue $\lambda = \pm 1$. Equation~(\ref{eq:rashba1deigvec}) shows that
\begin{equation}\label{eq:sigmaexpval1D}
\left< \psi_{k_x,\pm} \right| \hat{\sigma}_y \left| \psi_{k_x,\pm} \right> =
\mp \frac{k_x}{\left| k_x \right|},
\end{equation}
meaning that each spin-split parabolic subband has a well-defined spin.
\begin{figure}[t]
\centerline{\psfig{file=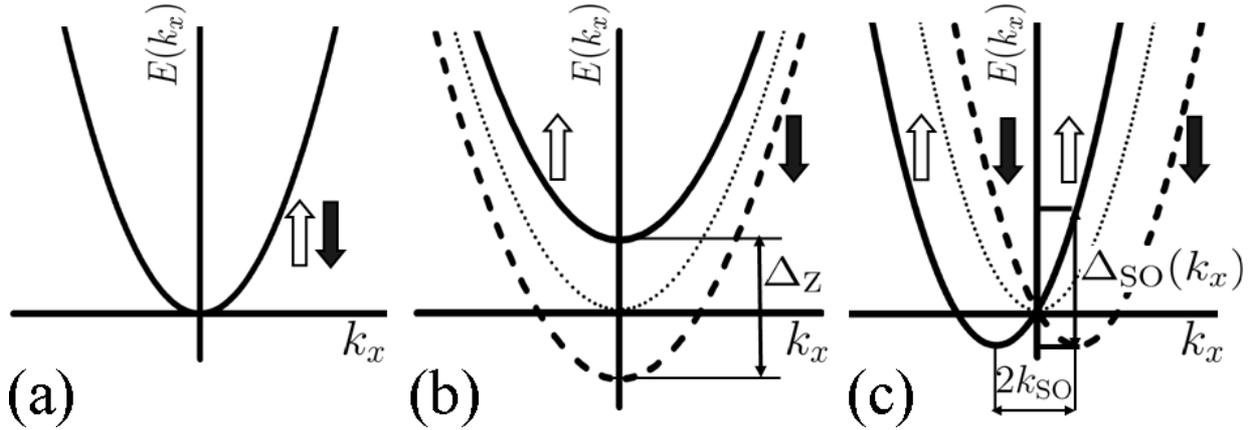,scale=0.65,angle=0}}
\caption{The one-dimensional energy-momentum dispersion for: a) infinite clean 1DEG, b) 1DEG in magnetic field, and c) 1DEG with the Rashba SO coupling. The SO-induced
spin splitting in (c) is signifies by the energy separation $\Delta_{\rm SO}$ at a given $k_x$. The complementary description of the SO-induced spin splitting is $k_{\rm SO}$ difference between the momenta of two electrons of opposite spins and chiralities in panel (c).}
\label{fig:dispersioncomparison}
\end{figure}

To elucidate the physical meaning of the SO-induced spin splitting, we compare in Fig.~\ref{fig:dispersioncomparison}
the Rashba dispersion with more familiar Zeeman splitting of 1DEG placed in the external (momentum-independent)
magnetic field. In the case of the Zeeman splitting, the spin-$\up$ and spin-$\dn$ subbands are shifted vertically
with respect to each other by the Zeeman energy $\Delta_{\rm Z}$. On the other hand, the Rashba spin-splitting depends on momentum
\begin{equation}\label{eq:1d_rashsplitting}
\Delta_{\rm SO}(k_x)=2\alpha k_x,
\end{equation}
so that $E_{+}(k_x)$ and $E_{-}(k_x)$ are shifted horizontally along the momentum axis rather than
along the energy axis. This ensures that the system remains  spin-unpolarized  in equilibrium, as
dictated by the time-reversal invariance and the fact that spin density is time-reversal odd quantity.
The spin splitting can also be described using  $k_{\rm SO}$ as the difference  between the momenta of two electrons of
opposite spins and chiralities, and different experiments (such as Shubnikov-de Haas, Raman scattering, or spin
precession) probe either $\Delta_{\rm SO}$ or $k_{\rm SO}$ illustrated in Fig.~\ref{fig:dispersioncomparison}(c).

From Eq.~(\ref{eq:sigmaexpval1D}) we see that the $y$-component of spin of the eigenstate
$\left| \psi_{k_x,\lambda} \right> $ is $\up$ for $\lambda=1$ and $k_x$ negative, and
$\dn$ for $\lambda=1$ and $k_x$ positive. The opposite is true for the other branch
$E_{\lambda=-1}(k_x)$ Thus, the eigenvalues of spin $\hat{\sigma}_y$  are good quantum numbers since in 1D $\hat{\sigma}_y$ commutes
with the Hamiltonian. Following this argument, it is easy to understand the relative spin orientations
of the states along each of the dispersion branches.
\begin{figure}[t]
\centerline{\psfig{file=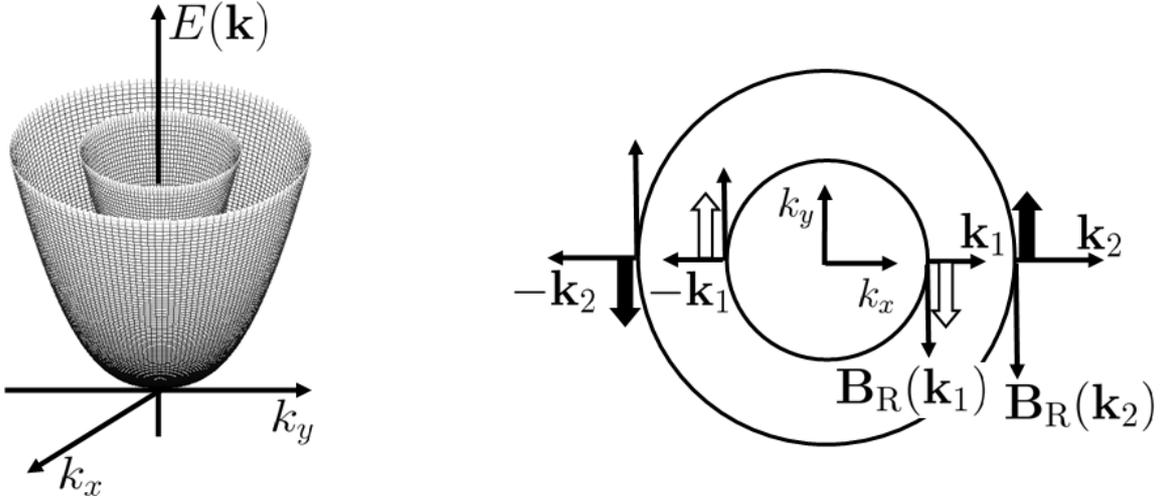,scale=0.65,angle=0}}
\caption{The energy-momentum dispersion $E(k_x,k_y)$ of the Rashba spin-split 2DEG.
The  states at the Fermi level lie on two concentric Fermi circles, $E_{+}(\mathbf{k})=E_{\rm F}$ for
the inner circle and $E_{-}(\mathbf{k})=E_{\rm F}$ for the outer circle, as shown in the right panel.
The radius of the inner circle is $k_1$ and the radius of the outer one is denoted by $k_2$. The Fermi momentum
$k_{\rm F}$ of the free particle is $k_1\leq k_{\rm F} \leq k_2$. For a given momentum
$\mathbf{k}$, the spin of an electron is oriented either parallel [$E_{+}(\mathbf{k})$ branch] or anti-parallel
[$E_{-}(\mathbf{k})$ branch] to the momentum-dependent effective magnetic field $\mathbf{B}_{\rm R}({\bf k})$.
The states of opposite momentum and spin on each Fermi circle are Kramers degenerate.
}
\label{fig:rashba2Ddispersion}
\end{figure}

As in 1D case, the Rashba Hamiltonian in 2D commutes with momentum,
chirality, and the time-reversal operator. Its eigenenergies
\begin{equation}\label{eq:rash2Deigen}
E_{\pm}(\mathbf{k})=\frac{\hbar^2\mathbf{k}^2}{2m^*}\pm \alpha \left| \mathbf{k} \right|,
\end{equation}
are plotted in Fig.~\ref{fig:rashba2Ddispersion}. They are labeled by a 2D wave vector  $\left| \mathbf{k} \right|=\sqrt{k_x^2+k_y^2}$
and the chirality eigenvalues $\lambda = \pm 1$ (i.e., the spin projection perpendicular to both $\mathbf{k}$ and
growth direction along the $z$-axis). The corresponding eigenstates are
\begin{equation}\label{eq:rash2Deigenstates}
\psi_{\mathbf{k},\pm}=\frac{e^{i \mathbf{k} \mathbf{r}}}{2\pi\hbar}\cdot \frac{1}{\sqrt{2}}
\left (
\begin{array}{c}
1 \\
\pm \displaystyle \frac{k_y-ik_x}{|{\bf k}|}
\end{array}
\right ).
\end{equation}
Finding the expectation value of the spin operator $\hbar \hat{\bm \sigma}/2$ in
eigenstate Eq.~(\ref{eq:rash2Deigenstates})
\begin{equation}\label{eq:sigmaexpval2D}
\left< \psi_{\mathbf{k},\pm} \left| \hat{\bm \sigma} \right| \psi_{\mathbf{k},\pm} \right> =
\pm \frac{k_y}{\left | \mathbf{k} \right| } \mathbf{e}_x \mp
 \frac{k_x}{\left | \mathbf{k} \right| } \mathbf{e}_y,
\end{equation}
demonstrates~\cite{Winkler2003} that no common spin quantization axis can be found for all eigenstates of the Rashba spin-split 2DEG ($\mathbf{e}_x$ and
$\mathbf{e}_y$ are the unit vectors within the $xy$-plane of 2DEG).

\subsection{Rashba coupling in quantum wires}\label{sec:rashba-so-coupling_qwires}

\subsubsection{Energy dispersion of Rashba spin-split transverse propagating subbands}\label{sec:subbands}

The Rashba Hamiltonian for a quantum wire  patterned within 2DEG (along the $x$-axis)
\begin{equation}\label{eq:rashba}
\hat{H}_{\rm R}^{\rm Q1D}=
\frac{\hat{\bf p}^2}{2m^*}  +
\frac{\alpha}{\hbar}\left(\hat{\bm \sigma}\times\hat{\bf p}\right) \cdot {\bf z}
+ V_{\rm conf}(y),
\end{equation}
describes quasi-one-dimensional electron gas (Q1DEG) of width $d$ whose lateral confinement is accounted by the
potential $ V_{\rm conf}(y)$.  The motion along the confinement direction is quantized, such that the energy-momentum
dispersion in the absence of SO coupling is split into spin degenerate quasi 1D subbands with quadratic dispersion $E_{\pm}(k_x)=E_n + \hbar^2k_x^2/2m^*$ in 1D wave vector $k_x$ labeling an eigenstate of the subband with index $n$.
\begin{figure}
\centerline{\psfig{file=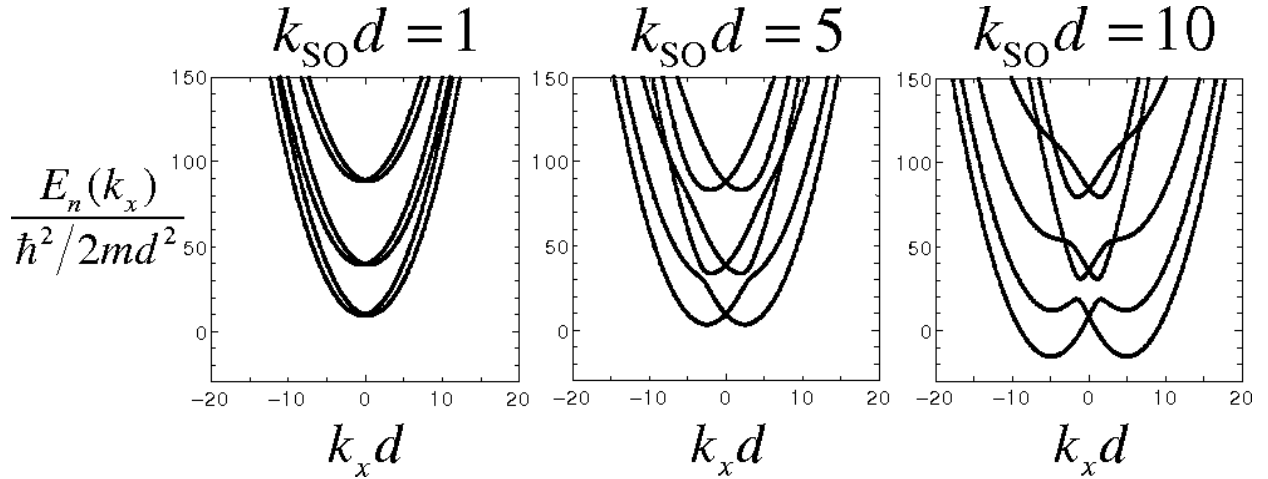,scale=0.65,angle=-90}}
\caption{The subband energy-momentum dispersion in the Rashba spin-split clean quantum wire. The parameter $k_{\rm SO}d$ ($d$ is the width of the wire)
defines the weak and strong SO coupling regimes. For large values of $k_{\rm SO}d$  the mixing of subbands is nontrivial giving rise to strong energy
dependence of the wire conductance.}
\label{fig:qwire_subbands}
\end{figure}
The effective mass Rashba Hamiltonian for a quantum wire is translationally invariant along
wire direction $x$, and can be separated into three terms
$\hat{H}=\hat{H}_{\rm sb}+\hat{H}_{\rm mix}+\hat{H}_{\rm 1D}$~\cite{Governale2004}:
\begin{subequations}\label{eq:rashba_wire}
\begin{eqnarray}
  \hat{H}_{\rm sb} & = & \frac{\hat{p}_y^2}{2m^*}+V_{\rm conf}(y), \\
  \hat{H}_{\rm mix} & = & -\frac{\hbar k_{\rm SO}}{m^*} \hat{\sigma}_x \hat{p}_y, \\
  \hat{H}_{\rm 1D} & = & \frac{\hbar^2}{2m^*}(k_x+k_{\rm SO}\hat{\sigma}_y)^2 +
\frac{\hbar^2k_{\rm SO}^2}{2m^*}.
\end{eqnarray}
\end{subequations}
The term $\hat{H}_{\rm sb}$ defines the eigenenergies $E_n$ of the transverse confining potential, while $\hat{H}_{\rm 1D}$
is the 1D translationally invariant Rashba term for the quantum wire. Therefore, in the hypothetical case where the second
term $\hat{H}_{\rm mix}$ is absent, the eigenenergies of the Hamiltonian Eq.~(\ref{eq:rashba_wire}) show only $k_x$-dependent splitting
\begin{equation}\label{eq:disp_rashba_wire}
E_{\pm}(k_x)=E_n+\frac{\hbar^2k_x^2}{2m^*}\pm \alpha  | k_x |.
\end{equation}
When the expectation values $E_{\rm mix}$ of $\hat{H}_{\rm mix}$ between the eigenstates of $\hat{H}_{\rm sb}+\hat{H}_{\rm 1D}$ are of the order of $\Delta E_n=E_{n+1}-E_n$, the mixing of the spin-split subbands becomes important. Thus, the ratio
\begin{equation}\label{eq:ksod}
  \frac{E_{\rm mix}}{\Delta E_n}\simeq \frac{\hbar k_{\rm SO}}{m^*}\frac{\pi\hbar}{d}
  \left( \frac{\pi^2\hbar^2}{m^*d^2}\right)^{-1}=\frac{d}{L_{\rm SO}}=\frac{k_{\rm SO}d}{\pi},
\end{equation}
gives a simple criterion to separate the strong and weak SO coupling regimes in Q1DEGs.
In the weak coupling regime $L_{\rm SO} \gg  d$ the subband mixing is negligible,
as illustrated in Fig.~\ref{fig:qwire_subbands}(a). In quantum wires with strong
Rashba effect, realized for  $L_{\rm SO} \lesssim d$ (or $k_{\rm SO} d \gtrsim \pi$)
hybridization of quasi-1D subbands becomes important and Eq.~(\ref{eq:disp_rashba_wire})
can not capture the non-parabolic energy dispersions in Fig.~\ref{fig:qwire_subbands}(b),(c).
Unlike the weak coupling regime where the Hamiltonian eigenstates are also eigenstates of
$\hat{\sigma}_y$ (i.e., their spins are polarized in-plane along the $y$-axis), in the
strong coupling regime no common spin quantum number can be assigned to states within a
given subband~\cite{Governale2004}.

\subsubsection{Spin precession in Rashba quantum wires}\label{sec:spin-precession_lso}

The lifting of spin degeneracy of each transverse subband of a quantum wire by the Rashba coupling means that
an electron injected at the Fermi energy $E_F$ can have two different wave vectors within the wire, $E_{+}(k_{x1})=E_{-}(k_{x2})=E_{\rm F}$.
The wave vector $k_{x1}$ labels the eigenstates of $E_+$ spin-split subband whose spinor
part is {\tiny $\frac{\displaystyle 1}{\displaystyle \sqrt{2}}\left( \begin{array}{c} 1 \\ i \end{array} \right) $}
[describing spin-$\up$ along the $y$-axis], while $k_{x2}$ labels the eigenstates of $E_-$
subband associated with the eigenspinor
{\tiny $ \frac{\displaystyle 1}{\displaystyle \sqrt{2}}{\left(\begin{array}{c} 1 \\  -i \end{array} \right)} $}
[describing spin-$\dn$ along the $y$-axis]. Thus, when electron with spin-$\up $ along the $z$-axis
is injected from a half-metallic ferromagnet into the Rashba quantum wire, as proposed in the
Datta-Das spin-FET device~\cite{Datta1990}, the outgoing wave function from a wire of length $L$ will be
\begin{equation}
\psi(L) \sim {\rm e}^{ik_{x1}L}
\left( \begin{array}{c} 1 \\ i \end{array} \right)
+
{\rm e}^{ik_{x2}L}
\left( \begin{array}{c} 1 \\ -i \end{array} \right).
\end{equation}
The probability to detect a particle with spin-$\up$ or with spin $\dn$ along the $z$-axis is given by
\begin{subequations}\label{eq:spin_prec_prob}
\begin{equation}\label{eq:spinup_prob_wire}
\left| \langle \left( 1 \mbox{ } 0 \right) \left|\right. \psi(L) \rangle \right|
=4\cos^2\frac{(k_{x1}-k_{x2})L}{2},
\end{equation}
and
\begin{equation}\label{eq:spindn_prob_wire}
\left| \langle \left( 0 \mbox{ } 1 \right) \left|\right. \psi(L) \rangle \right|
=4\sin^2\frac{(k_{x1}-k_{x2})L}{2},
\end{equation}
\end{subequations}
respectively. Such modulation of current by spin precession with angle $\Delta \theta =\Delta k_x L$, where
$\Delta k_x = k_{x2}-k_{x1}=2m^*\alpha/\hbar^2$ stems from  $E_{+}(k_{x2})-E_{-}(k_{x1})=0$,
would provide the basis for the operation of the envisioned spin-FET device~\cite{Datta1990}. It
also introduces in a transparent fashion the concept of the {\em spin precession length}
\begin{equation}\label{eq:lso}
L_{\rm SO}=\frac{\pi \hbar^2}{2m^*\alpha},
\end{equation}
along which the vector of the expectation values of spin precesses by an angle $\Delta \theta =\pi$
while propagating along the Rashba wire.

We emphasize that the energy independence of $L_{\rm SO}$ holds for weak SO coupling and narrow
wires. As discussed in Sec.~\ref{sec:subbands}, for strong SO coupling $d>L_{\rm SO}$,
intersubband mixing becomes important, so that Eq.~(\ref{eq:lso}) is inapplicable.
The $L_{\rm SO}$ scale appears as the characteristic length scales for various processes in
semiconductor spintronic systems. For example, in weakly disordered  systems $L_{\rm SO}$ also plays the role of a
characteristic scale for the D'yakonov-Perel' spin dephasing where nonequilibrium spin density decays due to
randomization  of $\mathbf{B}_{\rm R}(\mathbf{p})$ in each scattering event changing $\mathbf{p}$ and, therefore, the
direction of magnetic field around which spin precesses~\cite{Chang2004,Fabian2007}. It also sets a  mesoscopic scale
[$\sim 100$ nm in heterostructures with large $\alpha$~\cite{Nitta1997,Grundler2000}] at which one can expect the largest SHE response~\cite{Nikoli'c2005b,Nikoli'c2006,Moca2007}.

\subsection{Discrete representation of effective SO Hamiltonians}\label{sec:so-coupl-hamilt_latt_repr}

In order to perform numerical calculations on finite-size systems of arbitrary shape attached to external electrodes
it is  highly advantageous to discretize the effective SO Hamiltonian. Here we introduce a discretization scheme for
the Rashba Hamiltonian. In addition, we also discuss discretization scheme for the extrinsic SO Hamiltonian which
makes it possible to treat intrinsic and extrinsic SO coupling effect on equal footing in nanostructures~\cite{Nikoli'c2007,Dragomirova2008}.

The grid used for discretization is collection of points $\mathbf{m}=(m_x,m_y)$ on a square lattice
of constant $a$, where $m_x,m_y$ are integers. We use the following notation: $\psi_{\mathbf{m}}\equiv \left< \mathbf{m} \left| \right. \psi \right>$ for the wave function evaluated at point $\mathbf{m}$; and $A_{\mathbf{mm}^\prime} \equiv  \left< \mathbf{m} \left| \hat{A} \right| \mathbf{m}^\prime \right>$ for the matrix
element of an operator $\hat{A}$. In finite difference methods, one has to evaluate the derivatives of a function $f(x)$ on a grid
$\{ \ldots, x_{m-1}, x_m, x_{m+1}, \ldots \}\equiv \{ \ldots, (m-1)a, ma, (m+1)a, \ldots \}$:
\begin{equation}\label{eq:discretederiv}
  \left( \frac{df}{dx} \right)_m = \frac{f_{m+1}-f_{m-1}}{2a},
\end{equation}
where $f_m\equiv f(x_m)$. The second derivative can be computed as
\begin{equation}\label{eq:discretesecondderiv}
  \left( \frac{d^2f}{dx^2} \right)_m=\frac{f_{m+1}-2f_m+f_{m-1}}{a^2}.
\end{equation}
For a particular discretization scheme, the matrix elements of the derivative operators can be
expressed in the local orbital basis $\left| \mathbf{m} \right>$, which can be interpreted as
each site hosting a single $s$-orbitals as in the tight-binding models of solids. For
instance, using Eq.~(\ref{eq:discretederiv}) we get
\begin{equation}\label{eq:diffopdiscrete}
\left< m \left| \frac{d}{dx} \right| n \right> =
\frac{\left< m+1 \left| \right. n \right> - \left< m-1 \left| \right. n \right> }{2a} =\frac{\delta_{n,m+1}-\delta_{n,m-1}}{2a},
\end{equation}
where $\delta$ is the Kronecker symbol. The same can be done for the operator $d^2/dx^2$ using
Eq.~(\ref{eq:discretesecondderiv})
\begin{equation}\label{eq:secdiffopdiscrete}
\begin{split}
  \left< m \left| \frac{d^2}{dx^2} \right| n \right> & =
  \frac{\left< m+1 \left| \right. n \right>  -2 \left< m \left| \right. n \right>
    + \left< m-1 \left| \right. n \right> }{a^2} \\
  & =
  \frac{\delta_{n,m+1}-2\delta_{n,m}+\delta_{n,m-1}}{a^2}.
\end{split}
\end{equation}
It is worth mentioning that for a given grid, the differentiation operators matrix elements are
dependent on the discretization scheme used. For example, if  Eq.~(\ref{eq:diffopdiscrete}) is used
to get $d^2/dx^2=d/dx \cdot d/dx$ it  would lead to a different result than the one in
Eq.~(\ref{eq:secdiffopdiscrete}). This is due to the fact that these matrix elements are not computed but approximated,
and this depends on the selected discretization scheme.

The discrete one-particle operators are written in the second quantized notation as
\begin{equation}\label{eq:onebodysecquant}
\hat{A}=\sum_{m,n} \left< m \left| \hat{A} \right| n \right> \hat{c}_m^\dagger\hat{c}_n,
\end{equation}
which, together with Eqs.~(\ref{eq:diffopdiscrete}), (\ref{eq:secdiffopdiscrete}), leads to
\begin{subequations}
\begin{equation}\label{eq:diffopsecquantA}
\frac{d}{dx}  = \sum_m \frac{1}{2a}
\left( \hat{c}_m^\dagger \hat{c}_{m+1} -  \hat{c}_m^\dagger \hat{c}_{m-1} \right) ,
\end{equation}
\begin{equation}\label{eq:diffopsecquantB}
\frac{d^2}{dx^2} = \sum_m \frac{1}{a^2}
\left( \hat{c}_m^\dagger \hat{c}_{m+1} -2 \hat{c}_m^\dagger \hat{c}_{m} + \hat{c}_m^\dagger \hat{c}_{m-1} \right).
\end{equation}
\end{subequations}
Equations~(\ref{eq:diffopsecquantA}), (\ref{eq:diffopsecquantB}) can be used to
discretize the effective mass Hamiltonian of a quasielectron in the clean spin and charge interaction-free 2DEG
\begin{equation}
\hat{H}_{\rm free}=\frac{\hat{\bm p}^2}{2m^*} \otimes \mathbf{I}_{\rm S}=
-\frac{\hbar^2}{2m^*}\left(\frac{\partial^2}{\partial x^2}+\frac{\partial^2}{\partial y^2} \right)
\otimes \mathbf{I}_{\rm S},
\end{equation}
where the discretized version of the same Hamiltonian is
\begin{eqnarray}\label{eq:tbhclean}
\hat{H}_{\rm free}=\sum_{ {\bf
mm'} \sigma} t_{\mathbf{mm}'}\hat{c}_{{\bf m}\sigma}^\dag \hat{c}_{{\bf m'}\sigma}.
\end{eqnarray}
This is familiar tight-binding Hamiltonian on the square lattice with single $s$-orbital $\langle {\bf r}|\bf m\rangle=\phi({\bf r}-{\bf m})$
per site. Here $\hat{c}_{{\bf m}\sigma}^\dag$ ($\hat{c}_{{\bf m}\sigma}$) is the
creation (annihilation)  operator of an electron at site ${\bf m}=(m_x,m_y)$.
The spin-independent hopping matrix element
$t_{\mathbf{mm}'} $ is given by
\begin{equation}\label{eq:spinindephop}
t_{\mathbf{mm}'} =\left< \mathbf{m} \left|
-\frac{\hbar^2}{2m^*}\frac{\partial^2}{\partial \mathbf{r}^2}
\right| \mathbf{m}^\prime \right> =
\left\{
\begin{array}{cc}
-t_{\rm O} & \mbox{if } \mathbf{m}=\mathbf{m}^\prime \pm a \mathbf{e}_{x,y} \\
0         & \mbox{otherwise }.
\end{array}
\right.
\end{equation}
where $t_{\rm O}=\hbar^2/(2m^*a^2)$ is the orbital hopping parameter.

\subsubsection{Discretization of the Rashba SO Hamiltonian}\label{sec:discr-rashba-hamilt}
Following the general discretization procedure outlined above,
we recast the Rashba Hamiltonian in the local orbital basis representation as
\begin{eqnarray}\label{eq:tbh}
\hat{H}=\sum_{{\bf m}\sigma} \varepsilon_{\bf m} \hat{c}_{{\bf
m}\sigma}^\dag\hat{c}_{{\bf m}\sigma}+\sum_{{\bf
mm'}\sigma\sigma'} \hat{c}_{{\bf m}\sigma}^\dag t_{\bf
mm'}^{\sigma\sigma'}\hat{c}_{{\bf m'}\sigma'}.
\end{eqnarray}
While this Hamiltonian is of tight-binding type, its hopping parameters are non-trivial $2 \times 2$
Hermitian  matrices  ${\bf t}_{\bf m'm}=({\bf t}_{\bf mm'})^\dagger$ in the spin space.
The on-site potential $\ep_{\bf m}$ describes any static local potential,
such as the electrostatic potential due to the applied voltage or
the disorder which is usually simulated via a uniform random variable
$\varepsilon_{\bf m} \in [-W/2,W/2]$ modeling short range isotropic scattering off spin-independent
impurities.
\begin{figure}[t]
\centerline{\psfig{file=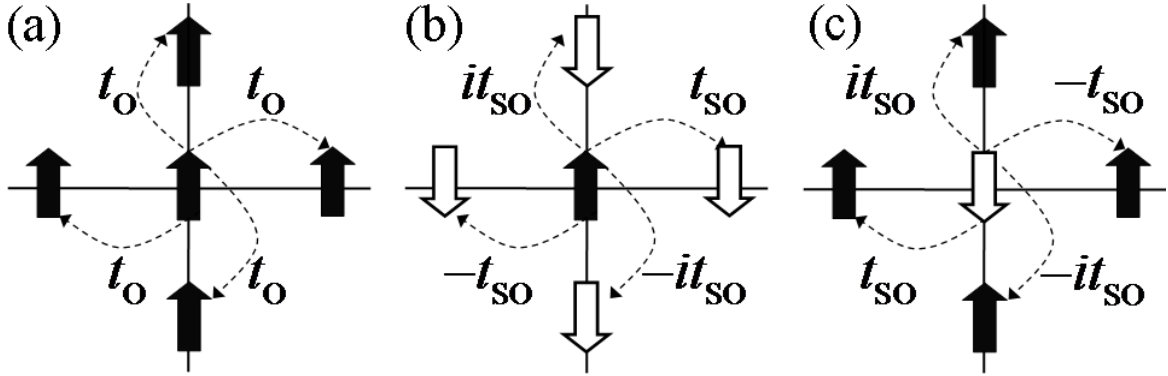,scale=0.6,angle=0}}
\caption{An illustration of the physical meaning of the orbital hopping $t_{\rm O}$ (a) and
the Rashba SO hopping $t_{\rm SO}$ for spin-$\up$ (b) and spin-$\dn$ (c) electrons
on the tight-binding lattice. As shown in panel (a), the probability amplitude for a spin to
hop between two sites without flipping is proportional to $t_{\rm O}$.
The $z$-spin $ |\!\! \up  \rangle_z$ at the central site $\mathbf{m}$ in panels (b), (c) flips by jumping to any of the four neighboring sites with
 probability amplitude  proportional to $t_{\rm SO}$.
}
\label{fig:rashba_hopping}
\end{figure}
The generalized
nearest neighbor hopping $t^{\sigma\sigma'}_{\bf mm'}=({\bf t}_{\bf mm'})_{\sigma\sigma'}$ accounts for the
Rashba coupling
\begin{eqnarray}\label{eq:hopping}
{\bf t}_{\bf mm'}=\left\{
\begin{array}{cc}
-t_{\rm O}{\bf I}_{\rm S}-it_{\rm SO}\hat{\sigma}_y &
({\bf m}={\bf m}'+{\bf e}_x)\\
-t_{\rm O}{\bf I}_{\rm S}+it_{\rm SO}\hat{\sigma}_x &  ({\bf m}={\bf m}'+{\bf e}_y)
\end{array}\right.,
\end{eqnarray}
through the SO hopping energy scale $t_{\rm SO}=\alpha/2a$ whose physical meaning is
illustrated in Fig.~\ref{fig:rashba_hopping}.  It is also useful to express the spin precession
length Eq.~(\ref{eq:lso})
\begin{equation}\label{eq:solength}
L_{\rm SO} = \frac{\pi t_{\rm O}}{2t_{\rm SO}} a,
\end{equation}
in terms of the lattice spacing $a$.

A direct correspondence between  the continuous
effective Rashba Hamiltonian Eq.~(\ref{eq:rashba}) (with quadratic and isotropic energy-momentum dispersion)
and its lattice version Eq.~(\ref{eq:tbh}) (with tight-binding dispersion)  is established  by
selecting the Fermi energy (e.g., $E_F=-3.8t_{\rm O}$)  of the injected
electrons to be close to the bottom of the band $E_b=-4.0t_{\rm O}$
(so that tight-binding dispersion reduces to the quadratic one), and by using
$t_{\rm O}=\hbar^2/(2 m^* a^2)$ for the orbital hopping  which yields the effective mass
$m^*$ in the continuum limit. For example, the InGaAs/InAlAs heterostructure employed in
experiments of \citeasnoun{Nitta1997}  is characterized by  the effective mass
$m^*=0.05m_0$ and the  width of the conduction band
$\Delta_b=0.9$ eV, which sets $t_{\rm O}=\Delta_b/8=112$ meV for the  orbital hopping
parameter on a square lattice (with four nearest neighbors of each site) and $a \simeq 2.6$ nm
for its  lattice spacing. Thus, the Rashba SO coupling of 2DEG formed in this  heterostructure,
tuned to a maximum value~\cite{Nitta1997} $\alpha=0.93 \cdot 10^{-11}$ eVm by the gate electrode,
corresponds to the SO hopping $t_{\rm SO}/t_{\rm O} \simeq 0.016$ in the lattice Hamiltonian Eq.~(\ref{eq:tbh}).

\subsubsection{Discretization of the extrinsic SO Hamiltonian}\label{sec:discr-extr-so}
The extrinsic SO coupling in disordered 2DEG is described by the Pauli Hamiltonian
\begin{equation}\label{eq:extr_hamil}
\hat{H} = \frac{\hat{p}_x^2+\hat{p}_y^2}{2m^*} + V_{\rm conf}(y) + V_{\rm disorder}(x,y)
+ \lambda \left(\hat{\vectg{\sigma}}\times\hat{\vect{p}} \right) \cdot \nabla V_{\rm disorder}(x,y),
\end{equation}
where $\lambda$ is the extrinsic SO coupling strength.
The fourth term in Eq.~(\ref{eq:extr_hamil}) can be rewritten as
\begin{equation}\label{eq:extrSO}
 \lambda \left(\hat{\vectg{\sigma}}\times\hat{\vect{p}} \right) \cdot \nabla V_{\rm disorder}(x,y) =
 -i\lambda \left[ \left( \partial_y V_{\rm disorder} \right) \partial_x -
\left( \partial_x V_{\rm disorder} \right) \partial_y \right] \hat{\sigma}_z .
\end{equation}
Equation~(\ref{eq:diffopdiscrete}) can then be used to find the discrete version of $\partial_x V_{\rm disorder}$
\begin{equation}\label{eq:discr_matrelem_extr}
\begin{split}
\left< \mathbf{m} \left| \partial_x V_{\rm disorder} \right| \mathbf{m}^\prime \right>  & =
\sum_{\mathbf{m}^{\prime\prime}}
\left< \mathbf{m} \left| \partial_x  \right| \mathbf{m}^{\prime\prime} \right>
\left< \mathbf{m}^{\prime\prime} \left| V_{\rm disorder} \right| \mathbf{m}^\prime \right> \\
& =
\frac{
V_{\rm disorder}(\mathbf{m}+a\mathbf{e}_x) \delta_{\mathbf{m}^\prime,\mathbf{m}+a\mathbf{e}_x}-
V_{\rm disorder}(\mathbf{m}-a\mathbf{e}_x) \delta_{\mathbf{m}^\prime,\mathbf{m}-a\mathbf{e}_x}
}{2a}.
\end{split}
\end{equation}
The final expression~\cite{Nikoli'c2007,Dragomirova2008} of the discretized version of Eq.~(\ref{eq:extr_hamil})
\begin{eqnarray}\label{eq:tbh_extr}
\hat{H}  & = &   \sum_{{\bf m},\sigma} \varepsilon_{\bf m} \hat{c}_{{\bf
m}\sigma}^\dag\hat{c}_{{\bf m}\sigma} - t_{\rm O} \sum_{ {\bf
mm'} \sigma} \hat{c}_{{\bf m}\sigma}^\dag \hat{c}_{{\bf m'}\sigma} \nonumber \\
&&  -  i  \lambda_{\rm SO} \sum_{{\bf m},\alpha\beta} \sum_{ij} \sum_{\nu\gamma} \epsilon_{ijz}
\nu \gamma (\varepsilon_{{\bf m} + \gamma {\bf e}_j} - \varepsilon_{{\bf m}+\nu {\bf e}_i}) \hat{c}_{{\bf
m},\alpha}^\dag \hat{\sigma}^z_{\alpha\beta} \hat{c}_{{\bf m}+\nu{\bf e}_i+\gamma{\bf e}_j,\beta},
\end{eqnarray}
introduces second neighbor hopping in the third term (in addition to usual nearest neighbor hopping in the second term). Here
$\lambda_{\rm SO}$ is the dimensionless extrinsic SO coupling strength $\lambda_{\rm SO}=\lambda \hbar/4a^2$ and $\epsilon_{ijz}$ stands for
the Levi-Civita totally antisymmetric tensor ($i,j$ denote the in-plane coordinate axes and dummy indices $\nu,\gamma$ take values $\pm 1$).  For
example, for the 2DEG in SHE experiments conducted by~\citeasnoun{Sih2005a}, the SO parameters are set to $\lambda_{\rm SO} \simeq 0.005$ and $t_{\rm SO} \simeq 0.003 t_{\rm O}$ for an additional Rashba term, assuming conduction bandwidth $\simeq 1$ eV and using the effective mass $m^* = 0.074 m_0$.

\section{SPIN CURRENT OPERATOR, SPIN DENSITY, AND SPIN ACCUMULATION IN THE PRESENCE OF INTRINSIC SO COUPLINGS}\label{sec:spin-curr-oper}

The theory of the intrinsic SHE in  infinite homogeneous systems is formulated in terms of  the spin current
density which is not conserved in media with SO coupling and, therefore, does not have well-defined experimental
measurement procedure associated with it. Moreover, these spin current densities can be non-zero even in thermodynamic
equilibrium~\cite{Rashba2003,Nikoli'c2006}. The conservation of charge implies the continuity equation in quantum mechanics for the
charge density $\rho=e |\Psi({\bf r})|^2$
\begin{equation} \label{eq:charge_continuity}
\frac{\partial \rho}{\partial t} + \nabla \cdot {\bf j} = 0,
\end{equation}
associated with a given wave function $\Psi({\bf r})$. From here one can extract the charge current density ${\bf j} = e {\rm Re} \, [ \Psi^{\dagger}({\bf r}) \hat{\bf v} \Psi({\bf r})]$
viewed as the quantum-mechanical expectation value [in the state  $\Psi({\bf r})$]
of the charge current density operator
\begin{equation}\label{eq:charge_current_op}
\hat{\bf j} = e \frac{\hat{n}({\bf r})\hat{\bf v} + \hat{\bf v}\hat{n}({\bf r})}{2}.
\end{equation}
This operator can also be obtained heuristically from the classical charge current density
${\bf j} = e n({\bf r}) {\bf v}$ via quantization procedure where the particle density $n({\bf r})$
and the velocity ${\bf v}$ are replaced by the corresponding operators and symmetrized
to ensure that $\hat{\bf j}$ is a Hermitian operator.

In SO-coupled systems $\hat{\bf j}$ acquires extra terms since the velocity operator
$i\hbar \hat{\bf v} = [\hat{\bf r},\hat{H}]$ is modified by the presence of SO terms
in the Hamiltonian $\hat{H}$. For example, for the Rashba SO Hamiltonian
Eq.~(\ref{eq:hso_rashba}) the velocity operator is $\hat{\bf v} = \hat{\bf p}/m^* -
(\alpha/\hbar) (\hat{\sigma}_y {\bf e}_x - \hat{\sigma}_x {\bf e}_y)$. The spin density
$S^i = \frac{\hbar}{2} [\Psi^\dagger ({\bf r}) \hat{\sigma}_i \Psi({\bf r})]$
then satisfies the following continuity equation
\begin{equation} \label{eq:spincont}
\frac{\partial S^i}{\partial t} + \nabla \cdot \mathbf{j}^i = F_{\rm S}^i.
\end{equation}
In contrast to the charge continuity equation Eq.~(\ref{eq:charge_continuity}), this contains the
spin current density
\begin{equation} \label{eq:spin_current_density}
{\mathbf{j}}^i= \frac{\hbar}{2} \Psi^\dagger({\bf r})
\frac{\hat{\sigma}_i \hat{\bf v} + \hat{\bf v} \hat{\sigma}_i}{2}\Psi({\bf r}),
\end{equation}
as well as a nonzero spin source term
\begin{equation} \label{eq:spin_source}
F_{\rm S}^i = \frac{\hbar}{2} {\rm Re} \, \left( \Psi^\dagger({\bf r})
\frac{i}{\hbar}[\hat{H},\hat{\sigma}_i] \Psi({\bf r}) \right).
\end{equation}
The nonzero $F_{\rm S}^i \neq 0$ term reflects non-conservation of spin in the presence of intrinsic
SO couplings which act as internal momentum-dependent magnetic field forcing spin into precession.
Thus, the plausible Hermitian operator of the spin current density
\begin{equation} \label{eq:spin_current_op}
\hat{j}^i_k = \frac{\hbar}{2} \frac{\hat{\sigma}_i \hat{v}_k + \hat{v}_k \hat{\sigma}_i}{2},
\end{equation}
is a well-defined quantity (a tensor with nine components where $\hat{j}^i_k$ describes
transport of spin $S_i$ in the $k$-direction, $i,k=x,y,z$) only when $\hat{\bf v}$ is spin
independent.

The lack of the usual physical justification for Eq.~(\ref{eq:spin_current_op}) in systems with intrinsic
SO couplings leads to an arbitrariness in the definition of the spin current~\cite{Shi2006}. Thus, different
definitions lead to ambiguities~\cite{Sugimoto2006} in the value of the intrinsic
spin-Hall conductivity $\sigma_{sH}={j}_y^z/E_x$ computed as the linear response to the applied
longitudinal electric field $E_x$ penetrating an infinite SO-coupled (perfect)
crystal. It also yields qualitatively different conclusions about the effect of impurities
on SHE~\cite{Nagaosa2008}, and does not allow us to connect~\cite{Nomura2005b} directly
the value of  $\sigma_{sH}$ to measured edge spin accumulation of opposite signs along
opposite lateral edges.

Under the time-reversal transformation, the mass, charge, and energy do not change
sign, while the velocity operator and the Pauli matrices change sign, $t \rightarrow -t \Rightarrow \hat{\bf v} \rightarrow -\hat{\bf v}$
and $t \rightarrow -t \Rightarrow \hat{\bm \sigma} \rightarrow - \hat{\bm \sigma}$.  Since the charge current density operator   Eq.~(\ref{eq:charge_current_op}) contains velocity, it changes sign under the time reversal $t \rightarrow -t \Rightarrow \hat{\bf j} \rightarrow -\hat{\bf j}$ and, therefore, has to vanish in the thermodynamic equilibrium [except in the presence of an external magnetic field  which breaks time-reversal invariance thereby allowing for circulating or diamagnetic charge currents even in thermodynamic  equilibrium~\cite{Baranger1989}]. On the other hand, the  spin current density operator Eq.~(\ref{eq:spin_current_op}) is the time-reversal invariant (or even) quantity $t \rightarrow -t \Rightarrow \hat{j}^i_k \rightarrow \hat{j}^i_k$: if the clock ran backward, spin current would continue to flow in the same direction. Thus, $\hat{j}^i_k$ can have non-zero expectation values even in thermodynamic equilibrium. This has been  explicitly demonstrated~\cite{Rashba2003} for the case of an infinite clean Rashba spin-split 2DEG where such {\em equilibrium} spin currents are polarized inside the plane. Therefore, the out-of-plane polarized spin current density has been considered as a genuine nonequilibrium SHE-induced response~\cite{Sinova2004,Hankiewicz2008}. However, within multiterminal finite-size Rashba coupled devices even out-of-plane polarized spin currents can have equilibrium nature~\cite{Nikoli'c2006}. This is essential information for the development of a consistent theory for transport (nonequilibrium) spin currents where the contributions from background (equilibrium) currents must be eliminated.

A plausible solution to these issues appears to be in defining a conserved quantity where non-conserved part Eq.~(\ref{eq:spin_source})
is moved to the right-hand side of  the continuity equation  Eq.~(\ref{eq:spincont}) and incorporated in the definition of a new spin
current operator~\cite{Shi2006}. However, this solution is of limited value since it cannot be used for realistic
devices with arbitrary boundary conditions due to sample edges and attached electrodes~\cite{Nagaosa2008}. In Sec.~\ref{sec:negf} we
discuss how to bypass these issues altogether by employing NEGF-based description of SHE in realistic finite-size
devices where spin transport is quantified through: ({\em i})  total spin currents flowing out of the sample through attached leads in which they
are conserved due to leads being ideal (i.e., spin and charge interaction free); ({\em ii}) local spin currents
whose sums in the leads is equal to the total currents of ({\em i}); and ({\em iii})  nonequilibrium spin density
near the boundaries (i.e., edge spin-Hall accumulation that is part of the definition of SHE) or within the
SO-coupled sample (which is not typically associated with SHE experimentally measured quantities, but represents
natural ingredient of theoretical modeling).

\section{NEGF APPROACH TO SPIN TRANSPORT IN MULTITERMINAL SO-COUPLED NANOSTRUCTURES}\label{sec:negf}

The NEGF theory~\cite{Leuween2006,Rammer2007,Haug2007} provides a powerful conceptual framework, as well as computational tools,
to deal with a variety of out-of-equilibrium situations in  quantum systems. This includes steady-state transport regime and
more general transient responses. Over the past three decades, it has become one of the major tools to study steady-state quantum
transport of charge currents through small devices in phase-coherent regime where electron is described by a single
wave function throughout the device. In this respect, NEGF offers efficient realization of Landauer's seminal
ideas to account for phase-coherent conduction in terms of the scattering matrix of the device~\cite{Datta1995}. Furthermore,
NEGF is much more complete framework making it possible to go beyond the paradigms of the Landauer-B\" uttiker scattering approach
by including many-body effects in transport where electron-electron interactions, electronic correlations, and electron-phonon
interactions introduce dephasing leading to incoherent electron propagation~\cite{Datta1995,Datta2005,Thygesen2008}.

The central concepts in NEGF description of quantum transport are: the sample Hamiltonian $\hat{H}$ and its matrix
representation ${\bf H}$ in a typically chosen local orbital basis (such as the discrete versions of the effective SO Hamiltonian discussed
in Sec.~\ref{sec:so-coupl-hamilt_latt_repr}); the retarded ${\bf \Sigma}_{\rm leads}$ and the lesser ${\bf \Sigma}^<$ self-energy matrices due to the
interaction of the sample with the electrodes; self-energy matrices due to many-body interactions within the sample ${\bf \Sigma}_{\rm int}$, ${\bf \Sigma}^<$;
and two independent Green functions---the retarded ${\bf G}$ and the lesser ${\bf G}^<$ one. The retarded Green function describes the density of available quantum-mechanical states, while the lesser one determines how electrons occupy those quantum states.\footnote{Both Green functions can be obtained from the contour ordered Green function defined for any two time values that lie along Kadanoff-Baym-Keldysh time contour~\cite{Leuween2006,Rammer2007,Haug2007}.}

In the case of phase-coherent calculations of total spin and charge currents, we only need to find
the retarded spin-resolved self-energies due to the electrodes ${\bf \Sigma}_{\rm leads}^\sigma$ (determining the escape rates of spin-$\sigma$
electrons into the  electrodes) and compute the retarded Green function elements connecting sites between interfaces where leads are attached
to the sample. For convenience, we term this usage of NEGF ``Landauer-B\"uttiker approach'' in Sec.~\ref{sec:multiprobe_LB}
since it allows to describe transport in multiterminal structures by constructing the transmission block of the
scattering matrix in terms of the (portion of) retarded Green function matrix.

For the computation of local quantities within the sample, we also need  to find matrix elements of the retarded Green function between any two
sites within the sample as well as ${\bf G}^<$. This requires to solve the Keldysh (integral or matrix) equation for ${\bf G}^<$. The same is true
for incoherent transport where dephasing takes place in the sample requiring ${\bf \Sigma}_{\rm int}$, ${\bf \Sigma}_{\rm int}^<$ self-energies to be computed
(typically in the self-consistent fashion together with ${\bf G}$ and ${\bf G}^<$), or for nonlinear transport where local charge density and the corresponding electric potential profile within the sample play a crucial role~\cite{Christen1996} in understanding the device current-voltage characteristics. We denote this procedure in Sec.~\ref{sec:multiprobe_LK} and Sec.~\ref{sec:multiprobel_LK_density} as the ``Landauer-Keldysh'' approach since here the full NEGF theory is applied to finite-size (clean or disordered) devices attached to semi-infinite electrodes, i.e., to the Landauer setup where such electrodes simplify boundary conditions for electrons assumed to escape to infinity through them to be thermalized in the  macroscopic reservoirs (ensuring steady-state transport).\footnote{The NEGF theory, which is often called ``Keldysh formalism,''~\cite{Leuween2006} has been initiated by the pioneering works of Schwinger, Baym, Kadanoff and Keldysh who considered infinite homogeneous systems out of equilibrium~\cite{Rammer2007,Haug2007}. Its present applications~\cite{Datta1995,Datta2005} to quantum transport in finite-size systems attached to semi-infinite electrodes can be traced to an early analysis of \citeasnoun{Caroli1971}.}

\subsection{Landauer-B\"uttiker approach to total spin currents in multiterminal nanostructures}\label{sec:multiprobe_LB}

\begin{figure}[t]
\centerline{
\psfig{file=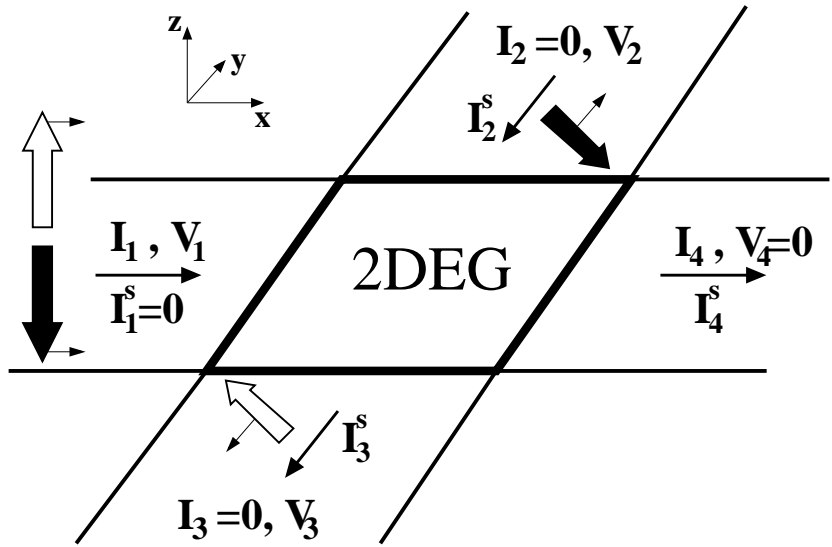,scale=0.40,angle=0}}
\caption{The four-terminal bridge for the detection of the mesoscopic SHE.
The central region is 2DEG, where electrons are confined within a semiconductor
heterostructure grown along the $z$-axis whose SIA  induces the Rashba
SO coupling. The four attached leads are clean, nonmagnetic, and without any SO coupling.
The unpolarized ($I_1^S=0$) charge current ($I_1 \neq 0$) injected through the longitudinal leads
induces spin-Hall current $I_2^{S_z}=-I_3^{S_z}$ in the transverse leads which act as the voltage probes
$V_2=V_3 \neq 0$, $I_2=I_3=0$.}
\label{fig:setup}
\end{figure}

The experiments on quantum Hall bridges in the early 1980s were posing a challenge for theoretical interpretation of
multiterminal transport measurements in the mesoscopic transport regime~\cite{Ando2003}. By viewing the current and voltage
probes on equal footing, \citeasnoun{Buttiker1986} has provided an elegant solution to these puzzles in the form of multiprobe formulas
\begin{equation}\label{eq:buttiker}
I_p = \sum_q (G_{qp} V_p - G_{pq}V_q)  = \sum_q G_{pq}(V_p - V_q).
\end{equation}
They relate charge current $I_p=I_p^\up + I_p^\dn$ in lead $p$ to the voltages $V_q$ in all other leads attached to the sample
via the conductance coefficients $G_{pq}$. To study the spin-resolved charge currents $I_p^\sigma$ ($\sigma=\up,\dn$) of individual
spin species $\up$, $\dn$ we imagine that each nonmagnetic lead in Fig.~\ref{fig:setup} consists of the two leads allowing only one spin species to propagate (as realized by, e.g., half-metallic ferromagnetic leads).
Upon replacement $I_p \rightarrow I_p^\sigma$ and $G_{pq} \rightarrow G_{pq}^{\sigma \sigma^\prime}$,
this viewpoint allows us to extract the multiterminal formulas for the spin-resolved charge
currents $I_p^\sigma$,  thereby obtaining the linear response relation for spin
current
\begin{equation}
I_p^S=\frac{\hbar}{2e} (I_p^\up - I_p^\dn),
\end{equation}
flowing through the lead $p$
\begin{eqnarray} \label{eq:spinbuttiker}
I_p^S  =
\frac{\hbar}{2e}\sum_q
\left[ (G_{qp}^{\up\up} + G_{qp}^{\dn\up} - G_{qp}^{\up\dn} -
G_{qp}^{\dn\dn}) V_p - (G_{pq}^{\up\up} + G_{pq}^{\up\dn} -
G_{pq}^{\dn\up} - G_{pq}^{\dn\dn}) V_q \right].
\end{eqnarray}
The spin-Hall conductance of the four-terminal bridge sketched in Fig.~\ref{fig:setup} is
then defined as
\begin{equation} \label{eq:gh}
G_{sH} = \frac{\hbar}{2e} \frac{I_2^S}{\Delta V}= \frac{\hbar}{2e} \frac{I_2^\uparrow - I_2^\downarrow}{V_1-V_4}.
\end{equation}
Below, we simplify the notation by introducing the labels
\begin{subequations}
\begin{eqnarray}
 G_{pq}^{\rm in} & = & G_{pq}^{\up\up}+G_{pq}^{\up\dn}-
G_{pq}^{\dn\up}-G_{pq}^{\dn\dn}, \\
 G_{pq}^{\rm out} & = & G_{pq}^{\up\up}+G_{pq}^{\dn\up}-
G_{pq}^{\up\dn}-G_{pq}^{\dn\dn}.
\end{eqnarray}
\end{subequations}
In fact, these coefficients have transparent physical interpretation:
$\frac{\hbar}{2e}G_{qp}^{\rm out} V_p$ is the spin current flowing from the lead $p$ with voltage
$V_p$ into other leads $q$ whose voltages are $V_q$, while $\frac{\hbar}{2e} G_{pq}^{\rm in} V_q$
is the spin current flowing from the leads $q \neq p$ into the lead $p$.

The standard charge conductance coefficients~\cite{Buttiker1986,Baranger1989,Datta1995} in the multiprobe
Landauer-B\" uttiker formalism  Eq.~(\ref{eq:buttiker}) are expressed in terms of the spin-resolved
conductances as $G_{pq}=G_{pq}^{\up\up}+G_{pq}^{\up\dn}+G_{pq}^{\dn\up}+G_{pq}^{\dn\dn}$.  Their introduction
in 1980s was prompted by the need to describe linear transport properties of a single sample (with specific
impurity arrangements and attached to specific probe configuration) by using measurable quantities
[instead of the bulk conductivity which is inapplicable to mesoscopic conductors~\cite{Baranger1989}].
They describe total charge current flowing  in and out of  the system in response to voltages applied
at its boundaries.

Regardless of the detailed microscopic physics of transport, conductance coefficients must satisfy the sum
rule $\sum_q G_{qp} = \sum_q G_{pq}$ in order to ensure the second equality in Eq.~(\ref{eq:buttiker}),
i.e.,  the charge current must be zero $V_q={\rm const.} \Rightarrow I_p \equiv 0$ in equilibrium.
On the other hand, the multiprobe spin current formulas Eq.~(\ref{eq:spinbuttiker})
apparently posses a nontrivial equilibrium solution $V_q={\rm const.} \Rightarrow I_p^S \neq 0$
[found by \citeasnoun{Pareek2004}] that would contradict the Landauer-B\" uttiker paradigm
demanding usage of only measurable quantities. However, when all leads are at the same potential,
a purely equilibrium nonzero term,  $\frac{\hbar}{2e} (G_{pp}^{\rm out} V_p - G_{pp}^{\rm in} V_p) = \frac{\hbar}{e}
(G^{\dn \up}_{pp} - G^{\up \dn}_{pp})V_p$, becomes relevant for $I_p^S$, canceling all other terms in Eq.~(\ref{eq:spinbuttiker})
to ensure that no {\em unphysical} total spin current $I_p^S \neq 0$ can appear in the leads of an unbiased ($V_q$=const.) multiterminal device~\cite{Souma2005a,Kiselev2005,Scheid2007}.

At zero temperature, the spin-resolved conductance coefficients
\begin{equation}\label{eq:spin_res_cond_coeff}
G_{pq}^{\sigma \sigma^\prime}=\frac{e^2}{h} \sum_{ij} |{\bf t}^{pq}_{ij,\sigma \sigma^\prime}|^2,
\end{equation}
where summation is over the conducting channels in the leads, are obtained from the Landauer-type formula
as the probability for spin-$\sigma^\prime$  electron incident in lead $q$ to be transmitted
to lead $p$  as spin-$\sigma$ electron.
The quantum-mechanical probability amplitude for this processes is given by the matrix elements
of the transmission matrix ${\bf t}^{pq}$, which is determined only by the wave functions
(or Green functions) at the Fermi energy~\cite{Baranger1989}.
The stationary states of the structure 2DEG + two leads supporting one or two  conducting
channels can be found exactly  by matching the wave functions in the leads to the eigenstates
of the Hamiltonian Eq.~(\ref{eq:rashba}), thereby allowing one to obtain the  charge conductance
from the Landauer transmission formula~\cite{Governale2004}. However, modeling of the full bridge geometry
with  two extra leads  attached in the transverse  direction, the existence of many open transverse propagating modes (``conducting
channels''), and possibly strong SO coupling regime when sample is bigger than the spin precession length ($L > L_{\rm SO}$),
is handled much  more efficiently through the NEGF formalism.\footnote{The multiterminal and
multichannel finite-size device can be also modeled using the random matrix theory for its scattering matrix~\cite{Bardarson2007}.
However, this method is strictly applicable only for weak SO coupling regime $L \ll L_{\rm SO}$~\cite{Aleiner2001}.}

For non-interacting particle which propagates through a finite-size
sample of arbitrary shape, the transmission matrices
\begin{eqnarray} \label{eq:transmission}
{\bf t}^{pq} &  =  & \sqrt{-\text{Im} \, {\bm \Sigma}_p \otimes {\bf I}_{\rm S} } \cdot
{\bf G}_{pq} \cdot \sqrt{-\text{Im}\, {\bm \Sigma}_q \otimes {\bf I}_{\rm S}}, \nonumber \\
\text{Im} \, {\bm \Sigma}_p &  = & \frac{1}{2i} \left( {\bm \Sigma}_p - {\bm \Sigma}_p^\dagger \right),
\end{eqnarray}
between different leads can be evaluated in a numerically exact fashion using
the real$\otimes$spin-space retarded Green operator $\hat{G}^r$
defined in the Hilbert space ${\mathcal H}_{\rm O} \otimes {\mathcal H}_{\rm S}$. Their matrix representation
in a basis $|{\bf m} \rangle \otimes |\sigma \rangle \in {\mathcal H}_{\rm O} \otimes {\mathcal H}_{\rm S}$
introduced by the Hamiltonian Eq.~(\ref{eq:tbh}) is obtained through  the matrix inversion
\begin{equation}\label{eq:green}
{\bf G} = \frac{1}{E {\bf I}_{\rm O} \otimes {\bf I}_{\rm S} - {\bf H} - \sum_{p=1}^4 {\bm \Sigma}_p \otimes {\bf I}_{\rm S}},
\end{equation}
Here $|\sigma \rangle$ are the eigenstates of the spin operator for the chosen spin quantization axis. The matrix elements $\langle {\bf m}^\prime,\sigma^\prime|\hat{G}^r| {\bf m},\sigma\rangle$ of ${\bf G}$ yield the probability amplitude for an electron to
propagate between two arbitrary locations ${\bf m}$ and ${\bf m}^\prime$ (with or without flipping its spin $\sigma$ during the motion)
inside an open conductor in the absence of inelastic processes.
Its submatrix ${\bf G}_{pq}$, which is required in Eq.~(\ref{eq:transmission}),
consists of those matrix elements which connects the layer of the sample attached
to the lead $q$ to the layer of the sample attached to the lead $p$. The sum of the self-energy matrices
$\sum_{p=1}^4 {\bm \Sigma}_p \otimes {\bf I}_{\rm S}$  accounts  for the ``interaction'' of an open system with the
attached four ideal semi-infinite leads $p$~\cite{Datta1995}.

\subsubsection{General expression for spin-Hall conductance}\label{sec:general}

Since the total charge current $I_p=I_p^\up +I_p^\dn$ depends only on the voltage
difference between the leads in Fig.~\ref{fig:setup}, we set one of them to zero
(e.g., $V_4=0$ is chosen as the reference potential) and apply voltage $V_1$ to the structure.
Imposing the requirement $I_2 = I_3 =0$  for the  voltage probes 2
and 3 allows us to get the voltages $V_2/V_1$ and $V_3/V_1$
by inverting the multiprobe charge current formulas Eq.~(\ref{eq:buttiker}). Finally, by
solving Eq.~(\ref{eq:spinbuttiker}) for $I_2^S$ we obtain the most general expression for
the spin-Hall conductance defined by Eq.~(\ref{eq:gh})
\begin{equation} \label{eq:gh_explicit}
G_{sH}=\frac{\hbar}{2e} \left[(G_{12}^{\rm out}+G_{32}^{\rm out}+ G_{42}^{\rm out})
\frac{V_2}{V_1} - G_{23}^{\rm in}\frac{V_3}{V_1} - G_{21}^{\rm in} \right].
\end{equation}
This quantity is measured in the units of the spin conductance quantum $e/4\pi$
(as the largest possible  $G_{sH}$ when transverse leads support only one open conducting channel),
which is  the counterpart\footnote{
Note that $(\hbar/2e)(e^2/h)=e/4\pi $, where $(e^2/h)$ is the natural unit
for spin-resolved charge conductance coefficients $G_{pq}^{\sigma\sigma^\prime}$.} of the familiar charge conductance quantum $e^2/h$.

In contrast to charge current, which is a scalar quantity, spin current has three components
because of the vector nature of spin
(i.e., different ``directions'' of spin correspond to different  quantum mechanical
superpositions of $|\!\! \up \rangle$ and $|\!\! \dn \rangle$  states).
Therefore, we can expect that, in general, the detection of spin transported through
the transverse leads of mesoscopic devices will find its expectation values to be
nonzero for all three axes. However, their flow properties
\begin{subequations}\label{eq:flow}
\begin{eqnarray}
I_2^{S_z} & = & - I_3^{S_z},\\
I_2^{S_x} & =& - I_3^{S_x}, \\
I_2^{S_y} & = &  I_3^{S_y},
\end{eqnarray}
\end{subequations}
show that only the $z$ and the $x$ components can represent the SHE response
for the Rashba SO-coupled  four-terminal bridge.
That is, if we connect the transverse leads 2 and 3 to each other
(thereby connecting the lateral edges of 2DEG by a wire),
only the spin current carrying $z$- and $x$-polarized spins
will flow through them, as expected from the general SHE phenomenology
where nonequilibrium spin-Hall accumulation detected in
experiments~\cite{Kato2004b} has opposite sign on the lateral edges of 2DEG.

Therefore, to quantify all nonzero components of the vector of transverse
spin current in the linear response regime, we can introduce three spin
conductances~\cite{Nikoli'c2005b}
\begin{subequations}\label{eq:sp_conductances_def_mutiprobe}
\begin{eqnarray}
G_{sH}^x & = & I_2^{S_x}/V_1, \\
G_{sp}^y & = & I_2^{S_y}/V_1, \\
G_{sH}^z & = & I_2^{S_z}/V_1,
\end{eqnarray}
\end{subequations}
 (assuming $V_4=0$).
They can be evaluated using the same general formula Eq.~(\ref{eq:gh_explicit})
where the spin quantization axis for $\up$, $\dn$ in spin-resolved charge
conductance coefficients is chosen to be the $x$, $y$, or $z$ axis, respectively.
For example,  selecting  $\hat{\sigma}_z  |\!\! \up \rangle = + |\!\! \up \rangle$ and
$\hat{\sigma}_z |\!\! \dn \rangle=  - |\!\! \dn \rangle$ for the basis
in which the Green operator Eq.~(\ref{eq:green}) is represented allows one to compute
the $z$-component of  the spin current $I_p^{S_z}$. In accord with
their origin revealed by Eq.~(\ref{eq:flow}), we denote $G_{sH}^z$ and $G_{sH}^x$
as the spin-Hall conductances, while $G_{sp}^y$ is labeled as the ``spin polarization'' conductance since
it stems from the polarization of 2DEG by the flow of unpolarized charge current in
the presence of SO couplings~\cite{Silsbee2004}.

\subsubsection{Symmetry properties of  spin conductances}\label{sec:symmetry}

Symmetry properties of the conductance coefficients with respect to the reversal of a bias
voltage or the direction of an external magnetic field play an essential role in our
understanding of linear response  electron transport in macroscopic and
mesoscopic conductors~\cite{Buttiker1986,Datta1995}.
For example, in the absence of magnetic field they satisfy  $G_{pq}=G_{qp}$
[which can be proved assuming a particular model for charge transport~\cite{Datta1995}].
Moreover, since the effective magnetic field ${\bf B}_{\rm R}({\bf p})$  of the Rashba SO
coupling depends on momentum, it does not break the time-reversal invariance thereby imposing
the following property of the spin-resolved conductance coefficients
$G_{pq}^{\sigma \sigma^\prime} = G_{qp}^{- \sigma^\prime -\sigma}$
in multiterminal SO-coupled bridges.

In addition, the ballistic four-terminal bridge in Fig.~\ref{fig:setup}
with no impurities possesses various geometrical symmetries.
It is invariant under rotations and reflections that interchange the leads,
such as:  ({\em i}) rotation $C_4$ ($C_2$) by an angle $\pi/2$ ($\pi$) around the $z$ axis
for a square (rectangular) 2DEG central region; ({\em ii}) reflection $\sigma_{vx}$ in the $xz$ plane;
and ({\em iii}) reflection $\sigma_{vy}$ in the $yz$ plane.
These geometrical symmetries, together with $G_{pq}=G_{qp}$ property,
specify $V_2/V_1 = V_3/V_1 \equiv 0.5$ solution for the voltages of the transverse
leads when $I_2 = I_3=0$  condition is imposed on their charge currents.

The device Hamiltonian containing the Rashba SO term commutes with the unitary
transformations which represent these symmetry operations in the Hilbert space
${\mathcal H}_{\rm O} \otimes {\mathcal H}_{\rm S}$: (i) $\hat{U}(C_2) \otimes \exp{(i\frac{\pi}{2}\hat{\sigma}_z)}$,
which performs the transformation
$\hat{\sigma}_x \rightarrow -\hat{\sigma}_x$,
$\hat{\sigma}_y \rightarrow - \hat{\sigma}_y$,
$\hat{\sigma}_z \rightarrow \hat{\sigma}_z$
and interchanges the leads 1 and 4 as well as the leads 2 and 3; (ii)
$\hat{U}(\sigma_{vx}) \otimes\exp{(i\frac{\pi}{2}\hat{\sigma}_y)}$,
which transforms the Pauli matrices
$\hat{\sigma}_x \rightarrow -\hat{\sigma}_x$,
$\hat{\sigma}_y \rightarrow \hat{\sigma}_y$,
$\hat{\sigma}_z \rightarrow - \hat{\sigma}_z$
and interchanges leads 2 and 3; and
(iii) $\hat{U}(\sigma_{vy}) \otimes\exp{(i\frac{\pi}{2}\hat{\sigma}_x)}$ which transforms
$\hat{\sigma}_x \rightarrow \hat{\sigma}_x$,
$\hat{\sigma}_y \rightarrow -\hat{\sigma}_y$,
$\hat{\sigma}_z \rightarrow -\hat{\sigma}_z$
and exchanges lead 1 with lead  4.
The Hamiltonian also commutes with the time-reversal operator.

The effect of  these symmetries on the spin-resolved charge conductance coefficients, and  the
corresponding spin conductances $G_{sH}^z$, $G_{sH}^x$ and $G_{sp}^y$ expressed in terms of them
through Eq.~(\ref{eq:gh_explicit}), is as follows.
The change in the sign of the spin operator means that  spin-$\up$ becomes  spin-$\dn$
so that, e.g.,  $G_{pq}^{\rm in}$ will be transformed into $-G_{qp}^{\rm in}$.
Also, the time-reversal implies changing the signs of all spin operators and
all momenta so that $G_{pq}^{\sigma \sigma^\prime} = G_{qp}^{- \sigma^\prime -\sigma}$
is equivalent to $G_{pq}^{\rm in}=-G_{qp}^{\rm out}$.
Thus, invariance with  respect to $\hat{U}(\sigma_{vy}) \otimes\exp{(i\frac{\pi}{2}\hat{\sigma}_x)}$
yields the identities
$G_{21}^{{\rm in},x}  =  G_{24}^{{\rm in},x}$, $G_{21}^{{\rm in},y} = -G_{24}^{{\rm in},y}$, and
$G_{21}^{{\rm in},z}  =  -G_{24}^{{\rm in},z}$.
These symmetries do not imply cancellation of $G_{23}^{{\rm in},x}$.
However, invariance with respect to
$\hat{U}(\sigma_{vx}) \otimes\exp{(i\frac{\pi}{2}\hat{\sigma}_y)}$ and
$\hat{U}(C_2) \otimes \exp{(i\frac{\pi}{4}\hat{\sigma}_z)}$
implies that $G_{23}^{{\rm in},y} \equiv 0$ and $G_{23}^{{\rm in},z} \equiv 0$.

These symmetry imposed conditions simplify the general formula Eq.~(\ref{eq:gh_explicit})
for spin conductances of a perfectly clean Rashba SO-coupled four-terminal bridge to
\begin{subequations} \label{eq:simple_rashba}
\begin{eqnarray}
G_{sH}^{x} & = & \frac{\hbar}{2e} \left(2G_{12}^{{\rm out},x}+G_{32}^{{\rm out},x}\right), \\
G_{sp}^{y} & = & \frac{\hbar}{2e} G_{12}^{{\rm out},y}, \\
G_{sH}^{z} & = & \frac{\hbar}{2e} G_{12}^{{\rm out},z}.
\end{eqnarray}
\end{subequations}
where we employ the result $V_2/V_1 = V_3/V_1 \equiv 0.5$ valid for a geometrically symmetric clean bridge.
Because this solution for the transverse terminal voltages is violated in disordered bridges,
its sample specific (for given impurity configuration) spin conductance cannot be computed from
simplified formulas Eq.~(\ref{eq:simple_rashba}), as it is often assumed in the recent mesoscopic
SHE studies~\cite{Sheng2006b,Ren2006}.

It insightful to apply the same symmetry analysis to the bridges with other types of SO couplings.
For example, if the Rashba term in the Hamiltonian Eq.~(\ref{eq:rashba}) is replaced by the
linear Dresselhaus SO term
$\frac{\beta}{\hbar} \left(\hat{p}_x \hat{\sigma}_x  - \hat{p}_y \hat{\sigma}_y  \right)$ due to
BIA~\cite{vZuti'2004}, no qualitative change in our analysis ensues
since the two SO couplings can be transformed into each  other by a unitary matrix
$(\hat{\sigma}_x + \hat{\sigma}_y)/\sqrt{2}$.
In this case, the spin-Hall response is signified by
$I_2^{S_z}  =  -I_3^{S_z}$ and $I_2^{S_y} = -I_3^{S_y}$ components of the transverse spin current,
while $I_2^{S_x} = I_3^{S_x}$.
For the Dresselhaus SO-coupled bridge, the general expression Eq.~(\ref{eq:gh_explicit}) simplifies to
\begin{subequations}
\begin{eqnarray}\label{eq:simple_dresel}
G_{sp}^{x} & = & \frac{\hbar}{2e} G_{12}^{{\rm out},x}, \\
G_{sH}^{y} & = & \frac{\hbar}{2e} \left( 2G_{12}^{{\rm out},y}+G_{32}^{{\rm out},y} \right), \\
G_{sH}^{z} & = & \frac{\hbar}{2e} G_{12}^{{\rm out},z}.
\end{eqnarray}
\end{subequations}
The qualitatively different situation emerges when both the Rashba and the linear Dresselhaus
SO couplings become  relevant in the central region of the bridge since in this case it is
impossible to find spin rotation which, combined with the spatial symmetry, would keep the
Hamiltonian invariant while only transforming the signs of its spin matrices.
Moreover, for such ballistic bridge the condition $I_2 = I_3 = 0$
leads to $V_2/V_1 = 1 - V_3/V_1$ solution for the voltages, whereas imposing the alternative
condition $V_2=V_3$  generates non-zero charge currents flowing through the transverse
leads 2 and 3 together with the spin currents [no simple relations akin
to Eq.~(\ref{eq:flow}) can be written in either of these cases].

\begin{figure}[t]
\centerline{\psfig{file=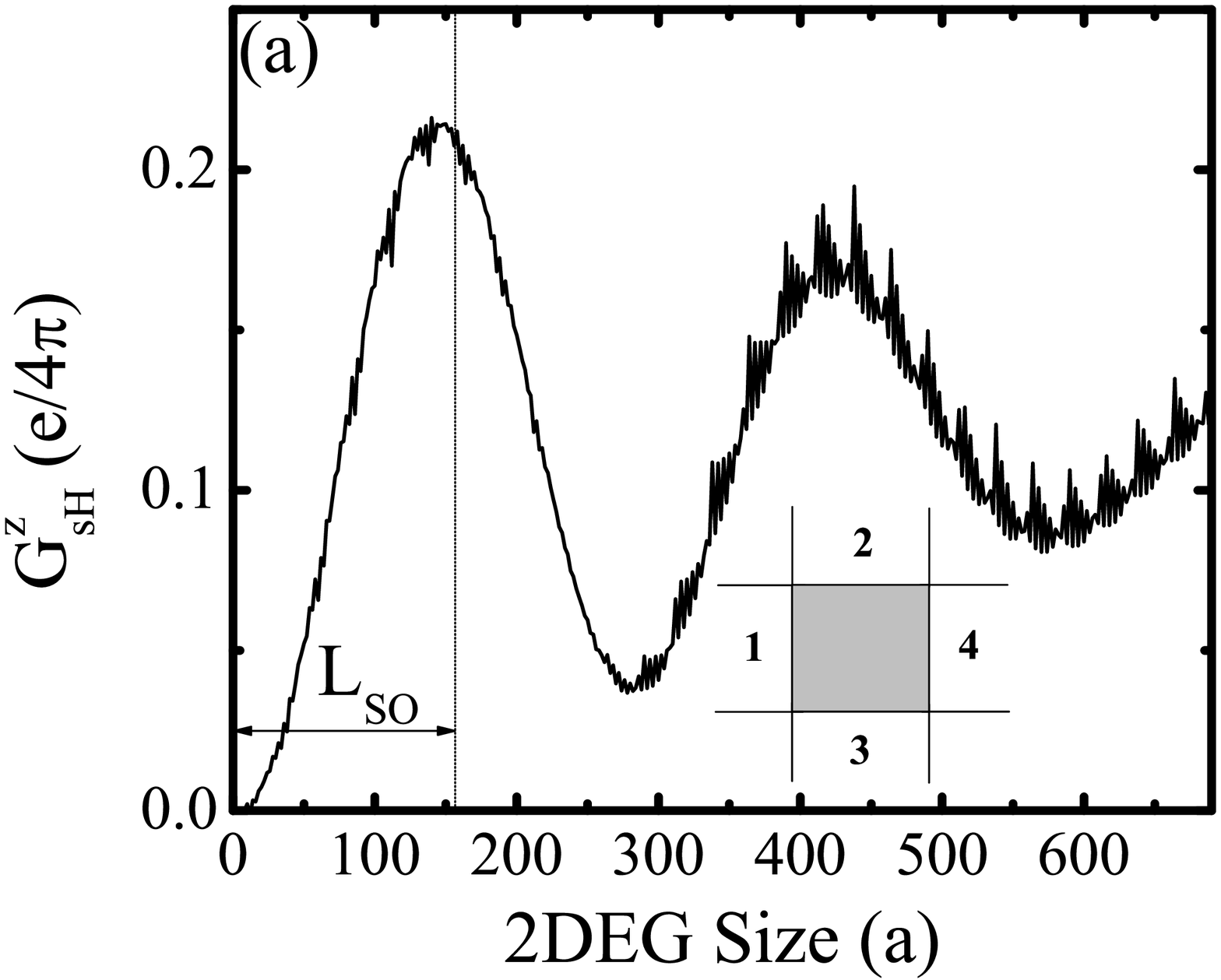,scale=0.37,angle=0} \psfig{file=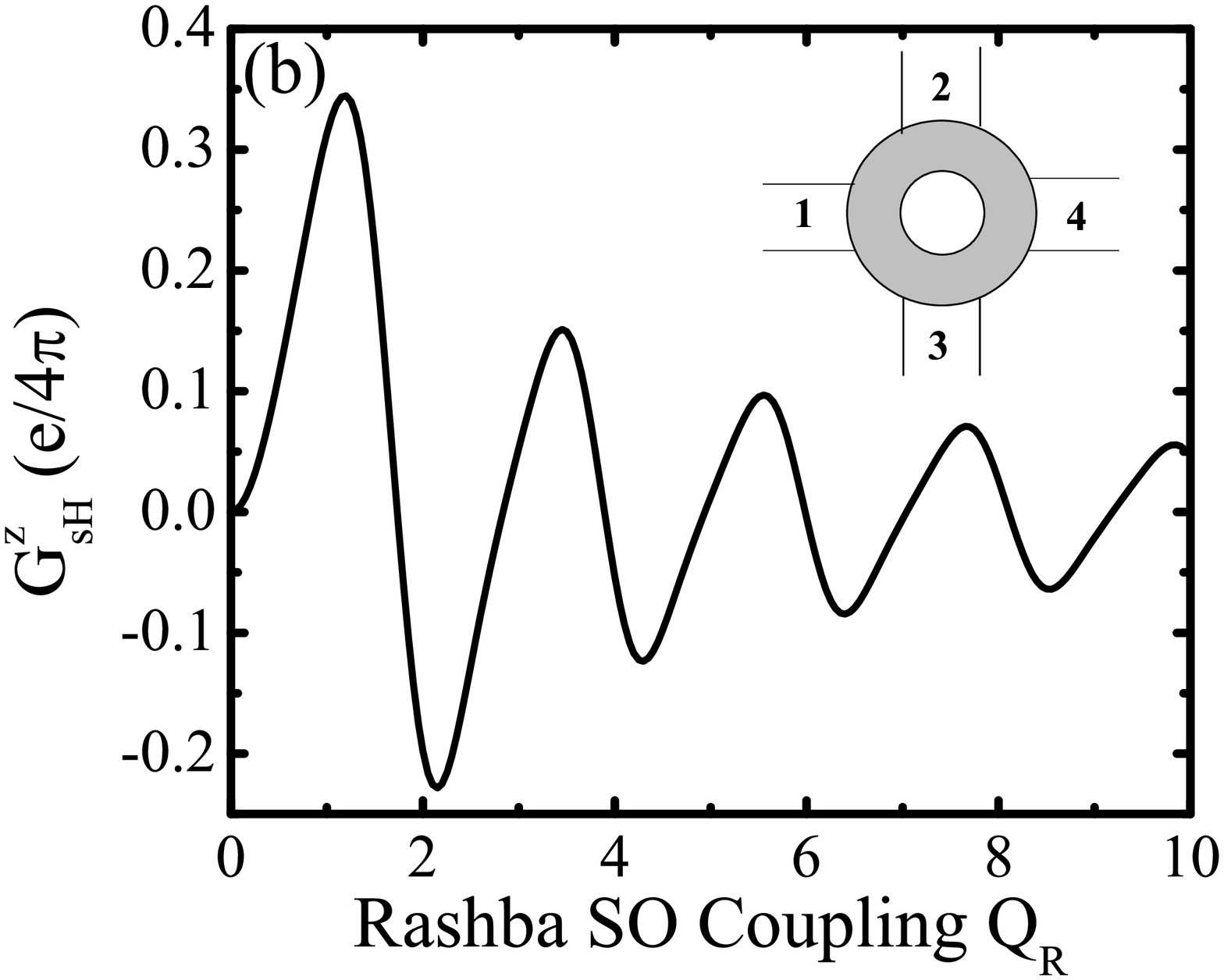,scale=0.37,angle=0}}
\caption{The spin-Hall conductance $G_{\rm sH}^z$ of: (a) Rashba SO-coupled square-shaped 2DEG as the function its size $L$
(in the units of the lattice spacing $a \simeq 3$ nm) for $t_{\rm SO}=0.01t_{\rm O}$ setting the spin precession length
$L_{\rm SO} \approx 157a$; and (b) single channel Aharonov-Casher ring as the function of the dimensionless Rashba SO
coupling $Q_R=(t_{\rm SO}/t_{\rm O}) N/\pi$ where the number of lattice sites discretizing the ring is $N=100$.}
\label{fig:meso_she}
\end{figure}

\subsubsection{Example: Transverse total spin currents in mesoscopic and quantum-interference-driven SHE nanostructures}\label{sec:lb_examples}

Figure~\ref{fig:meso_she} plots two examples of $G_{sH}^z$ computed for simply-connected (square-shaped)
2DEG and multiply-connected ring realized within 2DEG. In both cases, the Rashba SO coupling is present within
the device (gray area in the insets of Fig.~\ref{fig:meso_she}). Figure~\ref{fig:meso_she} confirms that
optimal device size to observe large mesoscopic SHE in multiterminal structures is indeed governed by
$L_{\rm SO}$~\cite{Nikoli'c2005b}. The computation of SHE conductance for very large square lattices in
Fig.~\ref{fig:meso_she}(a) is made possible by the usage of the recursive Green function algorithm for
four-terminal nanostructures discussed in Sec.~\ref{sec:recurs-green-funct}.

The spin conductance of the single-channel ring attached to four single-channel ideal leads in Fig.~\ref{fig:meso_she}(b) illustrates the {\em quantum-interference driven SHE} introduced by \citeasnoun{Souma2005a}. The ring represent a solid-state realization of the two-slit experiment since an electron entering the ring can
propagate in two possible directions---clockwise and counterclockwise. The superpositions of corresponding quantum states are sensitive to the acquired topological
phases in a magnetic [Aharonov-Bohm  effect] or an electric [Aharonov-Casher effect for particles with spin] external field whose changing generates
an oscillatory pattern of the ring conductance. Thus, unlike commonly discussed extrinsic and intrinsic SHE, whose essence can be understood using semiclassical arguments
and  wave packet propagation~\cite{Sinitsyn2008}, here the Aharonov-Casher phase difference acquired by opposite spin states during their cyclic evolution around the ring plays a crucial role. The spin conductance $G_{sH}^z$ becomes zero at specific values of the Rashba coupling when the destructive interference of opposite spins traveling in opposite directions
around the ring takes place~\cite{Souma2005a,Tserkovnyak2007}. The amplitude of such quasiperiodic oscillations of $G_{sH}^z$, which are absent in simply connected mesoscopic SHE devices  of Fig.~\ref{fig:meso_she}(a), gradually decreases at large Rashba coupling because of the reflection at the ring-lead interface.

\subsection{Landauer-Keldysh approach to local spin currents in multiterminal nanostructures}\label{sec:multiprobe_LK}

The theory of imaging of charge flow on nanoscale can be constructed efficiently within the framework of  lattice
models of mesoscopic devices and the corresponding bond charge currents~\cite{Baranger1989,Todorov2002}.
This makes it possible to obtain a detailed picture of charge propagation between two arbitrary sites of the
lattice~\cite{Nonoyama1998,Cresti2003,Metalidis2005,Nikoli'c2006,Zarbo2007}, thereby providing a way
to interpret recent scanning probe experiments. These experimental advances~\cite{Topinka2003} have brought
new insights into quantum transport by imaging its local features within a single sample,  rather than performing
conventional measurement of macroscopically averaged quantities. At the same time, scanning probe techniques
are becoming increasingly important in the quest for smaller electronic devices. For example, recent  imaging of  charge flow in
conventional $p-n$ junctions suggests that in structures shrunk below 50 nm individual positions of scarce dopants
will affect their function, thereby requiring to know precisely how charge carriers propagate on the  nanoscale~\cite{Yoshida2007}.
In the case of spin transport, Kerr rotation microscopy has made possible imaging of steady-state spin density driven by charge
flow through various SO-coupled semiconductor structures~\cite{Crooker2005a,Kato2005a}.

In this Section, we discuss NEGF-based tools that allow us to compute the spatial details of spin flow on the scale of
few nanometers by introducing the {\em bond spin currents}~\cite{Nikoli'c2006}. They represent the analog of bond charge currents,
as well as a lattice version of the spin current density conventionally employed in the studies of the intrinsic SHE in macroscopic
systems~\cite{Murakami2003a,Sinova2004}. Even though their sums over the cross sections within the sample change as we move from the
bottom to the top transverse electrode due to spin current non-conservation, they illustrate propagation of precessing spins  and within
the ideal transverse leads their sums over the cross section reproduce~\cite{Nikoli'c2006} conserved total spin currents discussed in
Sec.~\ref{sec:multiprobe_LB} using the Landauer-B\"{u}ttiker approach.

\subsubsection{Bond charge current operator in SO-coupled systems}

The charge conservation expressed through the familiar
continuity equation Eq.~(\ref{eq:charge_continuity}) yields a
uniquely determined bond charge current operator for quantum
systems  described on a lattice by a tight-binding-type of Hamiltonian
Eq.~(\ref{eq:tbh}). That is, the Heisenberg equation of motion
\begin{eqnarray}\label{eq:heisenberg}
\frac{d\hat{N}_{\bf m}}{dt}=\frac{1}{i\hbar}\left[\hat{N}_{\bf
m},\hat{H}\right],
\end{eqnarray}
for the electron number operator $\hat{N}_{\bf m}$ on site ${\bf m}$, $\hat{N}_{\bf m} \equiv \sum_{\sigma=\up,\dn} \hat{c}_{{\bf
m}\sigma}^\dag\hat{c}_{{\bf m}\sigma}$, leads to the charge continuity equation on the lattice
\begin{eqnarray}\label{eq:continuity}
&&e\frac{d\hat{N}_{\bf m}}{dt}+\sum_{k=x,y}\left(\hat{J}_{{\bf m},{\bf m}+{\bf e}_k}-\hat{J}_{{\bf m}-{\bf e}_k,{\bf m}}\right)=0.
\end{eqnarray}
This equation introduces the bond charge-current operator~\cite{Todorov2002}  $\hat{J}_{\bf mm'}$
which describes the particle current from site ${\bf m}$ to its nearest neighbor site ${\bf m}^\prime$.
The `bond' terminology is supported by a picture where current between two sites is represented
by a bundle of flow lines bunched together along a line joining  the two sites.

Thus, the spin-dependent Hamiltonian Eq.~(\ref{eq:tbh}) containing $2 \times 2$ hopping matrix
 defines the bond charge-current operator
$\hat{J}_{\bf mm'} =  \sum_{\sigma\sigma'} \hat{J}^{\sigma\sigma'}_{\bf mm'}$ which can be
viewed as the sum of four different {\em spin-resolved} bond charge-current operators
\begin{equation}\label{eq:J_charge_op1_spinresolved}
\hat{J}^{\;\sigma\sigma'}_{\bf mm'} =
\frac{e}{i\hbar}\left[\hat{c}_{{\bf m}'\sigma'}^\dag t_{ \bf
m'm}^{\sigma'\sigma}\hat{c}_{{\bf m}\sigma} -\mbox{H.c.} \right],
\end{equation}
where H.c. stands for the Hermitian conjugate of the first term. In particular, for the case of
$t_{ \bf mm'}^{\sigma\sigma'}$ being determined by the Rashba SO interaction Eq.~(\ref{eq:hopping}), we
can decompose the bond charge current operator into two terms, $\hat{J}_{\bf mm'} = \hat{J}^{\rm kin}_{\bf
mm'}+\hat{J}^{\rm SO}_{\bf mm'}$, having  transparent physical interpretation. The first term
\begin{equation}\label{eq:J_charge_op_kinetic}
\hat{J}^{\rm kin}_{\bf mm'} = \frac{e i t_{\rm O}}{\hbar}  \sum_{\sigma}
\left[ \hat{c}_{{\bf m}'\sigma}^\dag \hat{c}_{{\bf
m}\sigma} -\mbox{H.c.} \right],
\end{equation}
can be denoted as ``kinetic'' since it originates only from the kinetic energy
$t_{\rm O}$ and does not depend on the SO coupling energy $t_{\rm SO}$. On the other hand,
the second term
\begin{eqnarray}\label{eq:J_charge_op_so}
\hat{J}^{\rm SO}_{\bf mm'}&=& \left\{
\begin{array}{cc}
\displaystyle{ -\frac{4et_{\rm SO}}{\hbar^2}\hat{S}_{\bf mm'}^{y}
} & ({\bf m}={\bf m}^\prime+{\bf e}_x)\\
\displaystyle{ + \frac{4et_{\rm SO}}{\hbar^2}\hat{S}_{\bf mm'}^{x} }
& ({\bf m}={\bf m}^\prime+{\bf e}_y)
\end{array}
\right.
\nonumber \\
&=& \frac{4et_{\rm SO}}{\hbar^2}\left(({\bf m}'-{\bf
m})\times\hat{\vect{S}}_{\bf mm'}\right)_z
\end{eqnarray}
represents additional contribution to the intersite charge current flow due to
non-zero Rashba SO hopping $t_{\rm SO}$. Here we also introduce the
``bond spin-density'' operator
\begin{eqnarray}\label{eq:bond_spindensity_op}
\hat{\vect{S}}_{\bf mm'}&=& \frac{\hbar}{4} \sum_{\alpha\beta}
\left[\hat{c}_{{\bf m'}\alpha}^\dag
\hat{\vectg{\sigma}}_{\alpha\beta}\hat{c}_{{\bf m}\beta} +\mbox{H.c.}
\right],
\end{eqnarray}
defined for the bond connecting the sites ${\bf m}$ and ${\bf m}'$, which
reduces to the usual definition of the local spin density operator for
${\bf m}={\bf m}^\prime$ [see Eq.~(\ref{eq:spindensity})].

\subsubsection{Nonequilibrium bond charge current in SO-coupled systems} \label{sec:neq_bc}

The formalism of bond charge current makes it possible to compute physically
measurable~\cite{Topinka2003} spatial profiles of local charge current density
within the sample as the quantum-statistical average\footnote{
The quantum statistical average $\langle \hat{A} \rangle = {\rm Tr}[\hat{\rho}\hat{A}]$
is taken with respect to the  density matrix $\hat{\rho}$ that has evolved over sufficiently long time,
so that nonequilibrium state and all relevant interactions are fully established.}
of the bond charge-current operator in the nonequilibrium state~\cite{Caroli1971,Nonoyama1998,Cresti2003},
\begin{eqnarray}
\left<\hat{J}_{{\bf mm'}} \right>&=& \sum_{\sigma\sigma'}
\left<\hat{J}_{{\bf mm'}}^{\sigma \sigma'}\right>,\label{eq:J_charge_neq}
\\
 \left<\hat{J}_{{\bf
mm'}}^{\sigma\sigma'}\right>&=&\frac{-e}{\hbar}\int_{-\infty}^{\infty}
\frac{dE}{2\pi}\left[ t_{\bf m'm}^{\sigma'\sigma}G^{<}_{{\bf
mm'},\sigma\sigma'}(E)-t_{\bf mm'}^{\sigma\sigma'}G^{<}_{{\bf
m'm},\sigma'\sigma}(E) \right].
\label{eq:J_charge_neq_spinresolved}
\end{eqnarray}
Here the local charge current is expressed in terms of the nonequilibrium lesser Green function $G^{<}_{{\bf
m'm},\sigma'\sigma}(E)$.

The usage of the second quantized notation in Eq.~(\ref{eq:tbh}) facilitates the introduction
of NEGF expressions for the nonequilibrium expectation values~\cite{Caroli1971,Nonoyama1998}. We
imagine that at time $t^\prime = -\infty$ the sample
and the leads are not connected, while the left and the right longitudinal lead of a four-probe
device are in their own thermal equilibrium with the chemical potentials $\mu_L$ and $\mu_R$,
respectively, where $\mu_L=\mu_R+eV$. The adiabatic switching of the hopping parameter connecting
the leads and the sample generates time evolution of the density matrix of the
structure~\cite{Caroli1971}. The physical quantities are obtained as the nonequilibrium
statistical average $\left< \ldots \right>$  [with respect to the density matrix~\cite{Keldysh1965}
at time  $t^\prime=0$] of the corresponding quantum-mechanical operators expressed in terms of
$\hat{c}_{{\bf m}\sigma}^\dag$ and $\hat{c}_{{\bf m}\sigma}$. This will lead to the expressions of the type
$\left< \hat{c}^\dag_{{\bf m}\sigma} \hat{c}_{{\bf m}\sigma^\prime} \right>$, which define the lesser
Green function~\cite{Caroli1971,Nonoyama1998}
\begin{eqnarray} \label{eq:noneq_ev}
\left< \hat{c}^\dag_{{\bf m}\sigma}\hat{c}_{{\bf m}'\sigma^\prime} \right> &  = &
\frac{\hbar}{i}G^<_{{\bf m}'{\bf m},\sigma^\prime \sigma}(\tau=0) =
\frac{1}{2\pi i}\int_{-\infty}^{\infty}dE G^<_{{\bf m}'{\bf m},\sigma^\prime \sigma}(E).
\end{eqnarray}
Here we utilize the fact that the two-time correlation  function
[$\hat{c}_{{\bf m}\sigma}(t)=e^{i\hat{H}t/\hbar}\hat{c}_{{\bf m}\sigma}e^{-i\hat{H}t/\hbar}$]
\begin{equation}\label{eq:lesser}
G^<_{{\bf m}{\bf m}',\sigma\sigma'}(t,t')\equiv\frac{i}{\hbar}\left<\hat{c}^\dag_{{\bf
m}'\sigma'}(t')\hat{c}_{{\bf m}\sigma}(t)\right>,
\end{equation}
depends only on $\tau=t-t^\prime$ in stationary situations, so the time difference $\tau$
can be Fourier transformed to energy
\begin{equation}
G^<_{{\bf m}{\bf m}',\sigma\sigma'}(\tau) = \frac{1}{2\pi\hbar}\int_{-\infty}^{\infty}dE
G^<_{{\bf m}{\bf m}',\sigma\sigma'}(E)e^{iE\tau/\hbar},
\end{equation}
which will be utilized for steady-state transport studied here. We use the notation where
$\vect{G}^{<}_{\bf mm'}$ is a $2\times 2$ matrix in the spin space whose
$\sigma\sigma'$ element is $G^{<}_{{\bf mm'},\sigma\sigma'}$.

The spin-resolved bond charge current in Eq.~(\ref{eq:J_charge_neq_spinresolved})
describes the flow of charges which start as spin $\sigma$ electrons at the site ${\bf m}$ and end up as a
spin $\sigma'$ electrons at the site ${\bf m}'$ where possible spin-flips  $\sigma \neq \sigma'$
(instantaneous or due to precession) are caused by spin-dependent interactions. The decomposition of the
bond charge-current operator into the kinetic and SO terms  leads to a
Green function expression for the corresponding nonequilibrium bond charge currents
\begin{equation}
\left<\hat{J}_{\bf mm'} \right> = \left<\hat{J}_{\bf mm'}^{\rm kin}\right>
+\left<\hat{J}_{\bf mm'}^{\rm SO}\right>,
\end{equation}
with kinetic and SO terms given by
\begin{eqnarray}
\left<\hat{J}_{\bf mm'}^{\rm kin}\right> & = &
\frac{et_{\rm O}}{\hbar}\int_{-\infty}^{\infty} \frac{dE}{2\pi}
{\rm Tr}_{\rm S} \left[ \vect{G}^{<}_{\bf mm'}(E) -\vect{G}^{<}_{\bf m'm}(E)
\right]\label{eq:J_charge_neq2_kin}, \\
\left<\hat{J}_{\bf mm'}^{\rm SO}\right>&=& \frac{et_{\rm
SO}}{\hbar}\int_{-\infty}^{\infty} \frac{dE}{2\pi i}  {\rm Tr}_{\rm S} \left\lbrace \left[({\bf m}'-{\bf
m})\times \hat{\vectg{\sigma}} \right]_z \left[ \vect{G}^{<}_{\bf
mm'}(E)+\vect{G}^{<}_{\bf m'm}(E) \right] \right\rbrace \label{eq:J_charge_neq2_so}.
\end{eqnarray}
Note, however, that ``kinetic'' term is also influenced by the SO coupling through ${\bf G}^<$.
In the absence of the SO coupling, Eq.~(\ref{eq:J_charge_neq2_so}) vanishes and the bond charge
current reduces to the standard expression~\cite{Caroli1971,Nonoyama1998,Cresti2003}. The trace
${\rm Tr}_{\rm S}$ is performed in the spin Hilbert space. Similarly, we can also obtain the nonequilibrium
local charge density in terms of ${\bf G}^<$
\begin{eqnarray} \label{eq:charge_density}
e\left<\hat{N}_{\bf m}\right> & = & e\sum_{\sigma=\up,\dn}
\left<\hat{c}^\dag_{{\bf m}\sigma}\hat{c}_{{\bf m}\sigma}\right>
=  \frac{e}{2\pi i}\int_{-\infty}^{\infty}dE
\sum_{\sigma}G^<_{{\bf m}{\bf m},\sigma\sigma}(E) \nonumber \\
 & = & \frac{e}{2\pi i}\int_{-\infty}^{\infty}dE \, {\rm Tr}_{\rm S} [\vect{G}^<_{{\bf m}{\bf m}}(E)],
\end{eqnarray}
which is the statistical average value of the corresponding operator.

\subsubsection{Bond spin current operator in SO-coupled systems}\label{sec:bond-SC_SO_sys}

To mimic the plausible definition of the spin-current density operator ${j}^i_k$
in  Eq.~(\ref{eq:spin_current_op}), we introduce the  bond
spin-current operator for the spin-$S_i$ component as the symmetrized product
of the  spin-$\frac{1}{2}$  operator $\hbar \hat{\sigma}_i/2$  ($i=x,y,z$) and the bond
charge-current operator from Eq.~({\ref{eq:continuity}})
\begin{eqnarray}\label{eq:bond_spincurrent_op1}
\hat{J}^{S_i}_{\bf mm'} &\equiv& \frac{1}{4i} \sum_{\alpha\beta}
\left[\hat{c}_{{\bf m}'\beta}^\dag
\left\{\hat{\sigma}_i,{\bf t}_{ \bf
m'm}\right\}_{\beta\alpha}\hat{c}_{{\bf m}\alpha} -\mbox{H.c.}
\right].
\end{eqnarray}
By inserting the hopping matrix ${\bf t}_{ \bf m'm}$ Eq.~(\ref{eq:hopping}) of
the lattice SO Hamiltonian into this  expression we obtain its explicit form
for the case of the Rashba coupled system
\begin{eqnarray}\label{eq:bond_spincurrent_op2}
\hat{J}^{S_i}_{\bf mm'}
 =\frac{it_{\rm O}}{2}\sum_{\alpha\beta}\left(\hat{c}_{{\bf
m'}\beta}^\dag\left(\hat{\sigma}_i\right)_{\beta\alpha} \hat{c}_{{\bf
m}\alpha}-\mbox{H.c.}\right)
 + t_{\rm
SO} \hat{N}_{\bf mm'}\left[\vect{e}_i\times
(\vect{m}'-\vect{m})\right]_z,
\end{eqnarray}
which can be considered as the lattice version of Eq.~(\ref{eq:spin_current_op}).
Here we simplify the notation by using the ``bond electron-number operator'' $\hat{N}_{\bf mm'} \equiv \frac{1}{2}\sum_{\sigma}
\left(\hat{c}^\dag_{{\bf m}'\sigma}\hat{c}_{{\bf m}\sigma}+ {\rm H.c.} \right)$, which reduces to the standard electron-number
operator for ${\bf m}={\bf m}^\prime$.

\subsubsection{Nonequilibrium bond spin currents in SO-coupled systems} \label{eq:neq_bs}

Similarly to the case of the nonequilibrium  bond charge current in Sec.~\ref{sec:neq_bc},
the  nonequilibrium statistical average of the bond spin-current operator Eq.~(\ref{eq:bond_spincurrent_op2})
can be expressed using the lesser Green function ${\bf G}^<$ as
\begin{subequations}\label{eq:bond_spincurrent_full}
\begin{eqnarray}
\left<\hat{J}_{\bf mm'}^{S_i}\right>&=& \left<\hat{J}_{\bf
mm'}^{S_i{\rm (kin)}}\right>+ \left<\hat{J}_{\bf mm'}^{S_i{\rm
(so)}}\right>,\label{eq:bond_spincurrent} \\
\left<\hat{J}_{\bf mm'}^{S_i{\rm (kin)}}\right>&=&
\frac{t_{\rm O}}{2}\int_{-\infty}^{\infty} \frac{dE}{2\pi}
{\rm Tr}_{\rm S} \left[\hat{\sigma}_i \left(\vect{G}^{<}_{\bf m'm}(E) - \vect{G}^{<}_{\bf
mm'}(E) \right)
\right],\label{eq:bond_spincurrent_kin}
\\
\left<\hat{J}_{\bf mm'}^{S_i{\rm (SO)}}\right>&=&
\left[\vect{e}_i\times (\vect{m}'-\vect{m})\right]_z \frac{t_{\rm
SO}}{2}\int_{-\infty}^{\infty} \frac{dE}{2\pi i}
 {\rm Tr}_{\rm S} \left[ \vect{G}^{<}_{\bf
mm'}(E)+\vect{G}^{<}_{\bf m'm}(E)
\right]\label{eq:bond_spincurrent_so}.
\end{eqnarray}
\end{subequations}
Here we also encounter two terms which can be interpreted as the kinetic and the SO
contribution to  the bond spin current crossing from  site ${\bf m}$
to  site ${\bf m}'$. However, we emphasize that such SO contribution to
the spin-$S_z$ bond current is identically equal to zero, which
simplifies  the expression for this component to Eq.~(\ref{eq:bond_spincurrent_kin})
as the primary spin current response in the SHE.

\subsection{Landauer-Keldysh approach to local spin densities}\label{sec:multiprobel_LK_density}

Motivated by recent advances in Kerr rotation microscopy, which have made possible  experimental imaging
of steady-state spin polarization in various SO-coupled semiconductor
structures~\cite{Crooker2005a,Kato2005a}, we discuss in this Section a NEGF-based approach to computation of the
spatial profiles of $\left<\hat{\vect{S}}_{\bf m}\right>$. Unlike local spin current density, such {\em local flowing spin density}
is a well-defined and measurable quantity that offers insight into the spin flow in the nonequilibrium steady transport state.
For $\left<\hat{\vect{S}}_{\bf m}\right>$ computed at the lateral edges of the sample, we use the
term ``spin accumulation'' which was directly measured in the seminal spin-Hall
experiments~\cite{Kato2004b,Wunderlich2005}.

\subsubsection{Local spin density and its continuity equation} \label{sec:local_spin}

The local spin density in the lattice models is determined by the
local spin operator $\hat{{\bf S}}_{\bf m}=(\hat{S}^x_{\bf m},\hat{S}^y_{\bf
m},\hat{S}^z_{\bf m})$ at site ${\bf m}$ defined by
\begin{eqnarray}\label{eq:spindensity}
\hat{{\bf S}}_{\bf
m}=\frac{\hbar}{2}\sum_{\alpha\beta}\hat{c}^\dag_{{\bf
m}\alpha}\hat{\vectg{\sigma}}_{\alpha\beta} \hat{c}_{{\bf m}\beta}.
\end{eqnarray}
The Heisenberg equation of motion for each component $\hat{\bf S}^i$ ($i=x,y,z$)
of the spin density operator
\begin{equation}\label{eq:spinheisenberg}
\frac{d\hat{S}^i_{\bf m}}{dt}=\frac{1}{i\hbar}\left[\hat{S}^i_{\bf
m},\hat{H}\right],
\end{equation}
can be written in the form
\begin{eqnarray}\label{eq:spincontinuity}
&&\frac{d\hat{S}^i_{\bf
m}}{dt}+\sum_{k=x,y}\left(\hat{J}^{S_i}_{{\bf m},{\bf m}+{\bf
e}_k}-\hat{J}^{S_i}_{{\bf m}-{\bf e}_k,{\bf
m}}\right)=\hat{F}_{\bf m}^{S_i},
\end{eqnarray}
where $\hat{J}^{S_i}_{{\bf m}{\bf m}'}$ is the bond spin-current
operator given by Eq.~(\ref{eq:bond_spincurrent_op1}) so that the
second term on the left-hand side of Eq.~(\ref{eq:spincontinuity})
corresponds to the ``divergence'' of the bond spin current on
site ${\bf m}$. Here, in analogy with Eq.~(\ref{eq:spin_source}),
we also find the lattice version of the spin source operator $\hat{F}_{\bf m}^{S_i}$
whose  explicit form is
\begin{subequations}
\begin{eqnarray}
\hat{F}_{\bf m}^{S_x}&=&-\frac{t_{\rm SO}}{t_{\rm O}}\left(\hat{J}^{S_z}_{{\bf
m},{\bf m}+{\bf e}_x}+\hat{J}^{S_z}_{{\bf m}-{\bf e}_x,{\bf m}}
\right),
\\
\hat{F}_{\bf m}^{S_y}&=&-\frac{t_{\rm SO}}{t_{\rm O}}\left(\hat{J}^{S_z}_{{\bf
m},{\bf m}+{\bf e}_y}+\hat{J}^{S_z}_{{\bf m}-{\bf e}_y,{\bf m}}
\right),
\\
\hat{F}_{\bf m}^{S_z}&=&\frac{t_{\rm SO}}{t_{\rm O}}\left(\hat{J}^{S_x}_{{\bf
m},{\bf m}+{\bf e}_x}+\hat{J}^{S_x}_{{\bf m}-{\bf e}_x,{\bf m}}
+\hat{J}^{S_y}_{{\bf m},{\bf m}+{\bf
e}_y}+\hat{J}^{S_y}_{{\bf m}-{\bf e}_y,{\bf m}} \right).
\end{eqnarray}
\end{subequations}
The presence of the non-zero term $\hat{F}_{\bf m}^{S_i}$ on the right-hand side of  the
spin continuity Eq.~(\ref{eq:spincontinuity}) signifies, within the
framework of bond spin currents, the fact that spin is not conserved in SO-coupled
systems. The fact that the bond spin  current operator
Eq.~(\ref{eq:bond_spincurrent_op1}) appears in the spin continuity equation
Eq.~(\ref{eq:spincontinuity}) as its divergence implies that its definition in
Eq.~(\ref{eq:bond_spincurrent_op1}) is plausible. However, the presence of the
spin source operator $\hat{F}_{\bf m}^{S_i}$ reminds us that such definition cannot
be made unique~\cite{Shi2006}, in sharp contrast to bond charge current which is
is uniquely  determined by the charge continuity Eq.~(\ref{eq:continuity}).

Evaluating the statistical average of Eq.~(\ref{eq:spincontinuity}) in a steady state (which can
be either equilibrium or nonequilibrium), leads to the identity
\begin{eqnarray}
\sum_{k=x,y}\left(\left<\hat{J}^{S_i}_{{\bf m},{\bf m}+{\bf
e}_k}\right>-\left<\hat{J}^{S_i}_{{\bf m}-{\bf e}_k,{\bf
m}}\right>\right)=\left<\hat{F}_{\bf m}^{S_i}\right>.
\end{eqnarray}
In particular, for the spin-$S_z$ component  we get
\begin{eqnarray}
\sum_{k=x,y}\left(\left<\hat{J}^{S_z}_{{\bf m},{\bf m}+{\bf
e}_k}\right>-\left<\hat{J}^{S_z}_{{\bf m}-{\bf e}_k,{\bf
m}}\right>\right) =
\frac{t_{\rm SO}}{t_{\rm O}}\sum_{k=x,y}\left(\left<\hat{J}^{S_k}_{{\bf
m},{\bf m}+{\bf e}_k}\right>+\left<\hat{J}^{S_k}_{{\bf m}-{\bf
e}_k,{\bf m}}\right> \right),
\end{eqnarray}
which relates the divergence of the spin-$S_z$ current (left-hand side)
to the spin-source (right-hand side) determined by the sum of the longitudinal
component of the spin-$S_x$ current and the transverse component
of the spin-$S_y$ current.

Since no experiment has been proposed to measure local spin current density within the SO
coupled sample, defined through Eq.~(\ref{eq:spin_current_density}) or its lattice
equivalent Eq.~(\ref{eq:bond_spincurrent_full}), we can obtain additional information about the  spin fluxes
within the sample by computing
\begin{equation} \label{eq:spin_density}
\begin{split}
\left<\hat{\vect{S}}_{\bf m}\right>
& = \frac{\hbar}{2}\sum_{\alpha,\beta=\up,\dn}\hat{\vectg{\sigma}}_{\alpha\beta}
\left<\hat{c}^\dag_{{\bf m}\alpha}\hat{c}_{{\bf m}\beta}\right>
=   \frac{\hbar}{4\pi i}\int_{-\infty}^{\infty}dE\sum_{\alpha,\beta=\up,\dn}\hat{\vectg{\sigma}}_{\alpha\beta}
G^<_{{\bf m}{\bf m},\beta\alpha}(E) \\
& =\frac{\hbar}{4\pi i}\int_{-\infty}^{\infty}dE \, {\rm Tr}_{\rm S}
\left[\hat{\vectg{\sigma}}\vect{G}^<_{{\bf m}{\bf m}}(E)\right],
\end{split}
\end{equation}
as the nonequilibrium spin density driven by charge transport in the presence of SO couplings.

\subsection{Spin-resolved NEGFs for finite-size multiterminal devices}\label{sec:spin-resolv-land}
The spin-dependent NEGF formalism discussed in Sec.~\ref{sec:multiprobe_LK} does not actually depend on the details
of the external driving force which brings the system into a nonequilibrium state. That is, the system can be
driven by either the homogeneous electric field applied to an infinite homogeneous 2DEG or
the voltage (i.e., electrochemical potential) difference between
the electrodes attached to a {\em finite-size} sample. For example,
in the latter case, the external bias voltage only shifts the relative chemical
potentials of the reservoirs into which the longitudinal leads (employed to simplify
the boundary conditions) eventually terminate, so that the electrons do  not feel any
electric field in the course of ballistic propagation through clean 2DEG central region. The information
about these different situations is encoded into the lesser Green  function ${\bf G}^<$.

Here we focus on experimentally relevant spin-Hall devices where finite-size
central region (C),  defined on the $L \times L$ lattice, is attached to four external
semi-infinite leads of the same width $L$. The leads at infinity terminate into the reservoirs where electrons are brought into thermal equilibrium, characterized by the Fermi-Dirac distribution function $f(E-eV_p)$, to  ensure the steady-state
transport. In such multiterminal Landauer setup~\cite{Buttiker1986,Baranger1989,Datta1995}, current is limited by  quantum transmission
through a potential profile while power is dissipated non-locally in the reservoirs. The voltage in
each lead  of the four-terminal spin-Hall bridge is $V_p$ ($p=1,\ldots,4$), so that the
on-site potential $\varepsilon_{\bf m}$ within the leads has to  be shifted by $eV_p$.

The spin-dependent lesser Green function ${\bf G}^<$ defined in Eq.~(\ref{eq:lesser}) is evaluated within
the finite-size sample region as a $2L^2 \times 2L^2$ matrix in the site$\otimes$spin space through
the  spin-resolved matrix Keldysh equation~\cite{Keldysh1965}
\begin{eqnarray}\label{eq:keldysh}
{\bf G}^<(E) & = & {\bf G}(E){\bf \Sigma}^<(E) {\bf G}^\dagger(E),
\end{eqnarray}
which is valid for steady-state transport (i.e., when transients have died away).
Within the effective single-particle picture, the retarded Green function is
computed by inverting the Hamiltonian of an open system sample+leads
\begin{equation} \label{eq:retarded}
{\bf G}(E) =\left[E{\bf I}_C-{\bf H}_C - eU_{\bf m}-\sum_{p}{\bf \Sigma}_{p}(E - eV_p)\right]^{-1}.
\end{equation}
Here the retarded ${\bf \Sigma}_{p}(E)$ and the lesser ${\bf \Sigma}^<(E)$ self-energy matrices
\begin{eqnarray} \label{eq:self_energies}
{\bf \Sigma}_{p}(E)&=&{\bf H}^{\dag}_{pC}\left[\left(E+i0_+\right){\bf I}_p-{\bf H}^{\rm
lead}_{p}\right]^{-1} {\bf H}_{pC}, \label{eq:se_retarded}
\\
{\bf \Gamma}_{p}(E)&=&i\left[{\bf \Sigma}_{p}(E)-{\bf
\Sigma}^\dag_{p}(E)\right], \label{eq:se_gamma}
\\
{\bf \Sigma}^<(E) & = & i \sum_{p}{\bf \Gamma}_{p}(E-eV_p)f(E-eV_p), \label{eq:se_lesser}
\end{eqnarray}
are exactly computable in the non-interacting electron approximation and without any inelastic processes
taking place within the sample. They account for the ``interaction'' of the SO-coupled sample with
the attached leads, thereby generating a finite lifetime that electron spends within the 2DEG before escaping
through the leads toward the macroscopic thermalizing reservoirs. Here ${\bf I}_C$ is the $2L^2\times 2L^2$
identity matrix and  ${\bf I}_{p}$ is the identity matrix in the infinite
site$\otimes$spin space of the lead $p$. We  use the following Hamiltonian matrices
\begin{subequations}
\begin{eqnarray}
\left({\bf H}_{C}\right)_{{\bf m}{\bf
m}',\sigma\sigma'}&=&\left<1_{{\bf
m}\sigma}\right|\hat{H}\left|1_{{\bf m'}\sigma'}\right>,\;\;({\bf
m}, {\bf m'}\in C),
 \\
\left({\bf H}^{\rm lead}_{p}\right)_{{\bf m}{\bf
m}',\sigma\sigma'}&=&\left<1_{{\bf
m}\sigma}\right|\hat{H}\left|1_{{\bf m'}\sigma'}\right>,\;\;({\bf
m}, {\bf m'}\in p),
\\
\left({\bf H}_{pC}\right)_{{\bf m}{\bf
m}',\sigma\sigma'}&=&\left<1_{{\bf
m}\sigma}\right|\hat{H}\left|1_{{\bf m'}\sigma'}\right>,\;\;({\bf
m}\in p, {\bf m'}\in\mbox{C}),
\end{eqnarray}
\end{subequations}
where $\left|1_{{\bf m}\sigma}\right>$ is a vector in the Fock
space (meaning that the occupation number is one for the single
particle state $\left|{\bf m}\sigma\right>$ and zero otherwise)
and $\hat{H}$ is the Hamiltonian given in
Eq.~(\ref{eq:tbh}).

In the general case of arbitrary applied bias voltage, the gauge invariance of measurable quantities
(such as the current-voltage characteristics) with respect to the shift of electric potential everywhere
by a constant $V_{\rm c}$, $eV_p \rightarrow eV_p + eV_{\rm c}$ and $eU_{\bf m} \rightarrow eU_{\bf m} + eV_{\rm c}$,
is satisfied on the proviso that the retarded self-energies ${\bm \Sigma}_p(E-eV_p)$ introduced by each lead depend
explicitly on  the applied voltages at the sample boundary, while the computation of the retarded Green function
${\bf G}(E)$  has to include the  electric potential landscape\footnote{
The potential landscape $U_{\mathbf{m}}$
can be obtained from the Poisson equation with charge
density Eq.~(\ref{eq:charge_density}) as the source.} $U_{\bf m}$ within the
sample~\cite{Christen1996}.  However, when the applied bias is low, so that
linear response zero-temperature quantum transport takes place through  the sample
[as determined by ${\bf G}(E_F)$], the exact profile  of the internal potential
becomes irrelevant~\cite{Baranger1989,Nikoli'c1999}.

\subsubsection{How to introduce dephasing into SHE nanostructures}

Much of the applications of NEGF techniques to mesoscopic semiconductor and nanoscopic molecular
systems treat phase-coherent transport by relying on some type of the Landauer transmission formula~\cite{Rocha2006}.
On the other hand, applications demand devices operating at room temperature where device size is typically
much bigger than the dephasing length so that one seldom observes quantum interference effects in the
transmission due to multiple scattering off impurities and boundaries. Since such quantum coherence
effects are also seen in the linear response spin density in SO-coupled quantum wires, typically computed
at zero-temperature~\cite{Nikoli'c2005d,Reynoso2006}, to compare such calculations with experiments conducted
at finite temperature it is  desirable to introduce dephasing into NEGF formalism in a simple yet controllable
fashion (which leaves the  charge current conservation intact).

The most widely used example of the simple approach to inclusion of dephasing are fictitious B\"uttiker voltage probes~\cite{Buttiker1986},
such as one-dimensional  electrodes attached at each site ${\bf m}$ of the lattice and with their electrochemical potential
adjusted to ensure that the current drawn by each probe is zero.\footnote{Since current through B\"uttiker probes is fixed
to be zero, then for every electron that enters the lead and is absorbed by the corresponding reservoir another one
will come with phase memory washed out due to equilibration by inelastic effects assumed to take place
in the reservoir. Thus, this method introduces inelastic effects is a simple phenomenological way.} In NEGF formalism
this introduces additional self-energy whose matrix elements are $({\bf \Sigma}_{\rm deph})_{{\bf mm},\sigma \sigma} = -i\eta$
for all sites ${\bf m}$ within the sample except for those where the external electrodes are
attached~\cite{Golizadeh-Mojarad2007a}. However, while B\"uttiker probes coupled to each site can remove sharp features in
the transmission of the device, they also introduce additional scattering events thereby artificially enhancing the resistance.

A ``momentum conserving'' phenomenological model for dephasing, that is easily implemented within the NEGF formalism,
has recently  been proposed by~\citeasnoun{Golizadeh-Mojarad2007a}. Here the self-energies are taken as ${\bf \Sigma}_{\rm deph} (E) = d_i {\bf G}(E)$
and ${\bf \Sigma}_{\rm deph}^< (E)=d_i {\bf G}^<(E)$, with parameter $d_i$ controlling the strength of the elastic dephasing processes.
This choice is motivated by the general expressions for the first-order self-consistent Born approximation for the self-energies which ensure
the current conservation~\cite{Leuween2006}. The additional choice of the function multiplying matrix elements of NEGFs is taken to guarantee
momentum conservation, so that dephasing does not artifactually enhance the resistance~\cite{Golizadeh-Mojarad2007a}. This form of the retarded and lesser self-energies requires that Eq.~(\ref{eq:retarded}) be solved self-consistently [with the initial guess for ${\bf G}(E)$ being the solution with no dephasing] and then use such ${\bf G}(E)$  to solve the Keldysh equation Eq.~(\ref{eq:keldysh}) directly as the Sylvester equation of matrix algebra. This allows one to compare NEGF calculations~\cite{Golizadeh-Mojarad2007} to SHE experiments all the way to room temperature.  Another simple way to smear sharp features in the local spin density due to quantum coherence is  averaging over an energy interval~\cite{Nikoli'c2005d}, as shown in Fig.~\ref{fig:lk_examples}. For more details about different NEGF-based simple models of dephasing in quantum transport and their effect on momentum and spin relaxation see the chapter 3 by R. Golizadeh-Mojarad and S. Datta in this Volume.

\subsection{Example: Local spin density and spin currents in mesoscopic SHE nanostructures}\label{sec:lk_examples}

\begin{figure}[p]
\centerline{\psfig{file=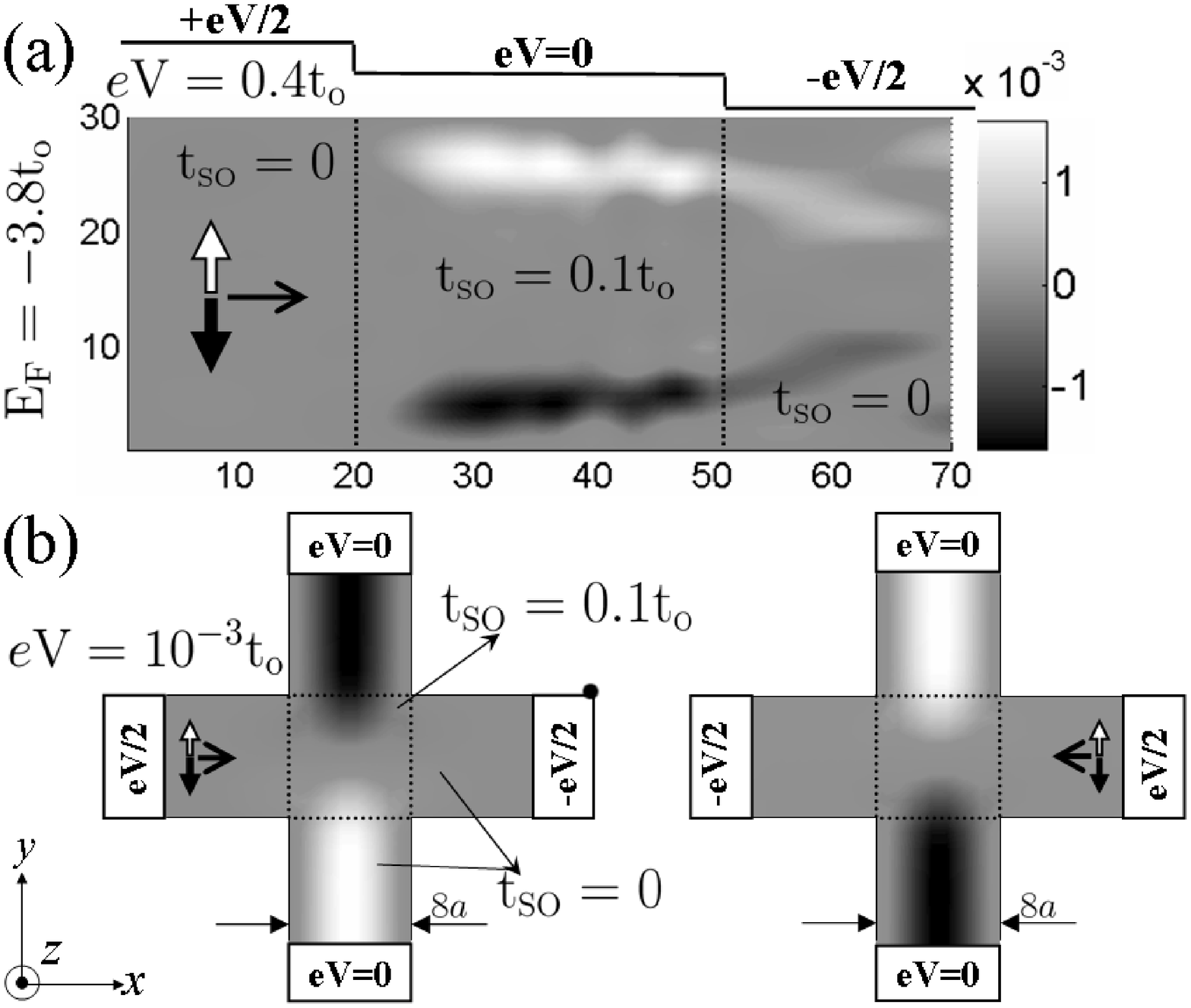,scale=0.45,angle=0}} \vspace{0.5cm}
\centerline{\psfig{file=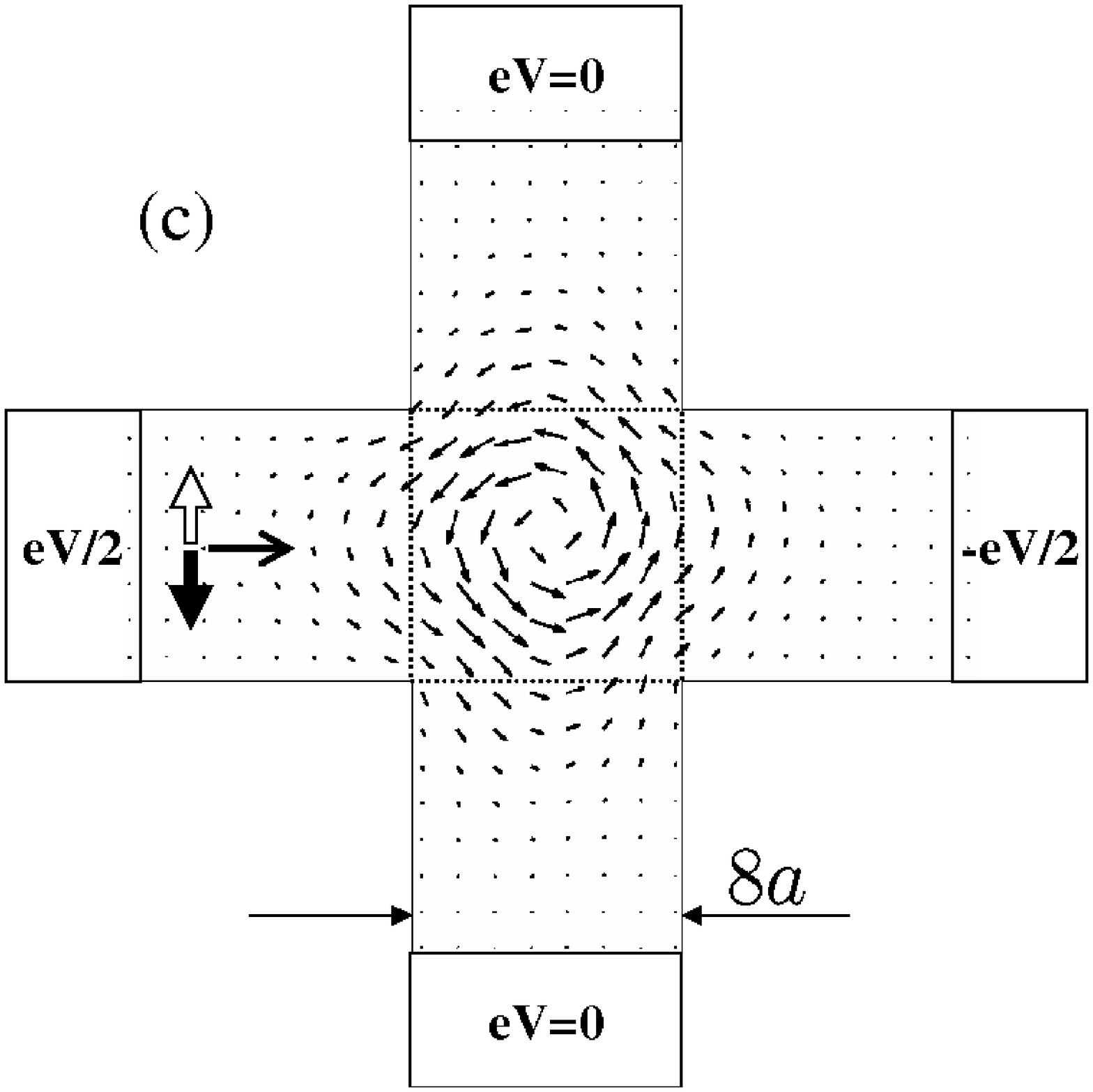,scale=0.32,angle=-90} \hspace{1.0cm} \psfig{file=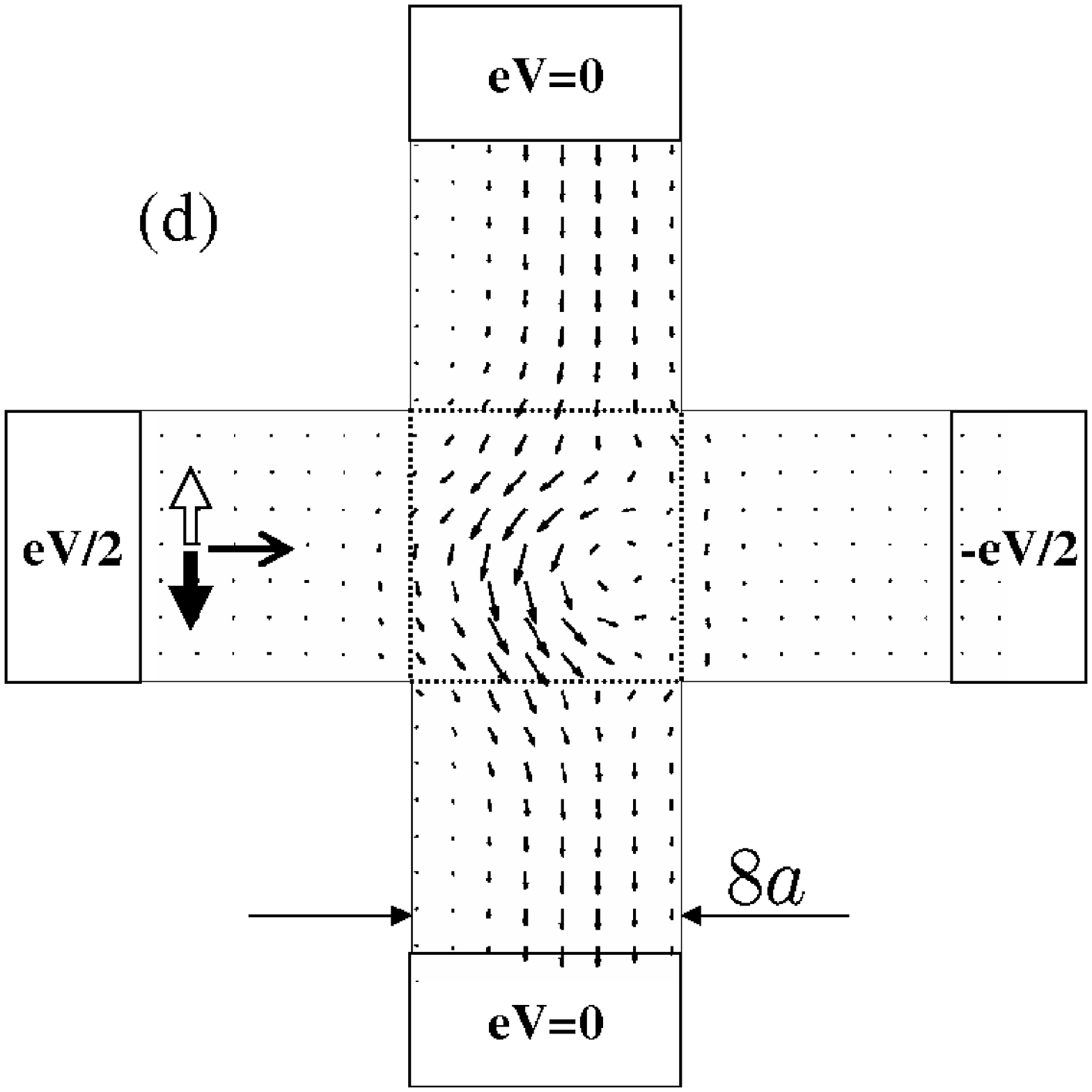,scale=0.32,angle=-90}}
\caption{(a) The out-of-plane component $\left< S^z_{\bf m} \right>$  of the nonequilibrium
spin accumulation induced by  nonlinear ($eV =0.4 t_{\rm O}$) quantum transport  of unpolarized charge current injected from the left lead into a two-terminal
clean 2DEG nanostructure (of size $L=30a > L_{\rm SO}$, $a \simeq 3$ nm)  with the Rashba SO coupling $t_{\rm SO}=0.1t_{\rm O}$ and spin precession length $L_{\rm SO} \approx 15.7a$.  (b) Lateral spin-$\up$ and spin-$\dn$ densities will propagate in opposite directions through the attached transverse ideal  ($t_{\rm SO} = 0$) electrodes to yield a linear response spin-Hall current $I_y^{S_z}$ flowing out of 2DEG ($L=8a < L_{\rm SO}$), which changes  sign upon reversing the bias voltage. (c), (d) The spatial distribution of  local spin currents in ballistic four-terminal 2DEG bridges with the Rashba SO coupling $t_{\rm SO}=0.1t_{\rm O}$ setting $L_{\rm SO} \approx 15.7a$ and linear response bias voltage $eV =10^{-3} t_{\rm O}$. The local spin current is the sum of equilibrium (persistent) spin current in (c), carried by the fully occupied states from $-4t_{\rm O}$ to $E_F - eV/2$, and the nonequilibrium (transport)  spin current in (d) carried by the partially occupied states around the Fermi energy from $\mu_R=E_F - eV/2$ (electrochemical potential of the right reservoir) to  $\mu_L=E_F + eV/2$ (electrochemical potential of the left reservoir).}
\label{fig:lk_examples}
\end{figure}

Figure~\ref{fig:lk_examples}(a) shows NEGF computation~\cite{Nikoli'c2005d} of spin density within the Rashba coupled 2DEG where all
essential features of the experimentally observed SHE in disordered semiconductor devices~\cite{Kato2004b} are
obtained but for perfectly clean sample, such as: the out-of-plane $\left< S^z_{\bf m} \right>$ component of the spin
accumulation develops two peaks of opposite signs at the lateral edges of the 2DEG; upon reversing the bias voltage, the edge peaks flip their sign;
around the left-lead-2DEG interface $\left< S^z_{\bf m} \right>$ is suppressed since unpolarized electrons are
injected at this contact.
By attaching two additional transverse leads at the lateral edges in Fig.~\ref{fig:lk_examples}(b), we show how spin-$\up$ and spin-$\dn$ densities will flow through
those leads in opposite directions to generate spin-Hall current, thereby providing an all-electrical semiconductor-based spin injector using ballistic
nanostructures with intrinsic SO couplings. No discussion of controversial spin currents within the sample is necessary to reach this conclusion. Note that simple picture of opposite spin densities flowing in opposite directions in Fig.~\ref{fig:lk_examples}(b) is obtained in samples smaller than $L_{\rm SO}$, while in larger samples these patterns are more complicated~\cite{Nikoli'c2006} due to multiple spin precession within the sample before spin has a chance to exit through the leads.

The spatial profile of local spin currents corresponding to  Fig.~\ref{fig:lk_examples}(b) is shown in Fig.~\ref{fig:lk_examples}(c),(d) where we separate the integration in Eq.~(\ref{eq:bond_spincurrent_kin}) for $S_z$ bond spin current into two parts $\left<\hat{J}_{\bf mm'}^{S_z}\right>=\left< \hat{J}_{\bf mm'}^{S_z({\rm eq})} \right> + \left< \hat{J}_{\bf mm'}^{S_z({\rm neq})}\right>$:
\begin{eqnarray} \label{eq:integration}
\left<\hat{J}_{\bf mm'}^{S_z}\right> & =  &
\frac{t_{\rm O}}{2}\int\limits_{E_b}^{E_F-eV/2} \frac{dE}{2\pi}{\rm Tr}_{\rm S} \left[\hat{\sigma}_z \left( \vect{G}^{<}_{\bf m'm}(E)-\vect{G}^{<}_{\bf mm'}(E) \right)\right] \nonumber \\
&& + \frac{t_{\rm O}}{2}\int\limits_{E_F - eV/2}^{E_F + eV/2} \frac{dE}{2\pi} {\rm Tr}_{\rm S} \left[\hat{\sigma}_z \left( \vect{G}^{<}_{\bf
m'm}(E)-\vect{G}^{<}_{\bf mm'}(E)\right)\right],
\end{eqnarray}
plotted as panels (c) and (d), respectively. The states from the band bottom $E_b$ to $E_F -eV/2$ are fully occupied, while states in the energy
interval from the electrochemical potential $E_F-eV/2$ ($eV>0$) of the right reservoir to the electrochemical potential $E_F+eV/2$ of the left reservoir
are partially occupied because of the competition between the left reservoir  which tries to fill them and the right reservoir which tries to deplete them.
The spatial distribution of the microscopic spin currents in Fig.~\ref{fig:lk_examples}(c) is akin to the vortex-like pattern of bond spin currents within
the device that would exist in equilibrium $eV_p={\rm const.}$~\cite{Nikoli'c2006}. These do not transport any spin between two points
in real space since their sum over any cross section is zero. Thus, Fig.~\ref{fig:lk_examples}(d) demonstrates that non-zero spin-Hall flux through the  transverse
cross sections is due to only the wave functions (or Green functions) around the Fermi energy. The same sum of $\left< \hat{J}_{\bf mm'}^{S_z({\rm neq})} \right>$ in (d) over arbitrary transverse cross section (orthogonal to the $y$-axis) defines the total spin current which, although not the same on different cross sections within the sample,  flows into the leads where it becomes conserved and it is measurable [e.g., via the inverse SHE~\cite{Hankiewicz2004a}].

\section{COMPUTATIONAL ALGORITHMS FOR REAL$\otimes$SPIN SPACE NEGFS IN MULTITERMINAL DEVICES}\label{sec:green-funct-appr}

While all formulas in Sec.~\ref{sec:spin-resolv-land} for the core NEGF quantities, ${\bf G}(E)$ and ${\bf G}^<(E)$, can be implemented by brute force operations on full matrices, this is typically restricted to small lattices of few thousands sites due to computational complexity of matrix operations. For example, for a system of size $L$ in $d$ dimensions, the computing time scales as $L^{3d}$ while the memory needed scales as $L^{2d}$. By separating system into slices described by much smaller Hamiltonian matrices, and by using the recursive Green function algorithm in serial or parallel implementation~\cite{Drouvelis2006}, the complexity can be reduced drastically to $L^{3d-2}$ scaling of the required computing time.

It is often considered~\cite{Kazymyrenko2008} that the recursive algorithm can be applied only to two-terminal devices and for the computation of total transport quantities in the leads. Some alternative algorithms with reduced computational complexity tailored for multiterminal structures have been proposed recently~\cite{Kazymyrenko2008}, and the recursive algorithm has also been extended to get local charge quantities within the sample~\cite{Cresti2003,Metalidis2005,Lassl2007}. We discuss in Sec.~\ref{sec:recurs-green-funct} how to extend the recursive scheme to four-terminal mesoscopic spin-Hall bridges in Fig.~\ref{fig:setup} that makes possible computation of the spin-Hall conductance on large lattice sizes shown in Fig.~\ref{fig:meso_she}(a). In addition, Sec.~\ref{sec:recurs-gf_spindens} shows how to obtain local spin densities within large two-terminal structures~\cite{Nikoli'c2005d} using the recursive-type approach. Since the starting point of any algorithm to obtain ${\bf G}(E)$ and ${\bf G}^<(E)$ for open finite-size system is computation of self-energies for the infinite part (ensuring continuous spectrum and dissipation) of the system  in the form of semi-infinite electrodes attached to the finite-size sample, we briefly review recent developments in numerical techniques used to compute the surface Green function of semi-infinite leads that generates matrices for ${\bm \Sigma}(E)$ and ${\bm \Sigma}^<(E)$.

\subsection{Numerical algorithms for computing self-energy matrices}\label{sec:lead-self-energy}

Several different numerical techniques are available to evaluate single-particle Green function of a semi-infinite electrode using a localized basis~\cite{Velev2004}. For example, the so-called Ando method~\cite{Ando1991}  computes the surface Green function of a homogeneous lead, consisting of repeating supercell described by the Hamiltonian  $\mathbf{H}_{i,i}$ where supercells are coupled by the Hamiltonian $\mathbf{H}_{i,i-1}$, from its exact Bloch propagating modes at energy $E$. This method offers high precision numerical evaluation of the self-energy and can be generalized~\cite{Khomyakov2005,Rocha2006} or accelerated~\cite{Sorensen2008} to complicated homogeneous periodic systems where  $\mathbf{H}_{i,i-1}$  is not invertible (required in the original Ando algorithm) due to atomistic structure of the leads or more complicated than square tight-binding lattice [such as the honeycomb one~\cite{Chen2007,Zarbo2007}].

An alternative and approximate method, applicable also to inhomogeneous  systems (when the lead does not consist of repeating supercells with identical $\mathbf{H}_{i,i}$), is the recursive (or continued fraction) algorithm for the surface Green function~\cite{Velev2004}. It connects Green function of a given layer to Green functions of neighboring layers, so that starting from the surface layer and repeating this process until the effect of all other layers on the surface layer is taken account yields the surface Green function (in practice, the recursive method is executed to finite depth into the lead where the number of considered neighboring layers affecting the surface layer is set by the convergence criterion).

\begin{figure}[t]
\centerline{\psfig{file=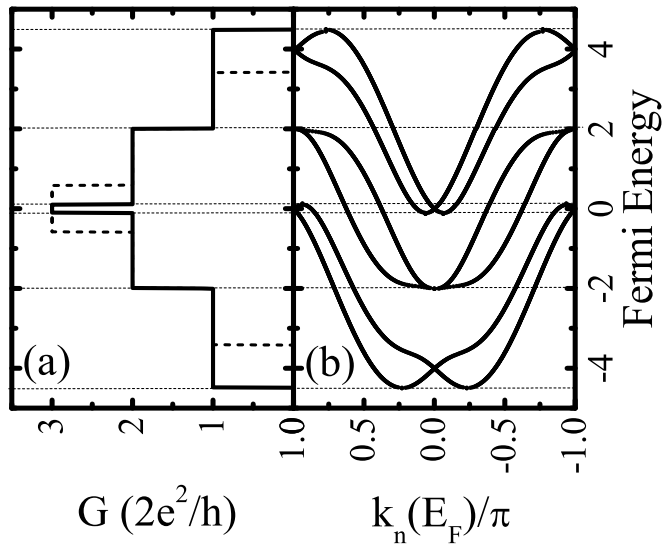,scale=0.35,angle=90}}
  \caption{The conductance quantization (a) in the Rashba spin-split infinite clean quantum wire modeled on the lattice with three sites per cross section. The wire is decomposed  into the finite-size central sample attached to two semi-infinite leads where the Rashba hopping $t_{\rm SO}=t_{\rm O}$ exists in all three segments. The dashed line in (a) corresponds to quantized conductance of the same wire in the absence of the SO coupling $t_{\rm SO}=0$. Panel (b) plots the Rashba-split subband dispersion  responsible for this unusual conductance quantization.}
\label{fig:rashba_lead}
\end{figure}

The Ando method can be executed analytically for simple square lattice lead with no SO coupling~\cite{Datta1995}. For comparing mesoscopic SHE calculations to the bulk SHE description, it is also useful to have a system where SO-coupled leads are attached to identical SO-coupled sample to form an infinite spin-split wire, while two additional ideal (with no spin interactions) leads are attached in the transverse direction~\cite{Sheng2006b}. The self-energies for semi-infinite  electrodes with the Rashba SO coupling can be obtained via the Ando method, with doubled size matrices $\mathbf{H}_{i,i}$ and $\mathbf{H}_{i,i-1}$ to include spin. We employ this method in Fig.~\ref{fig:rashba_lead} to provide a simple example of the conductance quantization through three-channel infinite spin-split quantum wire that can be used as a building block of the four-terminal calculations. The Rashba-split subbands define new conducting
channels whose spin polarization properties can be highly nontrivial for strong SO coupling~\cite{Governale2004}, as discussed in Sec.~\ref{sec:subbands}.
This is in contrast to standard wires where conducting channels are always separable states $|n\rangle \otimes |\sigma \rangle$, which simplifies the analysis of spin injection~\cite{Nikoli'c2005} while also introducing the scattering at the lead-sample interface which can reduce the spin-Hall conductance~\cite{Nikoli'c2005b,Sheng2006b}.

\subsection{Recursive Green function algorithm for total spin and charge currents in multiterminal nanostructures}\label{sec:recurs-green-funct}

\begin{figure}
\centerline{\psfig{file=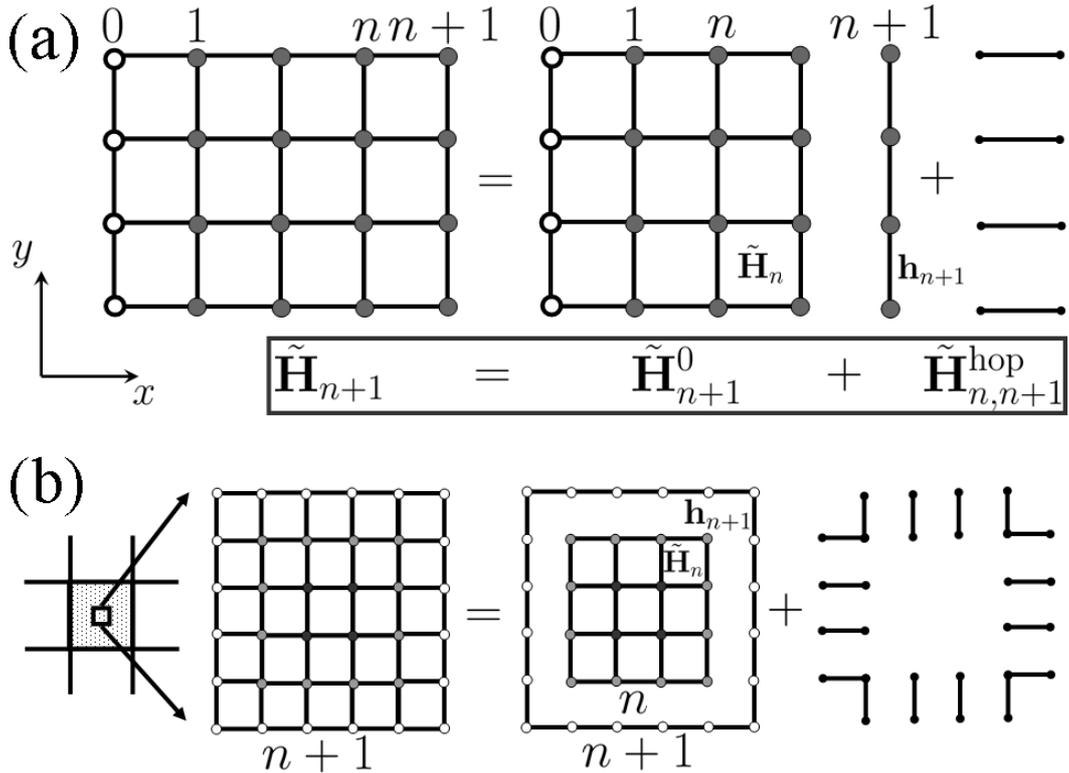,scale=0.55,angle=0}}
  \caption{Illustration of the recursive Green function computational algorithm for: a) two-terminal setup; and  b) four-terminal setup defined on the square tight-binding lattice.}
  \label{fig:recursive_lattice}
\end{figure}

For many transport calculations the full retarded Green function Eq.~(\ref{eq:retarded}) is not required. For instance, to
compute the transmission matrix~(\ref{eq:transmission}) from lead $q$ to lead $p$, one only needs the Green function matrix
elements $({\bf G})_{{\mathbf{mm}^\prime},{\sigma\sigma^\prime}}$ where the sites $\mathbf{m}$ belong to the lead $p$ edge
supercell and the sites $\mathbf{m}^\prime$ belong to the first supercell of the lead $q$. We first discuss the recursive algorithm~\cite{Drouvelis2006} that
evaluates only this portion $\mathbf{G}_{N+1,0}$  of the full retarded Green function matrix for the two-probe setup in  Fig.~\ref{fig:recursive_lattice}(a). The sample is
divided into $N$ layers, and to account for the effect of the leads, their self-energies are added onto the layers $0$ and $N+1$. These two
layers are the first principal layers of the  two leads. It is important to emphasize that adding the self-energies directly onto the sample
in more complicated cases than simple square lattice can lead to incorrect result for the transmission eigenvalues.

The recursive scheme starts by adding  the self-energy of the left lead ${\bm \Sigma}_L$ onto
its principal layer directly connected to the sample. The Hamiltonian of any of the $N+2$ blocks
is labeled by $\mathbf{h}_n$. The retarded Green function matrix of the $n$-th isolated block is
$\mathbf{g}_n=(E-\mathbf{h}_n-i\eta)^{-1}$. The hopping matrix connecting blocks $n$ and $n+1$ is
\begin{equation}\label{eq:hop_nnp1}
\mathbf{H}_{n,n+1}^{\rm hop}=
\left(
\begin{array}{cc}
0  & \mathbf{H}_{n,n+1} \\
\mathbf{H}_{n+1,n}                 & 0
\end{array}
\right).
\end{equation}
We also introduce $\tilde{\mathbf{H}}_n$ as the matrix which is the sum of the Hamiltonian of $n+1$ blocks and the
lead self-energy added to the $0$-th block. The corresponding retarded Green function matrix is $\mathbf{G}_n=(E-\tilde{\mathbf{H}}_n+i\eta)^{-1}$.
One can add the next isolated block $\mathbf{h}_{n+1}$ to $\tilde{\mathbf{H}}_n$ and obtain $\tilde{\mathbf{H}}_{n+1}^{(0)}=\tilde{\mathbf{H}}_n\oplus \mathbf{h}_{n+1}$.

Assuming that the Green function of $n$ blocks is known, we add the $n+1$-th block  to
$\tilde{\mathbf{H}}_n$ without turning on the coupling $\mathbf{H}_{n,n+1}^{\rm hop}$ between them.
The new Hamiltonian, is block diagonal
\begin{equation}\label{eq:Hnp1_0}
\tilde{\mathbf{H}}_{n+1}^{(0)}=
\left(
\begin{array}{cc}
\tilde{\mathbf{H}}_n & 0 \\
0                    & \mathbf{h}_{n+1}
\end{array}
\right),
\end{equation}
whose inversion leads to
\begin{equation}\label{eq:gf0_np1}
\mathbf{G}_{n+1}^{(0)}=
\left(
\begin{array}{cc}
\mathbf{G}_n & 0 \\
0                    & \mathbf{g}_{n+1}
\end{array}
\right).
\end{equation}
To find the Green function $\mathbf{G}_{n+1}$ it is necessary to add the matrix elements connecting
the $n+1$-th block to $\tilde{\mathbf{H}}_n$
\begin{equation}\label{eq:Hnp1}
\tilde{\mathbf{H}}_{n+1}=
\left(
\begin{array}{cc}
\tilde{\mathbf{H}}_n & \mathbf{H}_{n,n+1} \\
 \mathbf{H}_{n+1,n}            & \mathbf{h}_{n+1}
\end{array}
\right).
\end{equation}
The Green function of $n+2$ blocks can be found by using the Dyson equation
\begin{equation}\label{eq:dyson}
\mathbf{G}_{n+1}=\mathbf{G}_{n+1}^{(0)}+\mathbf{G}_{n+1}^{(0)}\mathbf{H}_{n,n+1}^{\rm hop}\mathbf{G}_{n+1}.
\end{equation}
All matrices in this equation are of the dimension $n+2$ in the block coordinate space. Equation~(\ref{eq:dyson}) can be used
to obtain the recurrence relationships for the Green function blocks
we are interested in. To get the conductance of the two-terminal device we need to compute
$\left< N+1 \left| \mathbf{G}_{N+1} \right| 0 \right> \equiv \mathbf{G}_{N+1}(N+1,0)$ which connects the
right and the left lead.

In the multiterminal case, we choose to start from the Hamiltonian of the block of sites in
the center of the sample and then add one-by-one additional blocks of the same shape until we reaches the
attached electrodes, as sketched in Fig.~\ref{fig:recursive_lattice}(b). Therefore, in this case we need the
Green function elements $\mathbf{G}_{N+1}(N+1,N+1)$. From Eq.~(\ref{eq:dyson}) we find
\begin{equation}\label{eq:dyson_p1}
\mathbf{G}_{n+1}(n+1,n+1)=\mathbf{G}_{n+1}^{(0)}(n+1,n+1)+
\left< n+1 \left|
\mathbf{G}_{n+1}^{(0)} \mathbf{H}_{n,n+1}^{\rm hop} \mathbf{G}_{n+1}
\right| n+1 \right>.
\end{equation}
By inserting the identity matrix in the subspace of block coordinates into Eq.~(\ref{eq:dyson_p1})
\begin{equation}\label{eq:dyson_p2}
\mathbf{I}=\sum_{m} \left|  m \right> \left< m  \right|,
\end{equation}
the second term in Eq.~(\ref{eq:dyson_p1}) becomes
\begin{equation}\label{eq:dyson_p3}
\left< n+1 \left|
\mathbf{G}_{n+1}^{(0)}
\sum_{m} \left|  m \right> \left< m  \right|
\mathbf{H}_{n,n+1}^{\rm hop}
\sum_{m} \left|  l \right> \left< l \right|
\mathbf{G}_{n+1}
\right| n+1 \right>.
\end{equation}
It is easy to see from Eqs.~(\ref{eq:hop_nnp1}),~(\ref{eq:gf0_np1}), and~(\ref{eq:Hnp1}) that
\begin{subequations}\label{eq:dyson_p4}
\begin{eqnarray}
\left< n+1 \right|  \mathbf{G}_{n+1}^{(0)}  \left| m \right>
& = & \delta_{n+1,m} \mathbf{g}_{n+1}, \\
\left< m \right|  \mathbf{H}_{n,n+1}^{\rm hop} \left| l \right>
& = & \delta_{m,n}\delta_{l,n+1} \mathbf{H}_{n,n+1}
+ \delta_{m,n+1}\delta_{l,n} \mathbf{H}_{n+1,n} .
\end{eqnarray}
\end{subequations}
Equation~(\ref{eq:dyson_p4}) can be substituted into Eq.~(\ref{eq:dyson_p3}) to yield the Dyson Eq.~(\ref{eq:dyson})
for $\mathbf{G}_{n+1}(n+1,n+1)$
\begin{equation}\label{eq:dyson_np1_np1}
\mathbf{G}_{n+1}(n+1,n+1)=\mathbf{g}_{n+1}+\mathbf{g}_{n+1} \mathbf{H}_{n+1,n}
\mathbf{G}_{n+1}(n,n+1),
\end{equation}
in which $\mathbf{G}_{n+1}(n,n+1)$ is an unknown quantity. By computing the matrix
elements of the terms in the Dyson Eq.~(\ref{eq:dyson}) between $\left| \!\, n \right>$ and
$\left| \!\, n+1 \right>$ we find that
\begin{equation}\label{eq:dyson_n_np1}
\mathbf{G}_{n+1}(n,n+1)=\mathbf{G}_{n}(n,n)\mathbf{H}_{n,n+1}\mathbf{G}_{n+1}(n+1,n+1).
\end{equation}
Finally, combining Eqs.~(\ref{eq:dyson_np1_np1}) and (\ref{eq:dyson_n_np1}) leads the following recurrence
relationship
\begin{equation}\label{eq:rec_np1_np1}
\mathbf{G}_{n+1}(n+1,n+1) =
\frac{\mathbf{I}_{n+1}}{ \mathbf{g}_{n+1}^{-1} -
\mathbf{H}_{n+1,n} \mathbf{G}_{n}(n,n) \mathbf{H}_{n,n+1} } .
\end{equation}
Here the matrix $\mathbf{I}_{n+1}$ is the identity matrix in the subspace of the $n+1$-th block and
the  denominator is the matrix to be inverted.

If the recursive scheme in Fig.~\ref{fig:recursive_lattice}(b) is used, we need to obtain
 $\mathbf{G}_{N+1}(N+1,N+1)$ from which the transmission matrix can be computed.
For the two-terminal case, the recursive scheme sketched in Fig.~\ref{fig:recursive_lattice}(a) is more
efficient computationally. Computing the matrix elements of the Dyson
Eq.~(\ref{eq:dyson}) between the states $\left| \! \, n+1 \right> $
and $\left| \! \, m \right>$ gives
\begin{equation}\label{eq:rec_np1_m}
\mathbf{G}_{n+1}(n+1,m)=
\mathbf{G}_{n+1}(n+1,n+1)\mathbf{H}_{n+1,n}\mathbf{G}_{n}(n,m)
\end{equation}
where $m<n+1$.

We summarize the general recursive Green function algorithm for the computation of (spin-resolved) transmissions, and therefore spin or charge conductances,  in the form of the following steps:
\begin{enumerate}\label{item:algorithm_rec}

\item Calculate the self-energies for the leads, as described in Section~\ref{sec:lead-self-energy}.

\item Identify the $N+2$ blocks for the recursive method as shown in Fig.~\ref{fig:recursive_lattice}
(the number of sites of each block might be different).

\item Initialize the recurrence by computing $\mathbf{G}_{0}(0,0)\equiv \mathbf{g}_0$, where
$\mathbf{g}_0=(E-\tilde{\mathbf{H}}_{0}-i\eta)^{-1}$ is the Green function of the first block.
This block has the self-energy added to it in the two-terminal case. In the multiterminal case, the
first block can be chosen to be a rectangle of the size $N_x^{(0)}\times N_y^{(0)}$ sites, as illustrated
in Fig.~\ref{fig:recursive_lattice}(b). A difficulty may arise in the multiterminal cases where the
matrix $E-\mathbf{h}_0$ can turn to be singular for certain choices of the shape of the sample or its
parameters.

\item Using the formulas~(\ref{eq:rec_np1_np1}) and (\ref{eq:rec_np1_m}) compute the Green functions
$\mathbf{G}_{N+1}(N+1,N+1)$ and $\mathbf{G}_{N+1}(N+1,0)$.

\item Obtain the retarded Green function matrix elements $\mathbf{G}_{pq}$ connecting each pair of
leads $p$ and $q$, either from $\mathbf{G}_{N+1}(N+1,0)$ in the two-terminal case, or

$\mathbf{G}_{N+1}(N+1,N+1)$ in the multiterminal case.
\item Compute the transmission matrix using  Eq.~(\ref{eq:transmission}).

\end{enumerate}

\subsection{Recursive Green function algorithm for local spin and charge densities}\label{sec:recurs-gf_spindens}

The computation of nonequilibrium spin and charge densities using Eq.~(\ref{eq:spin_density}) and~(\ref{eq:charge_density}), respectively,
requires to obtain $\mathbf{G}^<$. This has the highest computational complexity since one has to perform the matrix multiplication
in Eq.~(\ref{eq:keldysh}). Moreover, this matrix multiplication has to be done for all the energy points in  Eq.~(\ref{eq:spin_density}).
Here we introduce a method for recursive calculation of the spin and charge nonequilibrium densities within the NEGF formalism which
makes it possible to greatly reduce this computational complexity and, thereby, analyze spin densities in sizable samples~\cite{Nikoli'c2005d}.

In order to illustrate the essential steps of this method, it is enough to consider a simple two-probe setup in Fig.~\ref{fig:recursive_lattice}(a).
The two-probe system considered here is made of $N+2$ blocks, where the $0$-th block is the first left lead supercell onto which the left lead self-energy
is added, and $N+1$-th block is the first right lead supercell onto which the right lead self-energy is added. The voltage applied onto the left and right
lead is  $V_L$ and $V_R$, respectively. The lesser self-energy matrix in Eq.~(\ref{eq:keldysh}) can be rewritten as
\begin{eqnarray}\label{eq:sigma_less_2probe}
{\bm \Sigma}^{<}(E) & = & i{\bm \Gamma}_L(E-eV_L)f(E-eV_L)+i{\bm \Gamma}_R(E-eV_R)f(E-eV_R) \nonumber \\
& = & \left(
\begin{array}{ccc}
i{\bm \Gamma}_L(E-eV_L)f(E-eV_L)  & \dots & 0 \\
\dots & \dots & \dots \\
0     & \dots  & i{\bm \Gamma}_R(E-eV_R) f(E-eV_R)
\end{array}
\right).
\end{eqnarray}
where the only non-zero elements are the first and the last on the diagonal. This simplifies greatly the matrix multiplication in Eq.~(\ref{eq:keldysh}),
since only multiplication by the non-zero matrix elements of ${\bm \Sigma}^{<}$ is required
\begin{subequations}\label{eq:nonzero_sigma_less}
\begin{eqnarray}
{\bm \Sigma}^{<}(0,0) & = & i{\bm \Gamma}_L(E-eV_L)f(E-eV_L), \\
{\bm \Sigma}^{<}(N+1,N+1) & = & i{\bm \Gamma}_R(E-eV_R) f(E-eV_R).
\end{eqnarray}
\end{subequations}
In order to calculate the nonequilibrium spin density Eq.~(\ref{eq:spin_density}) or charge density
Eq.~(\ref{eq:charge_density}), only the diagonal elements of $\mathbf{G}^{<}(E)$ are required.
Using Eqs.~(\ref{eq:sigma_less_2probe}) and (\ref{eq:nonzero_sigma_less}) in Eq.~(\ref{eq:keldysh}),
leads to
\begin{eqnarray}
\mathbf{G}^{<}(m,m) & = & \mathbf{G}(m,N+1){\bm \Sigma}^{<}(N+1,N+1)\mathbf{G}^\dagger(N+1,m) \nonumber \\
&+& \mathbf{G}(m,0){\bm \Sigma}^{<}(0,0)\mathbf{G}^\dagger(0,m),
\end{eqnarray}
where $m$ labels the coordinates of the $m$-th block.

The retarded Green function matrices $\mathbf{G}(m,N+1)$ and $\mathbf{G}(m,0)$ can be found using the Dyson
Eq.~(\ref{eq:dyson}) introduced in Sec.~\ref{sec:recurs-green-funct}. To find $\mathbf{G}(m,N+1)$, one
starts the recurrence from the left lead and finds
\begin{equation}\label{eq:rec_g_m_np1}
\mathbf{G}_{n+1}(m,n+1) = \mathbf{G}_{n}(m,n) \mathbf{H}_{n,n+1} \mathbf{G}_{n+1}(n+1,n+1),
\end{equation}
where $m \leq n$. The Green function block $\mathbf{G}_{n+1}(n+1,n+1)$ can be obtained from
 Eq.~(\ref{eq:rec_np1_np1}).
To find $\mathbf{G}(m,0)$, the recurrence has to start from the right lead and we get
\begin{equation}\label{eq:rec_g_m_nm1}
\mathbf{G}_{n-1}(m,n-1)=\mathbf{G}_{n}(m,n) \mathbf{H}_{n,n-1} \mathbf{G}_{n-1}(n-1,n-1),
\end{equation}
where $m \geq n $. The system of equations is finally closed by using
\begin{equation}\label{eq:rec_g_nm1_nm1}
 \mathbf{G}_{n-1}(n-1,n-1) =
\frac{\mathbf{I}_{n-1}}{\mathbf{g}_{n-1}^{-1}-
\mathbf{H}_{n-1,n}\mathbf{G}_{n}(n,n)\mathbf{H}_{n,n-1}}.
\end{equation}
In this method, one has to compute recursively the retarded Green functions
$\mathbf{G}(m,N+1)$ and $\mathbf{G}(m,0)$ for all $m=\overline{0,N+1}$.
It is not as computationally costly as the full Green function matrix inversion, and it can
be extended to compute bond spin and charge currents. In that case, we also have to find $\mathbf{G}(m\pm 1,N+1)$ and
$\mathbf{G}(m\pm 1,0)$.

\section{CONCLUDING REMARKS}

The pure spin currents, which arise when equal number of spin-$\up$ and spin-$\dn$ electrons propagate in opposite directions so that net charge current carried by them is zero, have recently emerged as one of the major topics of both metal and semiconductor-based spintronics. This is due to the fact that they offer new realm to explore fundamental spin transport phenomena in solids, as well as to construct new generation of spintronic devices with greatly reduced power dissipation and new functionality in transporting either classical or quantum (such as flying spin qubits) information. Their generation (which is currently experimentally achieved mostly through spin pumping, non-local spin injection using lateral spin valves, and SHE), control, and demanding detection typically involves quantum-coherent spin dynamics.  In particular, the recently discovered SHE and the inverse SHE have offered some of the most efficient scheme to generate and detect pure spin currents, respectively. However, they have also posed numerous challenges for their theoretical description and device modeling since conventionally defined spin current operator does not satisfy standard continuity equation for spin when intrinsic SO couplings (which act as homogeneous internal momentum-dependent magnetic fields affecting the band structure and causing spin precession) are present. Even with modified spin current definitions, this fundamental problem makes difficult easy connection between theoretically computed spin conductivities of infinite homogeneous systems and usually experimentally measured  spin densities for devices with arbitrary boundaries and attached external electrodes.
An attempt to give up on the theoretical description of spin flow through spin current altogether, by using spin density within the diffusion equation formalism (which is expected to provide a complete and physically meaningful description of spin and charge diffusive transport in terms of position and time dependent spin and charge densities), leads to another set of obstacles. That is, the explicit solution of the diffusion equation strongly depends on the boundary conditions (extracted for charge diffusion from conservation laws) where spin non-conservation, manifested as ballistic spin precession at the edges, cannot be unambiguously matched with the diffusive dynamics in the bulk captured  by the diffusion equation for length scales much longer than the mean free path.

This chapter of the Handbook provides a summary of such fundamental problems in spin transport through SO-coupled semiconductors and possible resolution through spin-dependent NEGF formalism that can handle {\em both} ballistic and diffusive regimes, as well as phase-coherent and dephasing effects, in realistic multiterminal nanostructures operating at different mobilities and temperatures. The central quantities of this approach---total spin currents flowing out of the device through ideal (interaction free) electrodes, local spin densities within the sample and along the edges, and local spin currents (whose sums within the leads gives total spin currents)---can be used to understand experiments or model novel spintronics devices. For example, NEGF approach is indispensable~\cite{Hankiewicz2004a} for computational modeling of generator-detector experimental setups~\cite{Brune2008,Roth2009} where direct and inverse SHE multiterminal bridges are joined into a single circuit to sense the pure spin-Hall current through voltages induced by its flow through regions with SO interactions. The chapter also provides extensive coverage of relevant technical and computational details, such as: the construction of retarded and lesser nonequilibrium Green functions for SO-coupled nanostructures attached to many electrodes; computation of relevant self-energies introduced by the lead-sample interaction; and accelerated algorithms which make possible spin transport modeling in device whose size is comparable to the spin precession length of few hundreds of nm where SHE response to injected unpolarized charge current is expected to be optimal~\cite{Nikoli'c2005b,Moca2007}.

\bigskip

\addcontentsline{toc}{section}{Acknowledgments}
{\bf Acknowledgments}

\bigskip

We acknowledge numerous insightful discussions with \.{I}. Adagideli, S. Datta, J. Fabian, J. Inoue, S. Murakami, N. Nagaosa,
J. Nitta, K. Nomura, M. Onoda, E. I. Rashba, and J. Sinova. This work was supported by NSF Grant No. ECCS 0725566, DOE Grant No.
DE-FG02-07ER46374, and the Center for Spintronics and Biodetection at the University of Delaware.

\addcontentsline{toc}{section}{References}


\end{document}